\documentclass[a4paper,11pt]{article}
\pdfoutput=1 

\usepackage{jheppub} 
\usepackage[utf8]{inputenc}
\usepackage{setspace}

\usepackage{hyperref}
\definecolor{refkey}{gray}{0.45}
\definecolor{labelkey}{RGB}{155,48,48}
\definecolor{UI_blue}{RGB}{32, 64, 151}
\definecolor{UI_red}{RGB}{187, 62, 24}
\definecolor{UI_blue2}{RGB}{0, 84, 147}
\definecolor{UI_red2}{RGB}{159, 32, 66}
\definecolor{UI_gray}{RGB}{169, 169, 169}
\definecolor{UI_sepia}{RGB}{112, 66, 20}
\definecolor{UI_bittersweet}{RGB}{254, 111, 94}
\definecolor{UI_emerald}{RGB}{80, 200, 120}
\definecolor{UI_olivegreen}{RGB}{181, 179, 92}
\definecolor{UI_cadetblue}{RGB}{95, 158, 160}
\definecolor{UI_fuchsia}{RGB}{255, 0, 255}
\definecolor{UI_midnightblue}{RGB}{25, 25, 112}
\definecolor{UI_royalblue}{RGB}{0,35, 102}
\definecolor{UI_periwinkle}{RGB}{204, 204, 255}
\definecolor{UI_redorange}{RGB}{255, 83, 73}
\definecolor{UI_brickred}{RGB}{203,65,84}	
\definecolor{UI_forestgreen}{RGB}{34, 139, 34}
\definecolor{UI_tan}{RGB}{210,180,140}	
\definecolor{UI_burlywood}{RGB}{222,184,135}
\definecolor{UI_burlywood}{RGB}{192,64,0}
\definecolor{UI_darkorchid}{RGB}{153,50,204}

\usepackage{mathtools}
\usepackage{physics}
\usepackage{simplewick}
\usepackage[obeyFinal]{todonotes}

\def\Tr{\text{Tr}}

\def\beq{\begin{eqnarray}}\def\eeq{\end{eqnarray}}
\def\be{\begin{equation}}\def\ee{\end{equation}}

\def\mes[#1]{d^{3}{#1}}

\def\del{\partial}

\newcommand{\half}{\frac{1}{2}}

\def\del{\partial}

\def\order{\ensuremath{\mathcal{O}}}

\reversemarginpar

\author[a,b]{Upamanyu Moitra,}
\author[b]{Sunil Kumar Sake,}
\author[b]{Sandip P. Trivedi,}

\affiliation[a]{\it International Centre for Theoretical Physics,\\ Strada Costiera 11, Trieste 34151, Italy\\  }
\affiliation[b]{\it Department of Theoretical Physics,
	Tata Institute of Fundamental Research,\\  Colaba, Mumbai, India, 400005\\}

\emailAdd{umoitra@ictp.it}
\emailAdd{sunil.sake@tifr.res.in}
\emailAdd{sandip@theory.tifr.res.in}

\vspace{1cm}

\abstract{ We discuss JT gravity in AdS and dS space in the second order formalism. For the  pure dS JT theory without matter, we show that the path integral gives rise in general to the Hartle-Hawking wave function which describes an arbitrary number of disconnected universes produced  by tunnelling  ``from nothing", or to transition amplitudes which describe  the tunnelling of an initial state consisting of several  contracting universes  to a final state  of several expanding universes. These processes can be described by a hologram consisting of Random Matrix Theory (RMT) or, we suggest, after   some modification  on the gravity side, by   a hologram  with the  RMT being replaced by  SYK theory. In the presence of matter, we discuss the double trumpet path integral and argue that with suitable twisted boundary conditions, a divergence in the moduli space integral  can be avoided and the system can tunnel  from a contracting phase to an expanding one avoiding a potential big bang/big crunch singularity.  The resulting spectrum of quantum perturbations which are produced   can exhibit   interesting departures from scale invariance. We also  show that the divergence in moduli space can be avoided for suitable correlators which involve different  boundaries in the AdS/dS cases, and suggest that a hologram consisting of the SYK theory with additional matter could  get rid of these divergences   in general. Finally, we analyse the AdS double trumpet geometry and show that going to the micro-canonical ensemble instead of the canonical one, for the spectral form factor, does not get rid of the divergence in moduli space.}

\title{Aspects of Jackiw-Teitelboim Gravity in Anti-de Sitter and de Sitter spacetime}
\preprint{\parbox{3cm}{TIFR/TH/22-1}}

\begin{document}
	\maketitle
	\flushbottom
	\vskip 10pt
	
	\vskip 20pt
	\section{Introduction}
	\label{intro}
	Fascinating connections between a class of spin systems and  two dimensional gravity have been uncovered   in the past few years. The SYK model \cite{Sachdev:1992fk,Kitaev-talks:2015} is the best studied example of such  a spin system and  exhibits an interesting pattern of symmetry breaking, with its  low-energy dynamics being governed by Goldstone-like modes described  by the Schwarzian  action. The same  Schwarzian theory arises  in JT  gravity \cite{JACKIW1985343,Teitelboim:1983ux}  in two dimensional Anti-de Sitter space (AdS)  which in turn  has been shown to quite universally describe   the near-horizon dynamics of a wide
	class of near-extremal black holes, { \cite{nayak,Moitra:2019bub, Moitra:2018jqs, Ghosh:2019rcj}. }
	
	JT gravity in two dimensional de Sitter space (dS) is another interesting system to study. One can hope to use the simplicity of this theory for  understanding some of the conceptually interesting and deep questions of  de Sitter space in general.  In \cite{Maldacena:2019cbz,Cotler:2019nbi} it was proposed that  JT gravity in dS space can also be described by a version of Random Matrix theory which arises 
	in the study of the AdS case, \cite{Saad:2019lba},  and which is related to the low-energy sector of the SYK theory.

	This paper has two motivations. First, to study the  behaviour of JT gravity in dS space in more detail, including  the pure JT theory and also the theory with extra matter which gives rise to propagating degrees of freedom in the bulk.  Second, to analyse certain divergences which arise both in the AdS and dS cases for JT gravity coupled to matter, when we consider the theories at higher genus or with multiple boundaries, as has been discussed in earlier work, \cite{Moitra:2021uiv}, in more detail. 
	
	Some of the key results of the paper are as follows.

	We consider the pure JT theory in dS space and discuss the path integral in the second order formalism. We argue that there is a non-trivial amplitude in the quantum theory   for producing a single universe, or an arbitrary number of disconnected universes, by ``tunnelling from nothing".
	The path integral is carried out along  the contour  discussed in \cite{Maldacena:2019cbz}, and  involves  in general an intermediate ``-AdS$_2$" geometry  with two time-like directions, i.e. with signature $(0,2)$. It  gives  rise to the Hartle-Hawking (HH) wave function for producing one or several disconnected universes, or alternatively,  
 depending on the analytic continuation which is carried out in the vicinity of each boundary, it gives the transition amplitude for $n_-$ initial disconnected universes to tunnel to $n_+$ disconnected universes in the future, with $n_-,n_+$ taking arbitrary values. The higher genus or additional boundary amplitudes are  suppressed by a factor of $e^{\chi S_0}$, where $\chi=2-2H-B$ is the Euler character of the intermediate geometry and $B,H$ are the number of boundaries and handles. $S_0$ is related, upto a factor of $2$  to the entropy of the extremal black hole in $4$ dimensions from which the JT theory arises after dimensional reduction. This  quantisation of dS JT gravity  therefore leads us to the dizzying picture of a multiverse, where 
	  quantum tunnelling can change the number of universes.  These results also agree with those obtained in the first order formalism, \cite{Cotler:2019nbi}. 

We also discuss the proposal,  \cite{Maldacena:2019cbz,Cotler:2019nbi}, mentioned above, that  dS JT theory has a hologram consisting of Random Matrix theory (RMT) in the double scaled limit considered in \cite{Saad:2019lba}. The RMT hologram lives  on the various spatial boundaries of the disconnected universes involved in the HH wave function or the transition amplitudes, with the matrix giving rise to spatial translations along each boundary. We suggest that in fact this proposal can be further extended with the SYK theory being the hologram instead of RMT. The SYK model and RMT should approximately agree when the (renormalised) length of each boundary is big in units of the scale $J$,  which is the single energy scale that characterises the SYK model. This extension should correspond to adding extra degrees of freedom on the gravitational side which remain to be understood better \footnote{ See \cite{Gao:2021uro,Blommaert:2021fob} for some discussion on this issue in AdS space.}. In fact this is an interesting proposal to consider both in the context of 2 dim. gravity in dS and AdS space. For some discussion along these lines see also, \cite{Cotler:2019nbi}.

 Another topic we discuss in the paper is to  consider an orbifold of dS space, obtained by making an identification  along a spatial direction. 
	The  spatial direction shrinks to vanishing size at some time in the orbifold theory giving rise to an interesting toy model for a big crunch or big bang singularity. Once matter is added we argue, based on the method of images, that  the classical theory has a singularity  where the dilaton, $ \phi\rightarrow -\infty $. The  interpretation  that a sufficiently negative value of the dilaton corresponds to a singularity is   motivated by thinking about  JT theory as the dimensional reduction of a near extremal black hole in $4$ dim. dS space, with the dilaton being the radius of the $S^2$. One might wonder if quantum effects can cure this singularity. We study this question in the semi-classical limit where quantum effects of conformal matter are included and find that the singularity persists along  a spacelike locus in the resulting theory. 
	
	Next, we turn to the full quantum theory where the gravity-dilaton system is also quantised. After carrying out a path integral quantisation of the ``double trumpet" geometry  we find that the system can have  a well defined transition amplitude for  the universe to start  off as a contracting dS spacetime  in the far past and   tunnel to an  expanding dS phase in the far future.  The resulting final state wave function for the matter fields  depends on the initial conditions for matter. And the spectrum of perturbations, which is the analogue of CMB perturbations in this toy model,  shows interesting departure from scale invariance which can persist up to length scales much smaller than the size of the universe. This toy model  therefore   illustrates the possibility of a cosmology,  alternative to the big bang, where the expanding universe arises after   tunnelling  from an  initial smooth spacetime with potentially observable consequences in the spectrum of quantum perturbations which are produced at late times. 
	
	More precisely, we find  the  well defined transition amplitude, referred to in the previous paragraph,  arises only if the massless scalar fields we study have appropriate twisted boundary conditions along the spatial direction. In the quantum theory, the path integral involves a sum over a modulus, which is related to the size of the spatial direction,  and if  the matter boundary conditions are periodic instead, without a twist,  the integral over moduli space blows up, as was discussed earlier in \cite{Moitra:2021uiv},  and the path integral is ill-defined. With twisted boundary conditions too, while the divergence is avoided for the particular tunnelling amplitude we analyse, it is likely to arise at higher genus or in the case with additional boundaries. 
	 
 The divergence in moduli space, or their absence, is also a feature of the path integral with matter in AdS space,  instead of dS space, since the path integrals are closely connected to each other.  We have mentioned above in the discussion of the pure JT theory the possibility of a boundary hologram being the SYK model. In the presence of extra matter, to control the divergences in moduli space in general, we suggest that one can consider a further extension of this proposed hologram  where additional light matter is added to the SYK theory. We discuss how this can be done and argue that the resulting quantum mechanical system should then give finite results. A proper understanding of how this works  on the gravitational side is left for the future. 

Two other results we obtain are worth mentioning here. 
For massless scalar fields we find that even with periodic boundary conditions the divergence in moduli space can be avoided if one is calculating the path integral  with an appropriate  number of cross-boundary correlators connecting the different boundaries.  The analysis we carry out is  in AdS space and   is similar to what was found in {\cite{Stanford:2020wkf,Stanford:2021bhl}} with massive fields. We also analyse with considerable care the double trumpet  path integral
with matter in AdS space and argue that changing ensembles and going from the canonical to the microcanonical one, does not allow us to control this divergence and extract a finite result.

The paper is organised as follows.  In section \ref{basup} we elaborate on the classical solutions of JT gravity.  In particular, in  the JT theory in de Sitter  space with  a black hole we verify  that the second law of thermodynamics is satisfied if the matter satisfies the null energy condition, when one includes both the cosmological and black hole horizons. In section \ref{dssemicla} we study the semiclassical theory of JT gravity with matter in the orbifold backround mentioned above. In section \ref{moddivs}, we discuss the full quantisation of JT gravity in the presence of matter and the various related points concerning analytic continuations from AdS, multi-boundary generalization, connections to Random matrix theory etc. In section \ref{dswithmatter} we discuss the double trumpet geometry in the presence of matter with twisted boundary conditions and also correlations functions with insertions on both boundaries.  In \ref{diskdtsff}, we discuss that spectral form factor and its behaviour in the micro-canonical ensemble. Section \ref{conclusion} ends with a discussion about SYK theory coupled to additional matter as  a possible hologram to JT gravity with matter in dS and some open questions.   Appendix \ref{canquant} contains detailed discussion about attempts to canonically quantise the JT theory in AdS/dS space. Appendix \ref{amainadsdt},\ref{fdsdt} contains some important details related to section \ref{dswithmatter}.  including the calculation of the 4 point OTOC in the double trumpet geometry. 
	Appendix \ref{sffapp} is has more details concerning section \ref{diskdtsff}.

Some important references  pertaining to this paper include, \cite{Vilenkin:2021awm,Vilenkin:1986cy,Witten:2021nzp,Suzuki:2021zbe,Castro:2021fhc,Moitra:2021uiv,Narayan:2020pyj,Hartman:2020khs,Balasubramanian:2020xqf,Chen:2020tes,Witten:2020ert,Stanford:2020qhm,Betzios:2020nry,Mirbabayi:2020grb,Cotler:2019dcj,Fernandes:2019ige,Hikida:2021ese,Aguilar-Gutierrez:2021bns,Forste:2021roo,Blommaert:2020tht,PhysRevD.28.2960,Stanford:2017thb,Iliesiu:2020zld,Moitra:2019xoj,Hennauxjt,Louis-Martinez:1993bge,Strobl:1993yn,Saad:2018bqo,Eynard:2007fi,Mirzakhani:2006fta,GOLDMAN1984200,wolperteleform,wolpertonweil,Anninos:2021eit,Maldacena:2019cbz,Cotler:2019nbi,Saad:2019lba,Saad:2018bqo,Almheiri:2014cka,Jensen:2016pah,Maldacena:2016upp,Engelsoy:2016xyb,Harlow:2018tqv,Blommaert:2018iqz,Yang:2018gdb,Blommaert:2019hjr,Stanford:2017thb,Mertens:2017mtv,Kitaev:2017awl,nayak,Moitra:2019bub,Moitra:2019xoj,Moitra:2018jqs,Lin:2018xkj, Stanford:2019vob,Mertens:2019tcm,Iliesiu:2019xuh,Mertens:2019bvy,Blommaert:2020seb,Lin:2019qwu,Maldacena:2018lmt,Mefford:2020vde,Suh:2020lco,Saad:2019pqd,Xian:2019qmt,Cotler:2019dcj,Grumiller:2020fbb,Haehl:2017pak,PhysRevResearch.2.043310,Jafferis:2019wkd, Constantinidis:2008ty,Gaikwad:2018dfc, Anninos:2018svg, Anninos:2020cwo, Griguolo:2021wgy}.

	\section{De Sitter JT Gravity with Classical matter}
	\label{basup}
	In this section, we shall first analyze the classical behaviour of  JT gravity in   dS$_2$ spacetime. 
	Let us introduce the basic setup. To avoid repetition of various equations in the absence and presence of matter, we write most the of equations in general form in the presence of matter but restrict to matter-less case wherever needed by turning the off the stress tensor for them. 
	The action for Jackiw-Teitelboim (JT) model  in 2D de Sitter spacetime is given by,
	\begin{align}
		I_{JT}=\frac{-i}{16\pi G}\pqty{\int d^2 x\,\sqrt{-g}\,\phi (R-2)-2\int_{bdy}\sqrt{-\gamma}\phi K }\label{jtact}.
	\end{align}
	where $G$ is the 2D Newton's constant, $R$ is the Ricci scalar, $K$ is the extrinsic curvature of the boundary and $\phi$ is the dilaton field. The boundary term in the above action is the usual Gibbons-Hawking term added to render the variational principle well-defined for Dirichlet boundary  conditions on the metric and dilaton. { Note that a boundary term proportional to length of the boundary is added in AdS consistent with holographic renormalization, as a counterterm to cancel divergence that arise in the path integral when the length of boundary diverges. In de Sitter  context that we are interested here, we use eq.\eqref{jtact} to compute the wavefunction and such a counterterm is not to be added. }  Note we have chosen units where the cosmological constant is $2$. 
	
	We consider the JT theory along with conformally invariant massless scalar fields whose action denoted by $I_m$, is given by
	\begin{align}
		I_m=\sum_{k=1}^{N}\frac{i}{2}\int d^2x \sqrt{-g} (\nabla_\mu\varphi_k)^2\label{smatter}
	\end{align}
	where $N$ is the number of species of matter fields.
	The factor of $i$ inlcuded in the actions above is so that the path integral in terms of the action just becomes
	\begin{align}
		Z=\int [\mathcal{D}g]\,[\mathcal{D}\phi]\,[\mathcal{D}\varphi_k] \,\,e^{-I_{JT}-I_m}\label{pi}
	\end{align}
	The matter fields only couple to gravity and not to the dilaton $\phi$. So, the equation of motion obtained by varying the dilaton is given by 
	\begin{align}
		R-2=0\label{rdes}
	\end{align}
	which shows that spacetime is always dS space. 
	Working in the conformal gauge in which the metric takes the form,
	\begin{align}
		\text{d}s^2 =e^{2\omega(x^{+},x^{-})}\,\text{d}x^{+}\text{d}x^{-}\label{conmet}
	\end{align}
	eq.\eqref{rdes} becomes
	\begin{align}
		8\del_{+}\del_{-}\omega+2 e^{2\omega}=0\label{omegaeq}
	\end{align}
	with {	 one} of the the solutions being given by 
	\begin{align}
		e^{2\omega}=\frac{4}{(x^+ -x^-)^2}\Rightarrow ds^2=\frac{4}{(x^+-x^-)^2}dx^+ dx^-\label{confds}
	\end{align}
	Varying the JT action eq.\eqref{jtact} with respect to the metric gives
	\begin{align}
		\frac{2}{\sqrt{-g}}\frac{\delta I_{JT}}{\delta g^{\mu\nu}}=\frac{i}{8\pi G}(\nabla_{\mu}\nabla_{\nu}\phi-g_{\mu\nu}\nabla^{2}\phi- g_{\mu\nu}\phi)\label{jtvarmet}
	\end{align}
	Including the contribution from the matter fields, the equations of motion obtained by varying the metric are given by 
	\begin{equation}
		- \frac{1}{8\pi G}(\nabla_{\mu}\nabla_{\nu}\phi-g_{\mu\nu}\nabla^{2}\phi- g_{\mu\nu}\phi)= T^{(m)}_{\mu\nu} \label{geneqs},
	\end{equation}
	with $T^{(m)}_{\mu\nu}$ being the matter stress tensor, defined by
	\begin{align}
		iT^{(m)}_{\mu\nu}\equiv\frac{2}{\sqrt{-g}}\frac{\delta I_m}{\delta g^{\mu\nu}}\label{tmdef}
	\end{align}
	Note that the above definition of stress tensor might seem odd at first sight due to an extra factor of $i$, but a glance at eq.\eqref{smatter} shows that there should indeed be such a factor to compensate for a factor of $i$ in the definition of the action.
	In the conformal gauge eq.\eqref{conmet}, eq.\eqref{geneqs} becomes
	\begin{align}
		-e^{2\omega}\del_{\pm}\pqty{e^{-2\omega}\del_{\pm}\phi}&=8 \pi G  T^m_{\pm\pm} \label{ppmm},\\
		2\del_{+}\del_{-}\phi+e^{2\omega}\,\phi&=16\pi G  T^m_{+-} \label{pm}.
	\end{align}
	Let us first analyze the theory in the absence of matter. Working in the coordinate system eq.\eqref{confds} and setting $T^{(m)}_{\mu\nu}=0$, we get the solution for the dilaton as 
	\begin{align}
		\phi={a+b(x^+ + x^-)+ cx^+x^-
			\over ( x^+-x^-)}\label{nomatdilsolds}
	\end{align}
	where $a,b,c$ are arbitrary constants. Doing an SL(2,R) transformation of the the coordinates $x^+,x^-$, 
	we can get the dilaton to the form
	\begin{align}
		\phi=\frac{1- \mu\, x^+ x^-}{x^+ - x^-}, \qquad \mu=b^2-ac\label{dilafsl}
	\end{align}
	where $\mu$ is a real parameter which can be either positive, negative or zero. 
	
	We analyse all three cases below. Before doing so, let us note the following two facts. First,   dS$_2$ in global coordinates is given by 
	\begin{align}
		\label{gds}
		ds^2=-d \hat{\tau}^2 +\cosh^2\hat{\tau}\, d\hat{\theta}^2
	\end{align}
	with $\hat{\tau}\in [-\infty,\infty]$, and $\hat{\theta}\in [0,2\pi]$, spanning a circle of length $2\pi$. These coordinates cover  all of dS space. 
	The transformations eq.\eqref{dsrsr}\eqref{dsrsze}\eqref{dspcgl}, relates these to $x^\pm$ poincare coordinates above. 
	The transformation eq.\eqref{dsrsr} the relates the coordinates to  $\zeta^\pm= \hat{\theta} \pm \hat{r}_*$, with $\hat{r}_*\in [0,\pi], \hat{\theta}\in [0,2 \pi]$  in terms of which one can easily  obtain the Penrose diagram shown in \ref{dsglpenr}. Note that, as is well known, the Penrose diagram for global dS is a rectangle, and a light ray which starts at $\hat{\theta}=0$ at time $\hat{r}_*=0$ reaches the antipodal point of the  circle $\hat{\theta} = \pi$ asymptotically in the far future. 
	
	Second,   $2$ dim. JT gravity can be thought of as arising from higher dimensions after carrying out a dimensional reduction, with $\phi$ being related to the volume of the internal space. Keeping this interpretation  in mind we will impose  that a singularity  arises in the JT theory when $\phi$ becomes sufficiently negative in value. For example,  
	as was discussed in \cite{Maldacena:2019cbz},  the  JT theory is  obtained from  4 dim Einstein theory with a  positive cosmological constant after dimensional reduction,  by considering a black hole which is close to extremality, with the cosmological and black horizons close together. Assuming spherical symmetry, the 4 dim. metric can be written as 
	\begin{align}
		\label{4dm}
		ds^2=g_{\alpha\beta} dx^\alpha dx^\beta + \Phi_0^2(1+\phi) d\Omega^2
	\end{align}
	where $g_{\alpha\beta}$ is the two dim. metric and $\Phi_0$ is the value of the horizon  radius at extremality. 
	$\phi$ then plays the role of the dilaton in the 2 dim. JT theory which arises in the near-extremal limit. 
	
	We see in this case  that when
	\begin{align}
		\label{sufn}
		\phi=-1
	\end{align}
	the volume of the internal $S^2$ vanishes and in fact a curvature singularity  arises in the 4 dimensional theory at that locus.  With this example in mind, in our subsequent discussion for definiteness we will take the singularity  in JT theory  to be located at $\phi=-1$. No  important consequences will depend on this precise value.

	Let us now consider the three solutions found in Poincare coordinates in eq.(\ref{dilafsl}). For $\mu<0$ after a rescaling	$x^\pm \rightarrow- {x^\pm\over \sqrt{\abs{\mu}}}$ and using eq.\eqref{dspcgl},\eqref{dsrsr} to go from Poincare to global coordinates we find the dilaton to be 
	\begin{align}
		\label{valdilaa}
		\phi = -\sqrt{\abs{\mu} }  \sinh \tau
	\end{align}
	We see that when $\tau$ becomes sufficiently big a spacelike singularity arises. E.g., for  $\phi=-1$, the singularity occurs when 
	$\sinh\tau=1/\sqrt{\abs{\mu}}$.

	Next consider the case $\mu=0$. Here, a   singularity will occur along a curve $x^+-x^-= {\rm const}$ which is again spacelike.
	For the dilaton taking the form, $\phi=A/(x^+-x^-)$, which is  obtained by rescaling the coordinates in eq.\eqref{dilafsl}, the singularity with $\phi=-1$ occurs at $x^+-x^-=-1/A$, which is shown in Fig \ref{dsponpenrose}. 
	
	Finally, we take the case $\mu>0$. Here,  rescaling $x^\pm \rightarrow -{1\over \sqrt{\mu} } x^\pm$, and a further  change of variables to $(r, \theta)$, by the series of transformation eq.\eqref{milnepoinco},\eqref{milnezetaco},\eqref{rthmil},  allows us to recast the solution  as,
	\begin{align}
		\label{solaa}
		ds^2 & =  - {dr^2 \over { r}^2-\mu} + ( r^2-\mu) d \theta^2 \\
		\phi & =     r
	\end{align}
	
	The variables $( r,  \theta)$ do not cover all of dS$_2$ but can be extended to do so, patch -wise in standard fashion. 
	The variable $ \theta$ has the range, $ \theta\in [-\infty, \infty]$.  For $ r\in [-\sqrt{ \mu}, \sqrt{\mu}]$ we get the static patch of dS between a cosmological and black hole horizon, see Fig. \ref{dsbhshwaves}.  
	The region $ r<-\sqrt{\mu}$ lies inside the black hole horizon, while $ r>\sqrt{\mu}$ lies outside the cosmological horizon. 
	The black hole singularity, at $\phi=-1$, lies at $ r=-1$. Requiring this singularity  to be inside the black hole horizon gives the condition
	\begin{align}
		\label{conda2}
		\sqrt{\mu}<1
	\end{align}
	Bigger values of $\mu$ give rise to a naked singularity.  On the other hand, the smallest value $\mu$ can take is $\mu=0$ which gives the extremal or Nariai limit. 
	
	It is worth noting that the Penrose diagram for  this case can actually be extended infinitely to give a chain of universes connected across regions containing black holes, as shown in Fig.\ref{penrosefig}. 	
	
	Finally let us mention  that we can consider an orbifold of the solution eq.(\ref{solaa}) which is obtained by an identification along the $ \theta$ direction,
	which is an isometry. To describe this orbifold we first rescale the coordinates and reexpress the solution in 
	eq.(\ref{solaa}) as 
	\begin{eqnarray}
		\label{orba}
		ds^2 & = & -{dr^2\over r^2-1} + (r^2-1) d\theta^2 \\
		\phi & = & A r 
	\end{eqnarray}
	where $A=\sqrt{\mu}$. 
	Next let us make the identification 
	\be
	\label{ido}
	\theta \simeq \theta + b.
	\ee 
	This is done in the two Milne patches, corresponding to $r^2>1$, which are marked regions I and II  in figure 
	\ref{dsbhshwaves}.  The regions  III and IV, where $r^2<1$,  are not included in the spacetime anymore, this can be done consistently since no world line leaves the  resulting spacetime. We will discuss this orbifold and the resulting cosmology in greater detail below. 
	
	For now let us note that once the regions $r^2<1$ are removed from the spacetime it does not have a singularity where the dilaton takes value $\phi=-1$, more generally where the dilaton  becomes sufficiently negative. 
	However,  the resulting spacetime now has a conical singularity at $r=1$ where the $ \theta$ circle shrinks to zero size.
	This  spacetime can therefore be thought of as a simple prototype of a cosmology containing a big crunch/ big bang singularity. Let us also note that  while there are no closed time like curves in it, there are closed null curves, as will be discussed in subsection \ref{orfsubsec} briefly.

	\subsection{Some Aspects of Thermodynamics}
	\label{clasinfal}
	In this subsection, we shall discuss some aspects of thermodynamics in  dS$_2$. 
	We will work in the static patch, shown as region III in Fig \ref{dsbhshwaves}, which has metric 
	\begin{align}
		\label{mets}
		ds^2=-(\mu-r^2)  d \theta^2 + {dr^2\over \mu-r^2}
	\end{align}
	with a time-like Killing vector $\partial_\theta$ .  
	This region contains a cosmological horizon located at $r=r_c=\sqrt{\mu}$ and a black hole horizon at $r=r_b=-\sqrt{\mu}$. 
	The entropy of both  horizons is  proportional to the value  the  dilaton takes on them. Matching with the  $4$ dim theory, appendix \ref{pennhlim}, gives 
	\begin{align}
	\label{enth}
	S_{h}={\phi_h\over 4 G},
	\end{align}
	so we learn that the cosmological and black hole horizons have entropy $\pm {\sqrt{\mu}\over 4 G}$. Negative entropy is strange, in fact more correctly, 
	the entropy in the dimensionally reduced case for each horizon is given by   $S={(1+\phi) \over 4 G}$  but we will drop the constant term  in this section since we are  mostly interested in changes of  entropy. 
	It is easy to see that the cosmological and black hole horizons both have temperature 
	\begin{align}
	\label{valt}
	T={{\sqrt \mu}\over 2 \pi}.
	\end{align}
	Thus while the black hole horizon has a negative specific heat, {$dS/dT=-{ \pi\over 2 G}$}, the cosmological horizon has a positive specific heat, {$dS/dT={ \pi\over 2G}$}. There is no natural notion of mass one can associate with the black hole in this case; a ``thermodynamic" definition could be given so that  the first law  $dM=TdS$, is satisfied. This leads to the change in mass being related to the change in the parameter $\mu$ by,   {$\delta M = -{1\over 16 \pi G} \delta \mu$.}
	We remind the reader that the parameter $\mu$ takes values  $0<\mu<1$. 
	
	We will now consider what happens in the presence of matter. Let us summarise the behaviour here. We see below that when 
	matter (satisfying the weak energy condition) comes out of the past horizon of the black hole and falls into the cosmological horizon, the entropy of the black hole decreases and that of the cosmological horizon increases by the same amount, keeping the net change to be zero. Also, the temperature of the black hole goes up, since it has negative specific heat, while that of the cosmological horizon also goes up by the same amount, making them both equal in the future as well. When matter falls in from the cosmological horizon into the black hole the entropy of the black hole goes up and its temperature goes down, while the entropy  of the cosmological horizon goes  down, and its temperature also goes down by the same amount. In all  cases, whether matter comes out of the past horizon of the black hole or the past cosmological horizon, { the entropy of the future black hole or cosmological event horizons are non-decreasing functions of horizon affine time in accordance with the second law of thermodynamics. }
	
	In more detail, consider the black hole configuration in Poincare coordinates as
	\begin{align}
		ds^2=\frac{4 dx^+ dx^-}{(x^+ - x^-)^2}\nonumber\\
		\phi=\frac{1-\mu_0 x^+ x^-}{x^+ - x^-}\label{pcbhconfig}
	\end{align}
	These coordinates are related to $(r_*,\theta)$ \eqref{dsstametor} by\footnote{The Poincare coordinates actually do not cover the whole static patch and break down at $\theta+r_*=0$ where $x^+\rightarrow \pm \infty $. To be careful one can use the coordinates $1/x^+, x^-$ instead, or use the Poincare coordinates in coordinate patches where they are valid, and paste the resulting solutions together.}
	\begin{align}
		\label{chc}
		x^+= -{1\over\sqrt{\mu_0}} \coth ({\sqrt{\mu_0}(\theta+r_*)\over 2}), x^-=-{1\over \sqrt{\mu_0}} \tanh ({\sqrt{\mu_0}(\theta-r_*)\over 2})
	\end{align}
	Now consider general infalling classical matter satisfying the null energy condition such that
	\begin{align}
		T_{++}>0,\,\, T_{--}>0,\,\, T_{+-}=0\label{tconds}
		\end{align}

	\tikzset{every picture/.style={line width=0.75pt}} 
	\begin{figure}
		\centering

		\tikzset{every picture/.style={line width=0.75pt}} 
		
		\begin{tikzpicture}[x=0.75pt,y=0.75pt,yscale=-1,xscale=1]
			
			\draw   (153.07,52.95) -- (284.08,52.82) -- (283.71,183.81) -- (152.7,183.95) -- cycle ;
			\draw    (153.07,52.95) -- (283.71,183.81) ;
			\draw    (152.71,183.95) -- (284.07,52.82) ;
			\draw   (284.6,52.59) .. controls (288.65,54.92) and (292.53,57.13) .. (297.05,57.18) .. controls (301.57,57.23) and (305.5,55.09) .. (309.6,52.84) .. controls (313.7,50.6) and (317.62,48.46) .. (322.14,48.51) .. controls (326.67,48.55) and (330.54,50.77) .. (334.6,53.1) .. controls (338.65,55.43) and (342.52,57.65) .. (347.05,57.7) .. controls (351.57,57.74) and (355.49,55.6) .. (359.59,53.36) .. controls (363.69,51.11) and (367.62,48.97) .. (372.14,49.02) .. controls (376.66,49.07) and (380.54,51.29) .. (384.59,53.62) .. controls (388.65,55.95) and (392.52,58.16) .. (397.05,58.21) .. controls (400.45,58.25) and (403.52,57.04) .. (406.57,55.49) ;
			\draw   (283.27,183.09) .. controls (287.33,185.2) and (291.21,187.21) .. (295.73,187.25) .. controls (300.26,187.3) and (304.18,185.37) .. (308.27,183.34) .. controls (312.37,181.32) and (316.29,179.39) .. (320.81,179.43) .. controls (325.34,179.48) and (329.22,181.49) .. (333.27,183.6) .. controls (337.33,185.71) and (341.21,187.72) .. (345.73,187.77) .. controls (350.25,187.82) and (354.17,185.89) .. (358.27,183.86) .. controls (362.37,181.83) and (366.29,179.9) .. (370.81,179.95) .. controls (375.34,179.99) and (379.21,182) .. (383.27,184.12) .. controls (387.32,186.23) and (391.2,188.24) .. (395.73,188.28) .. controls (398.76,188.32) and (401.52,187.46) .. (404.25,186.28) ;
			\draw    (406.58,55.31) -- (404.22,187.29) ;
			\draw    (284.07,52.82) -- (404.22,187.29) ;
			\draw    (283.71,183.81) -- (406.58,55.31) ;
			\draw [color={rgb, 255:red, 244; green, 156; blue, 9 }  ,draw opacity=1 ]   (241.12,62.12) -- (339,160) ;
			\draw [shift={(289,110)}, rotate = 45] [fill={rgb, 255:red, 244; green, 156; blue, 9 }  ,fill opacity=1 ][line width=0.08]  [draw opacity=0] (8.93,-4.29) -- (0,0) -- (8.93,4.29) -- cycle    ;
			\draw [shift={(239,60)}, rotate = 45] [fill={rgb, 255:red, 244; green, 156; blue, 9 }  ,fill opacity=1 ][line width=0.08]  [draw opacity=0] (8.93,-4.29) -- (0,0) -- (8.93,4.29) -- cycle    ;
			\draw [color={rgb, 255:red, 9; green, 118; blue, 244 }  ,draw opacity=1 ]   (356.7,82.81) -- (258,176) ;
			\draw [shift={(308.44,128.38)}, rotate = 136.65] [fill={rgb, 255:red, 9; green, 118; blue, 244 }  ,fill opacity=1 ][line width=0.08]  [draw opacity=0] (8.93,-4.29) -- (0,0) -- (8.93,4.29) -- cycle    ;
			\draw [shift={(358.88,80.75)}, rotate = 136.65] [fill={rgb, 255:red, 9; green, 118; blue, 244 }  ,fill opacity=1 ][line width=0.08]  [draw opacity=0] (8.93,-4.29) -- (0,0) -- (8.93,4.29) -- cycle    ;
			\draw  (486.38,115.27) -- (518.24,146.12)(517.32,89.69) -- (486.48,121.54) (516.69,137.66) -- (518.24,146.12) -- (509.73,144.85) (508.86,91.24) -- (517.32,89.69) -- (516.04,98.2)  ;
			
			\draw (224,160.4) node [anchor=north west][inner sep=0.75pt]    {$T_{--}$};
			\draw (345,153.4) node [anchor=north west][inner sep=0.75pt]    {$T_{++}$};
			\draw (168.36,139.53) node [anchor=north west][inner sep=0.75pt]  [rotate=-313.07]  {$H_{1}$};
			\draw (246.69,119.92) node [anchor=north west][inner sep=0.75pt]  [rotate=-41.91]  {$H_{2}$};
			\draw (206.07,103.16) node [anchor=north west][inner sep=0.75pt]  [font=\scriptsize,rotate=-317.76]  {$r=\sqrt{\mu }$};
			\draw (174.71,40.56) node [anchor=north west][inner sep=0.75pt]  [font=\scriptsize,rotate=-0.59]  {$r=\infty $};
			\draw (326.41,35.15) node [anchor=north west][inner sep=0.75pt]  [font=\scriptsize,rotate=-0.63]  {$r=-1$};
			\draw (305.42,54.65) node [anchor=north west][inner sep=0.75pt]  [font=\scriptsize,rotate=-47.29]  {$r=-\sqrt{\mu }$};
			\draw (295.26,73.61) node [anchor=north west][inner sep=0.75pt]  [font=\scriptsize,rotate=-90.38]  {$r=0$};
			\draw (262,104) node [anchor=north west][inner sep=0.75pt]  [color={rgb, 255:red, 208; green, 2; blue, 27 }  ,opacity=1 ] [align=left] {III};
			\draw (212,141) node [anchor=north west][inner sep=0.75pt]  [color={rgb, 255:red, 208; green, 2; blue, 27 }  ,opacity=1 ] [align=left] {II};
			\draw (209,64) node [anchor=north west][inner sep=0.75pt]  [color={rgb, 255:red, 208; green, 2; blue, 27 }  ,opacity=1 ] [align=left] {I};
			\draw (163,101) node [anchor=north west][inner sep=0.75pt]  [color={rgb, 255:red, 208; green, 2; blue, 27 }  ,opacity=1 ] [align=left] {IV};
			\draw (522,74.4) node [anchor=north west][inner sep=0.75pt]    {$x^{+}$};
			\draw (523,131.4) node [anchor=north west][inner sep=0.75pt]    {$x^{-}$};

		\end{tikzpicture}
		\label{dsbhshwaves}
		\caption{The yellow line corresponds to a pulse of shockwave coming out from the past black hole horizon and falling into the future cosmological horizon. The blue line corresponds to a pulse of shockwave falling from the past Cosmological horizon to the future black hole horizon. }
	\end{figure}
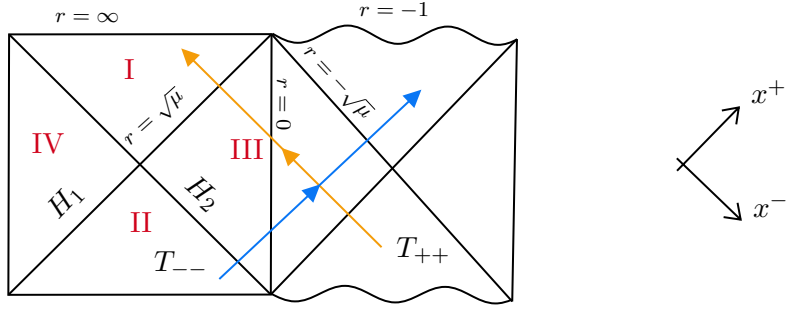

	The general solution to the equations governing dilaton eq.\eqref{ppmm},\eqref{pm} can be written as
	\begin{align}
		\phi=\frac{h(x^-)+k(x^+)}{x^+- x^-}+\half (h'(x^-)-k'(x^+))\label{dilinhk}
	\end{align}
	where the functions $h(x^-), k(x^+)$ satisfy
	\begin{align}
		h'''(x^-)=-16\pi G T_{--}\nonumber\\
		k'''(x^+)=16\pi G T_{++}\label{hkeq}
	\end{align} 
	Thus the general solution can be obtained simply by linearly superposing the response to the left and right-moving stress tensors. 
	The starting solution eq.(\ref{pcbhconfig}) can be obtained by taking,
	\begin{align}
		\label{stc}
		h(x^-)=1-\mu_0 (x^-)^2, k(x^+)=0,
	\end{align}
	
	{
		Now consider a shock wave consisting of right moving matter with the stress tensor, 
	\begin{align}
		\label{lmstress}
		T_{--}={\mu_-\over 8\pi G} \delta (x^-),
	\end{align}
	which falls into the black hole from the cosmological horizon moving along the trajectory  $x^-=0$,
	The resulting solution is given by 
	\begin{align}
		\label{resso}
		h(x^-)=1-\mu_0 (x^-)^2+\Theta(-x^-) \mu_-(x^-)^2
	\end{align}
	(with $k(x^+)=0$), where $\Theta(x)$ is the Heavyside theta function.
	We see that the value of $\mu_0$ decreases to $\mu=\mu_0-\mu_-$ once the shock wave passes. 
		It is easy to see that this corresponds to the ``mass" of the black hole going up, once the shock wave falls into it, with an increases in its entropy and a corresponding decrease in the cosmological horizon's entropy. 
	
	Similarly we can consider a shock wave which comes out of the past black hole horizon with stress tensor, 
	\begin{align}
		\label{strt}
		T_{++}= {\mu_+\over 8 \pi G} \delta(x^+-x^+_1)
	\end{align}
	(we take a non-zero value of $x^+_1$, since $\sqrt{\mu_0}|x^+|\ge 1$, eq.(\ref{chc}) ). 
	Starting with eq.(\ref{stc}) the solution is now given by 
	\begin{align}
		\label{laso}
		k(x^+)=\mu_+(x^+-x_1^+)^2 \Theta(x^+-x_1^+)
	\end{align}
	with $h(x^-)$ being unchanged from eq.(\ref{stc}). 
	The resulting value of the dilaton for $x^+>x_1^+$ is then 
	\begin{align}
		\label{resd}
		\phi={1+\mu_+x_1^2 -\mu_+ x_1^+(x^++x^-) -(\mu_0-\mu_+)x^+x^-\over x^+-x^-}
	\end{align}
	so that the discriminant, eq.\eqref{dilafsl}, takes the value, $b^2-ac=\mu_0 +\mu_+ (\mu_0^2 (x_1^+)^2-1)$ which is greater than $\mu_0$ since 
	from eq.\eqref{chc} $(\sqrt{\mu_0}\, x_1^+)^2=\coth^2\left({\sqrt{\mu_0}(r_{*1}+\theta_1)\over 2}\right)>1$. 	Comparing with the discussion in the  previous section we find that the entropy of the black hole  goes down at late times, and once the shock wave passes the cosmological horizon its entropy goes up by the same amount.

	}

	In the examples above we took the stress tensor to satisfy  the null energy condition and in both cases, as mentioned before the entropy of the future black hole and cosmological horizon increases. 
	It is easy to argue that this will be true in general. Consider a situation where initially matter falls in from the past cosmological horizon or out from the past black hole horizon for some time and then things settle down with the metric at late times being of the form eq.\eqref{mets} with a time-like killing vector. 
	Let the final black hole and cosmological event horizons be located at $x^+=x^+_{b}$ and $x^-=x^-_c$ respectively and let $\lambda_b$ and $\lambda_c$ be the affine parameters for the two horizons which satisfy,
{	\begin{align}
		{dx^-\over d\lambda_b}&=-(x^--x^+_b)^2 \nonumber\\
		{dx^+\over d\lambda_c} & = (x^+ -x^-_c)^2\label{affinparams}
	\end{align}}
	respectively. 
	
	Then using eq.\eqref{dilinhk}  and from the equations of motion eq.\eqref{hkeq} it is easy to see that 
	\begin{align}
		{d^2 S\over d\lambda_b^2} & =  (x^+_b - x^-)^2\del_{-}((x^+_b - x^-)^2\del_{-}S)=-2\pi (x^+_b-x^-)^4 T_{--}<0\label{selwcos}\\
		{d^2 S\over d\lambda_c^2} & =  (x^+ - x^-_c)^2\del_{+}((x^+ - x^-_c)^2\del_{+}S)= -2\pi (x_c^- - x^+)^4 T_{++}<0\label{selwbh}
	\end{align}
	Now in the far future  when $\lambda_{b,c}\rightarrow \infty$, since the time dependence ceases, $dS/d\lambda_c$ and $dS/d\lambda_b$ vanish. 
	It then follows that  $dS/d\lambda_c, dS/d\lambda_b$ must be non-negative at finite $\lambda_{b,c}$ respectively  and thus the entropy along the two future horizons must be a  non-decreasing function of affine time.

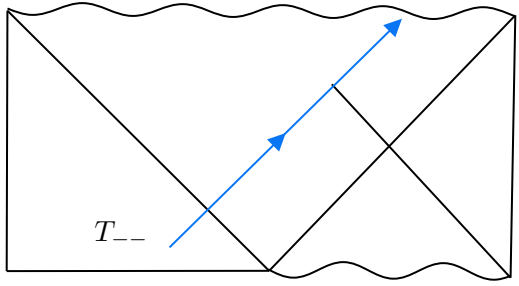
\begin{figure}[h!]
		\centering

		\tikzset{every picture/.style={line width=0.75pt}} 
		
		\begin{tikzpicture}[x=0.75pt,y=0.75pt,yscale=-1,xscale=1]
			
			\draw    (173.07,72.95) -- (303.71,203.81) ;
			\draw   (173.1,71.86) .. controls (177.16,73.49) and (181.04,75.04) .. (185.57,75.08) .. controls (190.09,75.13) and (194.01,73.66) .. (198.1,72.12) .. controls (202.19,70.57) and (206.11,69.1) .. (210.63,69.15) .. controls (215.15,69.2) and (219.04,70.75) .. (223.1,72.37) .. controls (227.16,74) and (231.04,75.55) .. (235.56,75.6) .. controls (240.09,75.64) and (244,74.17) .. (248.1,72.63) .. controls (252.19,71.09) and (256.1,69.62) .. (260.63,69.67) .. controls (265.15,69.71) and (269.03,71.26) .. (273.09,72.89) .. controls (277.15,74.52) and (281.04,76.07) .. (285.56,76.11) .. controls (290.09,76.16) and (294,74.69) .. (298.09,73.15) .. controls (302.19,71.6) and (306.1,70.13) .. (310.62,70.18) .. controls (315.15,70.23) and (319.03,71.78) .. (323.09,73.4) .. controls (327.15,75.03) and (331.04,76.58) .. (335.56,76.63) .. controls (340.08,76.67) and (344,75.2) .. (348.09,73.66) .. controls (352.18,72.12) and (356.1,70.65) .. (360.62,70.7) .. controls (365.15,70.74) and (369.03,72.29) .. (373.09,73.92) .. controls (377.15,75.55) and (381.03,77.09) .. (385.56,77.14) .. controls (390.08,77.19) and (393.99,75.72) .. (398.09,74.18) .. controls (402.18,72.63) and (406.1,71.16) .. (410.62,71.21) .. controls (415.14,71.26) and (419.03,72.81) .. (423.09,74.43) .. controls (424.26,74.91) and (425.42,75.37) .. (426.59,75.79) ;
			\draw   (303.27,203.09) .. controls (307.33,205.2) and (311.21,207.21) .. (315.73,207.25) .. controls (320.26,207.3) and (324.18,205.37) .. (328.27,203.34) .. controls (332.37,201.32) and (336.29,199.39) .. (340.81,199.43) .. controls (345.34,199.48) and (349.22,201.49) .. (353.27,203.6) .. controls (357.33,205.71) and (361.21,207.72) .. (365.73,207.77) .. controls (370.25,207.82) and (374.17,205.89) .. (378.27,203.86) .. controls (382.37,201.83) and (386.29,199.9) .. (390.81,199.95) .. controls (395.34,199.99) and (399.21,202) .. (403.27,204.12) .. controls (407.32,206.23) and (411.2,208.24) .. (415.73,208.28) .. controls (418.76,208.32) and (421.52,207.46) .. (424.25,206.28) ;
			\draw    (426.58,75.31) -- (424.22,207.29) ;
			\draw    (335,110) -- (424.22,207.29) ;
			\draw    (303.71,203.81) -- (426.58,75.31) ;
			\draw [color={rgb, 255:red, 9; green, 118; blue, 244 }  ,draw opacity=1 ]   (367.87,79.11) -- (254,192) ;
			\draw [shift={(312,134.5)}, rotate = 135.25] [fill={rgb, 255:red, 9; green, 118; blue, 244 }  ,fill opacity=1 ][line width=0.08]  [draw opacity=0] (8.93,-4.29) -- (0,0) -- (8.93,4.29) -- cycle    ;
			\draw [shift={(370,77)}, rotate = 135.25] [fill={rgb, 255:red, 9; green, 118; blue, 244 }  ,fill opacity=1 ][line width=0.08]  [draw opacity=0] (8.93,-4.29) -- (0,0) -- (8.93,4.29) -- cycle    ;
			\draw    (172.7,203.95) -- (303.71,203.81) ;
			\draw    (173.07,72.95) -- (172.7,203.95) ;
			
			\draw (215,177.4) node [anchor=north west][inner sep=0.75pt]    {$T_{--}$};

		\end{tikzpicture}
	
	\caption{A shockwave of appropriate strength causes the spacelike singularity originally hidden behind the black hole horizon to stretch outwards covering an entire spacelike slice. }
	\label{orbmat}
	\end{figure}

	We end this section with two comments. 
	First,  we considered here the effects of classical matter. One can also consider the behaviour of the generalised entropy, along an event horizon or a future $Q$ screen, in the semi-classical theory, which we describe below,  with matter being in various  quantum states. We leave such a more complete analysis for the future. 	Second, we discussed  above what happens in the presence of in-falling or outgoing matter in the static patch region of spacetime. It is also interesting to look at the behaviour in other regions. When matter is thrown in from the asymptotic past (the lower Milne patch shown as in Fig.\ref{dsbhshwaves} ,  the black holes goes towards extremality and eventually the cosmological and black hole merge. Any  excess matter above the extremal limit, when thrown in from past infinity,  causes a spacelike singularity to appear.  
	E.g., starting with a black hole with mass parameter $\mu$, if we throw in matter along a shock wave with a stress energy tensor of the form in eq.\eqref{lmstress} with $\mu_-=\mu_0$, we find that a singularity forms, stretching along an entire spacelike slice extending outside the infalling shockwave. This is shown in Fig. \ref{orbmat}, 
	spacelike singularity inside the black hole has extended over a spacelike slice outside the shockwave.

 In this sense then  dS JT gravity is ``prone" to forming singularities. The $\mu<0$ and $\mu=0$ solutions, eq.\eqref{dilafsl} already have spacelike singularities as discussed above near eq.\ref{valdilaa}. The $\mu>0$ solution has a singularity but hidden safely inside a black hole horizon. However once enough matter is sent in, past extremality, again a spacelike singualrity appears.

	\section{ Quantum matter and an Orbifold}
	\label{dssemicla}
	Here we consider the effects of quantum matter. More precisely we work in the semiclassical limit where we add $N$ massless scalar fields, 
	with the action eq.(\ref{smatter}), where 
	\begin{align}
		G\rightarrow 0,\, N\rightarrow \infty,\, GN=\text{const}\label{semiclslim}
	\end{align}
	In this limit the metric and dilaton continue to be described by their classical equations of motion, since their action in eq.\eqref{jtact} has a $1/G$ factor in front, while the matter fields which evolve in this classical background are quantum mechanical. 
	We see from eq.(\ref{smatter}) that the matter fields do not couple to the dilaton. Its equation of motion is therefore unchanged from eq.\eqref{rdes} and tells us that the geometry continues to be (locally) dS space. The equation of motion for the metric  are now given by 
	eq.\eqref{ppmm},\eqref{pm} with the stress tensor $T_{\mu\nu}$ being replaced by its expectation value $\langle T_{\mu\nu}\rangle $ in the appropriate quantum state for the matter fields. 
	
	First let us consider Poincare vaccum for the left and right movers, {with the conformal anomaly given by}
	\begin{align}
		T^\mu_\mu=\frac{NR}{24\pi}\label{confan}
	\end{align}
	Working with  Poincare coordinate system eq.\eqref{milnepoinmet}
	the general solution for the dilaton is given in general by 
	\begin{align}
		\label{solid}
		\phi={a+b(x^+ + x^-)+ c\,x^+x^-
			\over ( x^+-x^-)}+\frac{GN}{3},
	\end{align}
	Comparing with eq.\eqref{nomatdilsolds} we see that the effect of matter is only to shift the value of the dilaton. 
	The discussion in section \ref{basup} about the three kinds of solutions, eq.\eqref{dilafsl} therefore carries over here too once we account for this shift. 
	Note that in obtaining eq.(\ref{solid}) we used the fact that if the metric is given by 
	\begin{align}
		\label{metf}
		ds^2=-f(\zeta^+,\zeta^-) \,d\zeta^+ \,d\zeta^-
	\end{align}
	in some set of conformal coordinates $(\zeta^+,\zeta^-)$, the stress tensor in the $\zeta^+, \zeta^-$ vacua is given by 
	\begin{align}
		\label{stressv}
		T_{\zeta^\pm\zeta^\pm}= -\frac{1}{12\pi}f^{\frac{1}{2}}\del_{\zeta^\pm}^2f^{-\frac{1}{2}},
	\end{align}
	see eq.\eqref{txvac} and therefore vanishes in the Poincare vaccuum. 
	
	Now consider dS space containing a black hole, after suitable rescaling this is the solution in which the  metric is  given  as  in eq.(\ref{orba}). 
	Starting with this case we saw in the previous section that when matter meeting the null energy condition is thrown in, the black hole evolves towards extremality and if   additional matter is thrown, past that point, it leads to a spacelike singularity.
	The quantum stress tensor however does not have to meet the null energy condition and it is interesting to ask about the more general behaviour that can then result. 
	
	We will in particular be interested in the ``lower Milne wedge" region shown in Fig \ref{dsbhshwaves} which is bounded by two cosmological horizons, and in what happens when matter is thrown in from this region in the far past. 
	Well defined coordinates on the two cosmological  horizons where {{$r\rightarrow 1$}} are given by Kruskal coordinates, see eq.\eqref{fUVinfull} and Fig.\ref{eefig3},
	\begin{align}
		\label{kcoord}
{	X^+_K=-e^{\theta+r_*}, \ \ X^-_K=  e^{r_*-\theta}}
	\end{align}
	at $H_1$ where $\theta\rightarrow \infty, r_*\rightarrow -\infty$, $X^-_K=0$, while at $H_2$ where $\theta\rightarrow -\infty, r_*\rightarrow -\infty$, 
	$X^+_K=0$. 
	It is easy to see that the Kruskal and Poincare coordinates are related by SL(2,R) transformations and therefore the stress tensor components in both the Kruskal and Poincare vacua are the same and vanish. As a result, in particular the stress tensors in these vacua are well behaved at the two cosmological horizons. 
	
	Now let us consider more general states. The metric eq.\eqref{solaa} has an isometry along the $\partial_\theta$ direction and an interesting class of states are those for which $\langle T_{\mu\nu}\rangle $  is invariant under translation along this  direction. Imposing  that the Lie derivative of the stress tensor along this direction vanishes and that it is conserved leads to a two parameter family
	of allowed stress tensors, which in Poincare coordinates becomes :
	\begin{align}
		T_{x^+x^+}& =  {{\mathcal A} \over (1-(x^+)^2)^2} \label{qspp}\\
		T_{x^-x^-}& =  {{\mathcal B} \over (1-(x^-)^2)^2}\label{qsmm}
	\end{align}
	where ${\mathcal A}, {\mathcal B}$ are free to vary taking values which can be  both positive and negative. 
	
	The corresponding components in Kruskal coordinates, obtained using eq.\eqref{kporel},  are 
	\begin{align}
		T_{X^+_K X^-_K}&={{\mathcal  A}\over 4(X^+_K)^2} \nonumber\\
		T_{X^-_K X^-_K}&={{\mathcal  B}\over 4(X^-_K)^2}\label{stressinkrusk}
	\end{align}
	We see if ${\mathcal A}$ ,  ${\mathcal B}$ are  non-vanishing the corresponding components in Kruskal coordinates blows up at either $H_1$ or $H_2$ and one therefore expects that the back reaction becomes big in the  vicinity of  the locus $r\rightarrow 1$. 
	
	In fact the solution for the dilaton is easy to obtain. For simplicity let us take the case where  ${\cal B}={\cal A}$. In poincare coordinates the solution  is then given by 
	\begin{align}
		\phi=\frac{GN}{3}+4\pi G \mathcal{A}+4\pi G \mathcal{A}\pqty{\frac{x^+ x^- -1}{x^+ -x^-}}(\tanh^{-1}x^+ -\tanh^{-1}x^-)+\frac{c_1 x^+ x^-+c_2(x^+ +x^-)+c_3}{x^+ - x^-}\label{dilsolsemi}
	\end{align}
	The last  term dependent on $c_1,c_2,c_3$ is simply the solution to the homogeneous equations without the stress tensor source turned on. By an appropriate choice of these constants we get a  solution which preserves the symmetry under $\partial_\theta$ translations with the dilaton given   in terms of $r_*$ defined in eq.\eqref{milnepoinco},  by,
	\begin{align}
		\phi=\frac{GN}{3}+4\pi G \mathcal{A}-4\pi G \mathcal{A}\,r_*\coth r_*-{\tilde c} \coth r_*\label{dilsonback} 
	\end{align}
	
	In the asymptotic past, where $r_*\rightarrow 0^-$ with ${\tilde c}>0$ we see that the solution meets the boundary condition that $\phi \rightarrow \infty$. 
	When we come close to the region, 
	$r\rightarrow 1$, $r_*\rightarrow -\infty$, 
	\be
	\label{vladi}
	\phi\rightarrow -4\pi G \mathcal{A} |r_*|.
	\ee
	And for  $\mathcal{A}>0$, when the null energy condition is satisfied, we see that  $\phi\rightarrow -\infty$ and  it follows that  a spacelike singularity forms as {$r_*\rightarrow -\infty$}, as we would have expected from the previous discussion of classical matter. But when  ${\cal A}<0$ and the quantum stress tensor violates the null energy condition, something different and quite interesting happens- now $\phi\rightarrow + \infty$. 
	The reader will recall that $\phi$ arises as the radius of the internal space when we obtain JT gravity from dimensional reduction; $\phi \rightarrow +\infty$ therefore would imply that the internal volume is diverging and the spacetime is de-compactifying. However this happens for finite affine or proper time for geodesics in the resulting spacetime and  not  in an asymptotic dS region, suggesting that perhaps in the  underlying theory one might be able to go past the region near $r=1$. Unfortunately, a proper analysis  requires us to go beyond  the realm of validity of the JT theory, and we will have to leave the study of this fascinating possibility, in a more complete model, for the future. 
	
	\subsection{ The Orbifold}
	\label{orfsubsec}
	We now turn to studying an orbifold of dS space and its behaviour in the presence of quantum matter. 
	The  orbifold we discuss was  introduced above in eq.(\ref{orba}) and eq.(\ref{ido}).
	More precisely we start  with dS space consisting of the four regions - the two Milne regions,  I, II, and the ``Rindler" regions III and IV, see Fig.\ref{dsbhshwaves}, but after carrying out the orbifold identification, eq. (\ref{ido}) we only retain the Milne regions I and II in the spacetime, see Fig.\ref{compthepenro}. 
	We can do this because no world lines leave the regions I and  II, once the orbifold identification eq.(\ref{ido}) has been made. 
	As $r\rightarrow 1$ we note that the length along the spatial $\theta$ direction between $0\le \theta\le b$ shrinks to zero size, making this spacetime an interesting prototype for a big crunch/big bang singularity. 
	
	The nature of the underlying spacetime becomes clearer in Kruskal coordinates $X^+_K,X^-_K$, eq.\eqref{kcoord}.  In the lower Milne patch,  
	the metric and dilaton are given in these coordinates by 
	\begin{align}
		\label{metdilo}
		ds^2 & =  {4 \,dX^+_K dX^-_K\over (1+X^+_K X^-_K)^2} \nonumber \\
		\phi & =  A{(1-X^-_K X^+_K) \over 1+ X^+_K X^-_K}
	\end{align}
	The boost eq.(\ref{ido}) under which the points are identified acts by taking 
	\begin{align}
		\label{rnb}
		X^+_K\rightarrow X^+_K e^b, \ \ X^-_K\rightarrow X^-_K  e^{-b}
	\end{align}
	We see that the locus $r=1$ is given by $X^+_K X^-_k=0$. Thus the spacetime has   null closed curves, along, $X_K^-=0$ and $X_K^+=0$ and an orbifold singularity at $X^+_K=X^-_K=0$ 
	
	Before considering  quantum states for matter  let us briefly consider the effects of adding classical matter.  One might intuitively expect that the back reaction becomes big near the orbifold point,  where space is shrinking, and  a singularity arises in its vicinity. This ties in with what we observed in the previous section for dS space without the orbifold, namely that  
	excess matter beyond the extremal limit causes a spacelike singularity to appear at which spacetime terminates. Intuitively,  one would expect any matter in the orbifold theory to have 	``multiple images" and these images to amplify the effect of the  matter leading to a singularity near the orbifold point.

	We have carried out a preliminary analysis   which supports this intuition. 
	For concreteness let us  take a  shock wave travelling along the $x^+$ direction (right moving) from the asymptotic past. One can calculate its effect on the dilaton by the method of images as discussed in appendix \ref{obcm}. One finds after adding the images that  the total mass being added to the system is very high, in fact it diverges, see appendix \ref{obcm}. And this leads to the conclusion that a spacelike singularity, in effect a big crunch at which spacetime ends, should appear. However, this conclusion is a bit preliminary, since it could be that  the method of images might itself be  perhaps breaking down.  and we leave a more complete analysis of this issue for the future. For a discussion of the use of the method of images for studying back reaction in such orbifold models in higher dimensions  see \cite{Horowitz:2002mw} and for related discussion, see
	\cite{Liu:2002ft,Liu:2002kb}. 
	
	Next,  let us consider some quantum states. Like above we  consider states which preserve the $\partial_\theta$ symmetry of spacetime, the resulting stress tensor should then also preserve this symmetry and be of the form discussed above in eq.\eqref{qspp},\eqref{qsmm}. 
	One would expect that the orbifold identification which has lead to the spacelike direction $\theta$ becoming compact would result in an extra contribution to the stress tensor due to the Casimir effect and this contributions  scales inversely with $b$. The Casimir  contribution is also expected to  violate the null energy condition, i.e. the resulting contribution to $T_{++}$ or $T_{--}$ would be negative. 	
	
	In appendix \ref{sttrn} we describe one state for which this Casimir effect is calculated. The resulting stress tensor has the form, eq.\eqref{qspp},\eqref{qsmm}, for $T_{++}$ and similarly for $T_{--}$ with ${\mathcal A}={\mathcal B}$ and its value being given by 
	\be
	\label{vala}
	{\cal A}=-{1\over 12\pi}\left(1+ {4 \pi^2\over b^2}\right)
	\ee
	Note, that in the $b\rightarrow \infty$ limit this is the stress energy in the state which is the vacuum for the modes, $\log(-X_K^+),  \log(X_K^-)$, i.e. the ``Schwarzschild" modes. We see that as $b\rightarrow 0$ the Casimir effect gets more pronounced and ${\cal A}$ more negative. 
	From the analysis above it follows that in this case $\phi$ will diverge near the orbifold singularity where $r\rightarrow 1$. 
	It would be wonderful to  be able to investigate the behaviour of the  higher dimensional version of such a  model more completely.
	
	In the following section we will turn to quantising the full theory, including the metric and dilaton sector. It will turn out that the path integral, for appropriate topology, involves a  sum over all values of $b$. After carrying out the path integral  we will find in some cases that a transition from the far past to an expanding universe in the future is indeed possible.

	\tikzset{every picture/.style={line width=0.75pt}} 
	\begin{figure}
		
		\centering
		\label{compthepenro}
		\begin{tikzpicture}[x=0.75pt,y=0.75pt,yscale=-1,xscale=1]
			
			\draw    (360,118) -- (488,280) ;
			\draw [fill={rgb, 255:red, 214; green, 73; blue, 9 }  ,fill opacity=1 ]   (488,118) -- (360,280) ;
			\draw   (360,118) -- (488,118) -- (488,280) -- (360,280) -- cycle ;
			\draw   (373,129) .. controls (407,182.33) and (441,182.33) .. (475,129) ;
			\draw   (374,123) .. controls (407,161.67) and (440,161.67) .. (473,123) ;
			\draw   (475.8,270.58) .. controls (442.4,216.87) and (408.41,216.48) .. (373.8,269.42) ;
			\draw   (472.74,275.55) .. controls (440.18,236.52) and (407.18,236.14) .. (373.75,274.43) ;
			\draw  [color={rgb, 255:red, 65; green, 117; blue, 5 }  ,draw opacity=1 ][fill={rgb, 255:red, 184; green, 233; blue, 134 }  ,fill opacity=0.46 ] (424,199) -- (394,280.5) -- (424,280.5) -- cycle ;
			\draw  [color={rgb, 255:red, 65; green, 117; blue, 5 }  ,draw opacity=1 ][fill={rgb, 255:red, 184; green, 233; blue, 134 }  ,fill opacity=0.46 ] (424,199) -- (452.32,117.89) -- (422.33,118.52) -- cycle ;
			
			\draw (440.26,181.96) node [anchor=north west][inner sep=0.75pt]  [font=\footnotesize,rotate=-312.37] [align=left] {r=$\displaystyle 1$};
			\draw (413,104) node [anchor=north west][inner sep=0.75pt]  [font=\footnotesize] [align=left] {r=$\displaystyle \infty $};

			%
		\end{tikzpicture}	
	\end{figure}

	\section{Quantisation  of JT dS gravity}
	\label{moddivs}
	We now turn to quantising the full theory including the metric and dilaton sector. In fact, in this section we will only consider pure JT gravity, without matter. 
	The JT theory can be quantised in the first order formalism where it gives rise to BF theory,  see e.g.,\cite{Saad:2019lba} or  in the second order formalism, 
	\cite{Moitra:2021uiv}, where one works directly  with   $g_{\mu\nu}, \phi$.  
	
	Let us  summarise some of the key points in the previous literature  and also in the discussion below at the outset here. With one boundary, the path integral, subject to  appropriate boundary conditions, gives rise to the Hartle-Hawking  wave function of the universe, which describes the universe being ``born"  by tunnelling out of ``nothing", \cite{PhysRevD.28.2960}. When we are dealing with multiple boundaries the path integral, depending on the contour chosen and the analytic continuation carried out, either gives rise to the HH wave function for producing several disconnected universes, or transition amplitudes for some number, $n$,   of universes in the past to evolve  to $m$ universes universes in the future.
	
	We will be  interested here  in asymptotic boundaries in dS space, which can only be reached along geodesics at diverging proper or affine time.
	This corresponds to obtaining the wave function at ``late times" when the universe has a diverging size, or to transition amplitudes from the far past to the far future. 
	The metric in the vicinity of such an asymptotic boundary takes the form, 
	\begin{align}
		\label{assf}
		ds^2\simeq r^2 d\theta^2-{dr^2\over r^2}
	\end{align}
	We will be interested in boundaries of fixed length ${\hat l}_i$ where the dilaton $\phi_{B_i}$  takes specified boundary value.
	Both ${\hat l}_i,\phi_{B_i}$ at an asympytotic boundary diverge and we can write them as 
	\begin{align}
		\phi_{B_i}={1\over J  \epsilon_i}, {\hat l}_i={l_i \over \epsilon_i},\,\epsilon_i\rightarrow 0. \quad\label{dbasymcond}
	\end{align}
	Here $\epsilon_i$ is a cut-off  defined for each boundary independently and   we will be interested in the  limit $\epsilon_i\rightarrow 0$. 
	In eq.(\ref{dbasymcond}) we see that the ratio $\phi_{B_i}/{\hat l}_i={ 1\over Jl_i} $ remains finite  as $\epsilon_i\rightarrow 0$;    $l_i$ can be thought of as a ``renormalised" length which is also  finite. 
	
	In general we are interested in the multi-boundary case and at arbitrary genus. The path integral will then depend on the values $\phi_{B_i}, {\hat l}_i$ at each boundary. 
	
	 The JT  path integral  is evaluated by rotating the contour for the dilaton to be along the imaginary direction, by taking $ \phi \rightarrow i  \phi$. This gives rise to a constraint that localises the metric to have constant curvature $R=2$. 
	The  metric path integral is then done using   a spacetime of constant curvature with  $(0,2)$ signature and analytically continuing it to the dS spacetime of signature $(1,1)$, eq.\eqref{assf}.
	The degrees of freedom which remain are the moduli, and the   boundary reparametrization modes, at each boundary. These are    valued in $\text{Diff} (S^1)/\text{SL}(2,R)$ and described by a Schwarzian action. {Each set of these reparametrisation modes  are  integrated with a measure which follows from the ultra-local measure for metric deformations \cite{Moitra:2021uiv} and which exactly agrees with the measure considered earlier in the literature, \cite{Stanford:2017thb,Saad:2019lba} }.
Finally, as mentioned above,  in the vicinity of each boundary which lies  in the  asymptotic -AdS$_2$ region where the metric can be taken to have the form
\begin{align}
		\label{assr}
		ds^2\simeq-\pqty{r^2 d\theta^2 + {dr^2\over r^2}}
	\end{align}
for $r\gg 1$, 
we need to continue the metric from signature $(0,2)$ to $(1,1)$ arriving at dS space in the asymptotic region where the metric is given by 
eq.(\ref{assf}). 
\subsection{Analytic continuations}
	\label{transamp}
	In fact two analytic continuations are possible to go from a metric eq.(\ref{assr}) to eq.(\ref{assf}). 
	at each boundary and we turn to describing them in some detail next. 
	Note that the coordinate $r$ in eq.(\ref{assr}) can locally be taken to be positive at the boundary of interest,
	To continue to dS$_2$ in the vicinity of that boundary we can then either take 
	\begin{align}
		\label{firsta}
		r\rightarrow + i r
	\end{align}
	or 
	\begin{align}
		\label{seconda}
		r\rightarrow - i r
	\end{align}
	
	As was mentioned above  the dilaton path integral is done by rotating $\phi\rightarrow i \phi$, the bulk term in the JT action then gives rise to a delta function,   $\delta (R-2)$,
	which restricts the sum over metrics to constant curvature ones.  The only remaining term in the action in eq.\eqref{jtact} then is the boundary term which in -AdS$_2$  takes the form, 
	\begin{align}
		\label{btr}
		S_{JT,\partial}= -{\phi_B \over 8 \pi G} \int_\partial ds\, K
	\end{align}
	The extrinsic curvature $K = \nabla_\mu n^\mu$ where $n^\mu$ is the outward drawn normal, then goes like 
	$K=1+\text{corrections}$,  and this gives for the boundary located at $r_B$ 
	\begin{align}
		\label{btra}
		S_{JT,\partial}= -{\phi_B \over 8 \pi G} (2 \pi r_B)
	\end{align}
	at leading order. Now taking the continuation eq.(\ref{firsta}) or eq.(\ref{seconda}) gives rise to the action 
	\begin{align}
		\label{acta}
		S_{JT,\partial}=\mp  {i\phi_B \over 8 \pi G} (2 \pi r_B)\simeq  \mp  {i\phi_B {\hat l} \over 8 \pi G},\qquad r\rightarrow \pm ir
	\end{align}
	where we used the fact that the length of the boundary is $\hat{l}\simeq 2\pi r_B$. Since the path integral in our conventions involves the weighting factor $\Psi \sim e^{-S}$ we get the phase factor in the wave function
	\begin{align}
		\label{pfa}
		\Psi \sim e^{\pm  {i\phi_B {\hat l} \over 8 \pi G}}
	\end{align}
	with the $+$ arising for eq.(\ref{firsta}) and the $-$ sign for eq.(\ref{seconda}). 
	From eq.(\ref{dbasymcond}) we also see that at each boundary the phase factor in eq.(\ref{pfa}) diverges. 
	
	What is the physical interpretation of these two analytic continuations? 
	If we carry out a minisuperspace quantisation in the second order formalism for one universe,   the momentum conjugate to the length ${\hat l}$ is \cite{Maldacena:2019cbz} $\pi_{\hat l} = - \frac{\dot{\phi}}{8\pi G} $, see eq.\eqref{conjmom}, eq.(\ref{pilin}) . Canonical commutation relations mean that $\pi_{\hat l}=  -i \partial_{\hat {l}}$ . Now if the direction of time is taken towards the increasing value of the dilaton, $\dot{\phi}>0$, and therefore we learn that for the wave function describing an expanding universe 
	$\pi_{\hat l} \ln \Psi=-i \partial_{\hat l} \ln \Psi <0$.
	This tells us that for one boundary carrying out the analytic continuation eq.(\ref{seconda}) which gives from eq.(\ref{pfa})
	$\Psi \sim e^{{-i \phi_B {\hat l}\over 8 \pi G}}$,   leads to  the amplitude for   a universe which is expanding 
	whereas taking the continuation as in eq.(\ref{firsta}), which gives  $\Psi \sim e^{{+i \phi_B {\hat l}\over 8 \pi G}}$, leads to   the amplitude for a contracting universe\footnote{Also, while we are not considering matter in this section, as discussed in \cite{Maldacena:2019cbz},  the phase factor, $e^{-i \phi_B {\hat l} \over 8 \pi G}$, with a Klein Gordon norm, $ \langle \bar{\Psi}, \Psi\rangle= \int dl [{\bar \Psi} \partial_l \Psi-\partial_l  {\bar \Psi} \Psi]$, gives rise to  a sensible positive norm $|\Psi_{{\rm matter}}|^2$ for the matter part of the wave function, which we will use in {section {\ref{dswithmatter}}}.}. The divergence in the  phase factor  in eq.(\ref{pfa}) can be understood in the minisuperspace approximation as arising from  the fact that for a large expanding or contracting universe one is in the classical regime, where the WKB approximation is valid. 
	
	Similarly, when we consider the canonical quantisation in the first order formalism as described in appendix \ref{canquant}, we find that one can take $e^1_\theta$ to be a physical clock  $t_{\text{phys}}$ in the system.  For an expanding universe (in  the future where $t$ defined in eq.(\ref{metindscomp})  satisfies the condition $ t>0$), $e^1_\theta =  \frac{\hat{l}}{2\pi}$, where as for a contracting universe, $t<0 $ in eq.(\ref{metindscomp}), $e^1_\theta= -\frac{\hat{l}}{2\pi}$. Thus we see that the continuation eq.(\ref{seconda})  gives rise to a state where $ i \partial_{ t_{\text{phys}}} \ln \Psi>0$   in the future whereas the continuation eq.(\ref{firsta}) gives rise to a state for which  $i \partial_{t_{\text{phys}}}\ln  \Psi>0$ in the past. 
	
	In an analogous manner when we have several boundaries and we continue $n$ of them using the continuation eq.(\ref{firsta}) and $m$ using 
	eq.(\ref{seconda}) we would be describing a transition amplitude to start in the far past with $n$ contracting  universes which  tunnel in the far future to $m$ expanding universes. If all the $n$ boundaries are continued in the same way we have  the HH wave function for producing $n$ expanding universes, eq.(\ref{seconda}),  or $n$ contracting universes, eq.(\ref{firsta}). 
	
\begin{figure}
		\centering

		\tikzset{every picture/.style={line width=0.75pt}} 
		
		\begin{tikzpicture}[x=0.75pt,y=0.75pt,yscale=-1,xscale=1]
			
			\draw  [draw opacity=0] (156.11,56.03) .. controls (156.11,56.03) and (156.11,56.03) .. (156.11,56.03) .. controls (156.07,64.28) and (141.64,70.84) .. (123.87,70.67) .. controls (106.11,70.5) and (91.75,63.67) .. (91.79,55.41) .. controls (91.79,55.36) and (91.79,55.3) .. (91.79,55.24) -- (123.95,55.72) -- cycle ; \draw   (156.11,56.03) .. controls (156.11,56.03) and (156.11,56.03) .. (156.11,56.03) .. controls (156.07,64.28) and (141.64,70.84) .. (123.87,70.67) .. controls (106.11,70.5) and (91.75,63.67) .. (91.79,55.41) .. controls (91.79,55.36) and (91.79,55.3) .. (91.79,55.24) ;
			\draw  [draw opacity=0] (91.79,102.31) .. controls (91.79,102.31) and (91.79,102.31) .. (91.79,102.31) .. controls (91.76,94.05) and (106.13,87.25) .. (123.89,87.1) .. controls (141.65,86.96) and (156.08,93.54) .. (156.11,101.79) .. controls (156.11,101.85) and (156.11,101.9) .. (156.11,101.96) -- (123.95,102.05) -- cycle ; \draw   (91.79,102.31) .. controls (91.79,102.31) and (91.79,102.31) .. (91.79,102.31) .. controls (91.76,94.05) and (106.13,87.25) .. (123.89,87.1) .. controls (141.65,86.96) and (156.08,93.54) .. (156.11,101.79) .. controls (156.11,101.85) and (156.11,101.9) .. (156.11,101.96) ;
			\draw   (153.22,79.02) .. controls (153.22,66.32) and (154.51,56.02) .. (156.11,56.02) .. controls (157.71,56.02) and (159,66.32) .. (159,79.02) .. controls (159,91.73) and (157.71,102.02) .. (156.11,102.02) .. controls (154.51,102.02) and (153.22,91.73) .. (153.22,79.02) -- cycle ;
			\draw   (88.9,79.31) .. controls (88.9,66.61) and (90.2,56.31) .. (91.79,56.31) .. controls (93.39,56.31) and (94.68,66.61) .. (94.68,79.31) .. controls (94.68,92.01) and (93.39,102.31) .. (91.79,102.31) .. controls (90.2,102.31) and (88.9,92.01) .. (88.9,79.31) -- cycle ;
			\draw  [draw opacity=0] (280.48,43.16) .. controls (280.48,43.16) and (280.49,43.16) .. (280.49,43.16) .. controls (288.74,43.04) and (295.58,57.34) .. (295.76,75.1) .. controls (295.94,92.86) and (289.4,107.36) .. (281.15,107.48) .. controls (281.09,107.48) and (281.04,107.48) .. (280.98,107.48) -- (280.82,75.32) -- cycle ; \draw   (280.48,43.16) .. controls (280.48,43.16) and (280.49,43.16) .. (280.49,43.16) .. controls (288.74,43.04) and (295.58,57.34) .. (295.76,75.1) .. controls (295.94,92.86) and (289.4,107.36) .. (281.15,107.48) .. controls (281.09,107.48) and (281.04,107.48) .. (280.98,107.48) ;
			\draw  [draw opacity=0] (328.04,106.55) .. controls (328.04,106.55) and (328.04,106.55) .. (328.03,106.55) .. controls (319.78,106.74) and (312.69,92.51) .. (312.19,74.76) .. controls (311.7,57) and (317.99,42.45) .. (326.24,42.25) .. controls (326.29,42.25) and (326.35,42.25) .. (326.41,42.25) -- (327.14,74.4) -- cycle ; \draw   (328.04,106.55) .. controls (328.04,106.55) and (328.04,106.55) .. (328.03,106.55) .. controls (319.78,106.74) and (312.69,92.51) .. (312.19,74.76) .. controls (311.7,57) and (317.99,42.45) .. (326.24,42.25) .. controls (326.29,42.25) and (326.35,42.25) .. (326.41,42.25) ;
			\draw   (303.53,45.59) .. controls (290.83,45.85) and (280.51,44.76) .. (280.48,43.16) .. controls (280.45,41.57) and (290.72,40.07) .. (303.42,39.82) .. controls (316.12,39.56) and (326.44,40.65) .. (326.47,42.25) .. controls (326.5,43.84) and (316.23,45.34) .. (303.53,45.59) -- cycle ;
			\draw   (305.1,109.89) .. controls (292.4,110.15) and (282.08,109.06) .. (282.05,107.46) .. controls (282.02,105.87) and (292.29,104.37) .. (304.99,104.11) .. controls (317.69,103.86) and (328.01,104.95) .. (328.04,106.55) .. controls (328.07,108.14) and (317.8,109.64) .. (305.1,109.89) -- cycle ;
			\draw  [draw opacity=0] (158.11,176.03) .. controls (158.11,176.03) and (158.11,176.03) .. (158.11,176.03) .. controls (158.07,184.28) and (143.64,190.84) .. (125.87,190.67) .. controls (108.11,190.5) and (93.75,183.67) .. (93.79,175.41) .. controls (93.79,175.36) and (93.79,175.3) .. (93.79,175.24) -- (125.95,175.72) -- cycle ; \draw   (158.11,176.03) .. controls (158.11,176.03) and (158.11,176.03) .. (158.11,176.03) .. controls (158.07,184.28) and (143.64,190.84) .. (125.87,190.67) .. controls (108.11,190.5) and (93.75,183.67) .. (93.79,175.41) .. controls (93.79,175.36) and (93.79,175.3) .. (93.79,175.24) ;
			\draw  [draw opacity=0] (93.79,222.31) .. controls (93.79,222.31) and (93.79,222.31) .. (93.79,222.31) .. controls (93.76,214.05) and (108.13,207.25) .. (125.89,207.1) .. controls (143.65,206.96) and (158.08,213.54) .. (158.11,221.79) .. controls (158.11,221.85) and (158.11,221.9) .. (158.11,221.96) -- (125.95,222.05) -- cycle ; \draw   (93.79,222.31) .. controls (93.79,222.31) and (93.79,222.31) .. (93.79,222.31) .. controls (93.76,214.05) and (108.13,207.25) .. (125.89,207.1) .. controls (143.65,206.96) and (158.08,213.54) .. (158.11,221.79) .. controls (158.11,221.85) and (158.11,221.9) .. (158.11,221.96) ;
			\draw   (155.22,199.02) .. controls (155.22,186.32) and (156.51,176.02) .. (158.11,176.02) .. controls (159.71,176.02) and (161,186.32) .. (161,199.02) .. controls (161,211.73) and (159.71,222.02) .. (158.11,222.02) .. controls (156.51,222.02) and (155.22,211.73) .. (155.22,199.02) -- cycle ;
			\draw   (90.9,199.31) .. controls (90.9,186.61) and (92.2,176.31) .. (93.79,176.31) .. controls (95.39,176.31) and (96.68,186.61) .. (96.68,199.31) .. controls (96.68,212.01) and (95.39,222.31) .. (93.79,222.31) .. controls (92.2,222.31) and (90.9,212.01) .. (90.9,199.31) -- cycle ;
			\draw  [draw opacity=0] (374.6,169.1) .. controls (370.07,204.08) and (348,230.73) .. (321.25,230.99) .. controls (294.22,231.25) and (271.56,204.48) .. (266.83,169.04) -- (320.69,155.54) -- cycle ; \draw   (374.6,169.1) .. controls (370.07,204.08) and (348,230.73) .. (321.25,230.99) .. controls (294.22,231.25) and (271.56,204.48) .. (266.83,169.04) ;
			\draw  [draw opacity=0] (349.82,167.25) .. controls (347.6,195.38) and (335.66,216.86) .. (321.09,217.01) .. controls (306.62,217.15) and (294.42,196.17) .. (291.68,168.35) -- (320.64,157.14) -- cycle ; \draw   (349.82,167.25) .. controls (347.6,195.38) and (335.66,216.86) .. (321.09,217.01) .. controls (306.62,217.15) and (294.42,196.17) .. (291.68,168.35) ;
			\draw   (362.26,172.84) .. controls (355.41,172.75) and (349.88,170.95) .. (349.9,168.83) .. controls (349.92,166.71) and (355.49,165.06) .. (362.34,165.15) .. controls (369.19,165.24) and (374.73,167.03) .. (374.7,169.16) .. controls (374.68,171.28) and (369.11,172.93) .. (362.26,172.84) -- cycle ;
			\draw   (279.23,172.04) .. controls (272.38,171.95) and (266.85,170.15) .. (266.87,168.03) .. controls (266.89,165.9) and (272.46,164.25) .. (279.31,164.34) .. controls (286.16,164.43) and (291.7,166.23) .. (291.68,168.35) .. controls (291.65,170.48) and (286.08,172.12) .. (279.23,172.04) -- cycle ;
			\draw    (164,79) .. controls (181.37,80.93) and (183.84,69.82) .. (181.3,53.77) ;
			\draw [shift={(181,52)}, rotate = 79.99] [color={rgb, 255:red, 0; green, 0; blue, 0 }  ][line width=0.75]    (10.93,-3.29) .. controls (6.95,-1.4) and (3.31,-0.3) .. (0,0) .. controls (3.31,0.3) and (6.95,1.4) .. (10.93,3.29)   ;
			\draw    (166,203) .. controls (183.37,204.93) and (185.84,193.82) .. (183.3,177.77) ;
			\draw [shift={(183,176)}, rotate = 79.99] [color={rgb, 255:red, 0; green, 0; blue, 0 }  ][line width=0.75]    (10.93,-3.29) .. controls (6.95,-1.4) and (3.31,-0.3) .. (0,0) .. controls (3.31,0.3) and (6.95,1.4) .. (10.93,3.29)   ;
			\draw    (81,201) .. controls (66.45,204.88) and (61.31,198.41) .. (62.84,178.85) ;
			\draw [shift={(63,177)}, rotate = 95.44] [color={rgb, 255:red, 0; green, 0; blue, 0 }  ][line width=0.75]    (10.93,-3.29) .. controls (6.95,-1.4) and (3.31,-0.3) .. (0,0) .. controls (3.31,0.3) and (6.95,1.4) .. (10.93,3.29)   ;
			\draw    (83,82) .. controls (68.52,80.07) and (62.43,90.25) .. (63.82,106.24) ;
			\draw [shift={(64,108)}, rotate = 263.29] [color={rgb, 255:red, 0; green, 0; blue, 0 }  ][line width=0.75]    (10.93,-3.29) .. controls (6.95,-1.4) and (3.31,-0.3) .. (0,0) .. controls (3.31,0.3) and (6.95,1.4) .. (10.93,3.29)   ;
			\draw   (210,72) -- (231.6,72) -- (231.6,68) -- (246,76) -- (231.6,84) -- (231.6,80) -- (210,80) -- cycle ;
			\draw   (208,184) -- (229.6,184) -- (229.6,180) -- (244,188) -- (229.6,196) -- (229.6,192) -- (208,192) -- cycle ;
			
			\draw (165.25,37) node [anchor=north west][inner sep=0.75pt]  [font=\scriptsize,xslant=-0.08] [align=left] {$\displaystyle r\rightarrow -ir$};
			\draw (262,116) node [anchor=north west][inner sep=0.75pt]   [align=left] {Past to Future};
			\draw (256,242) node [anchor=north west][inner sep=0.75pt]   [align=left] {Nothing to two universes};
			\draw (165.25,160) node [anchor=north west][inner sep=0.75pt]  [font=\scriptsize,xslant=-0.08] [align=left] {$\displaystyle r\rightarrow -ir$};
			\draw (43.25,160) node [anchor=north west][inner sep=0.75pt]  [font=\scriptsize,xslant=-0.08] [align=left] {$\displaystyle r\rightarrow -ir$};
			\draw (45.25,113) node [anchor=north west][inner sep=0.75pt]  [font=\scriptsize,xslant=-0.08] [align=left] {$\displaystyle r\rightarrow ir$};

		\end{tikzpicture}
		\caption{AdS to dS double trumpet analytic continuations}
		\label{adsdsdtfig}
	\end{figure}
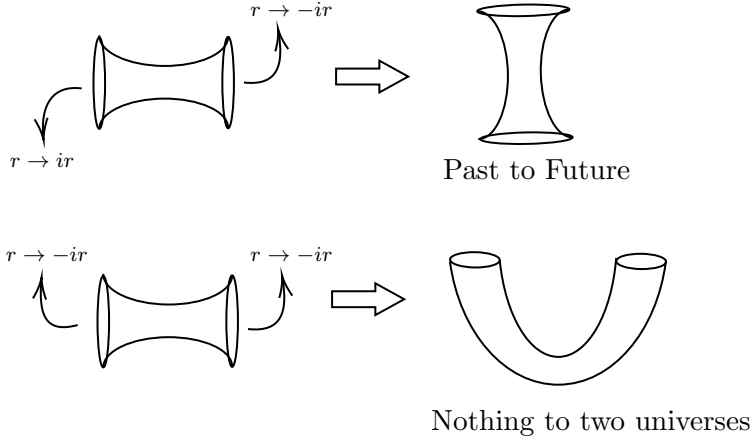

	The different continuations are illustrated for two boundaries in Fig.\ref{adsdsdtfig}. 
	
	Let us end this subsection with one comment. In AdS space   the  path integral with one or more boundaries  corresponds to the partition function or to correlations functions of the partition function of the  boundary theory(ies). And a  divergent  term related to the one   we have  discussed above  is absent, after a counterterm is added in the action as per the procedure of holographic renormalisation. This counterterm is   local in the boundary theory and   we are legitimately allowed to add it, as per standard renormalisation theory, when calculating the partition function. However in the dS case we are calculating a wave function (or a transition amplitude) and the divergent term cannot be cancelled and in fact has physical significance, as we have discussed above. 
	
\subsection{One and Two Boundary Cases}
\label{onetwob}
	Let us give some more details for  the one  and two boundary cases now. 
	
	{\bf One Boundary: }
	The one boundary case gives the wave function for one universe with dilaton $\phi_B$ and length ${\hat l}$, denoted  $\Psi[{\hat l }, \phi_B]$.
	
	This  case is special and  one can either consider a contour with   metric of $(0,2)$ signature, as mentioned above, or $(2,0)$ signature, in evaluating the path integral with one boundary. The dilaton integral forces the metric to be of constant positive curvature and 
	the $(0,2)$ and $(2,0)$ signature contours involve the same  metric
	\begin{align}
	\label{metaa}
	ds^2=-\left((r^2-1) d\theta^2+{dr^2\over r^2-1}\right),
	\end{align}
	with $r$ taking values $r>1$ and $r<1$ for the $(0,2)$ and $(2,0)$ cases respectively. 
	The resulting path integral gives rise to the Hartle-Hawking  (HH) wave function which is obtained by continuing the result  to dS$_2$ space. 
	
	To continue to dS$_2$ from  -AdS$_2$ case we go to  $r\gg 1$ and can then take $r\rightarrow -i r$, eq.(\ref{seconda}).
	In the $(2,0)$ case we continue at $r=0$ by taking $r\rightarrow -i r$.
	Both calculations give the same answer, as was discussed in section 5 of \cite{Moitra:2021uiv}  and gives the wave function  corresponding to the branch of the HH wave function which describes an expanding universe,   
	\begin{align}
	\label{val}
	\Psi_{\text{HH}}^{\text{exp}}[{\hat l }, \phi_B]={\cal N} e^{S_0}  \left({\phi_B \over {\hat l} }\right)^{3/2} e^{-{i \phi_B {\hat l}\over 8 \pi G}+\frac{i\pi\phi_B}{4G\hat{l}}}
	\end{align}
	{where $S_0$ is the half the 4D extremal entropy, discussed in appendix \ref{pennhlim}}. This result also agrees with what is obtained in the first order formalism. 
	From the discussion  in the previous subsection  it follows that this  wave function corresponds to  a universe which is expanding. 
	Note that we have not fixed the  overall normalisation ${\cal N}$ of the wave function. This normalisation is uncertain partly because the overall normalisation of the measure in the path integral is ambiguous and also because there is a phase factor whose value needs to be decided carefully \footnote{With the definition of the measure and our conventions as chosen in \cite{Moitra:2021uiv} we get $\mathcal{N}={({-32\pi i G })^{-\frac{3}{2}}}$. This gives rise to the phase factor of $\mathcal{N}$,   referred to above  eq.\eqref{valp}, to be $\theta= {\frac{3\pi}{4}}$or $\theta=-\frac{\pi}{4}$..}. 
	
	Alternatively continuing the $(0,2)$ or $(2,0)$ metrics by taking $r \rightarrow +i r$ as given in eq.(\ref{firsta})  gives rise to the wave function for a contracting universe, as also follows from the discussion above, leading to, 
	\begin{align}
	\label{valtwo}
	\Psi_{\text{HH}}^{\text{con}}[{\hat l }, \phi_B]=
	( {\cal N} )^* e^{S_0}  \left({\phi_B \over {\hat l} }\right)^{3/2} 
	    e^{ {i \phi_B {\hat l}\over 8 \pi G}-\frac{i\pi\phi_B}{4G\hat{l}}}
	\end{align}
	Note that while the overall normalisation of this branch cannot be determined either,  it can be shown to be  the complex conjugate of the expanding branch wave function\footnote{Any  phase in the overall normalisation of the measure will be common to both branches and is a standard phase ambiguity in the wave function in Quantum Mechanics.}.  Accordingly we have denoted it by  $({\cal N})^*$. Thus we see that the expanding and contracting branches of the HH wave function are complex conjugates of each other. 
	
	We also remind the reader that we have only carried out the path integral in the asymptotic dS limit where both the dilaton and the length diverge, meeting the condition, eq.(\ref{dbasymcond}). 
	Also,  strictly speaking, the wave function in both branches contains an extra factor of the ratio of  determinants ${\sqrt{\det'P^\dagger P}\over \det({ \nabla}^2+2)}$, see eq.(5.40),(5.41) of \cite{Moitra:2021uiv}. This ratio, depending on how the determinants are regulated, could go like $e^{c {\hat l}}$. Such a term, for the asymptotic dS case,   can be absorbed by shifting $\phi_B$ by a constant \footnote{When there are more than one boundaries this shift (by the same constant) will take care of such a term at all boundaries.}. 
	
	Adding the expanding and contracting branches we see from above that the result we obtain is 
	\be
	\label{resb}
	\Psi_{\text{HH}}({\hat l},\phi_B)=\Psi_{\text{HH}}^{\text{exp}}({\hat l},\phi_B)+\Psi_{\text{HH}}^{\text{con}}({\hat l}, \phi_B),
	\ee
	 which is  real. 
	
	The $(2,0)$ signature contour is  an $S^2$ hemisphere which is continued to dS space at the equator, this is the  $2$ dimensional analogue of the $S^4$ instanton for dS$_4$ which was considered by Hartle and Hawking   in defining their wave function \cite{PhysRevD.28.2960}.  In that case the boundary condition imposed was that $\Psi$ should vanish for $\sqrt{h}<0$, where $\sqrt{h}$ is the volume element of the $3$-geometry in ADM gauge,  and this also gave rise to a real wave function (its being real is tied to the CPT invariance of the state). 
	
	From the discussion in appendix \ref{canquant} it follows that in the minisuperspace approximation the wave function which agrees with the asymptotic limit, eq.(\ref{resb}) is given by, 
	\begin{align}
	\label{asmini}
	\Psi({\hat l}, \phi_B)= {e^{S_0}\frac{|{\cal N}|}{4\sqrt{G}}} \left( {{\hat l} \phi_B^2\over {\hat l}^2 - 4\pi^2} \right) \left[e^{i \alpha} H^{(2)}_2\left({\phi_B\sqrt{{\hat l}^2- 4\pi^2}\over 8 \pi G}\right) + 
	e^{-i\alpha} H^{(1)}_2\left({\phi_B\sqrt{{\hat l}^2- 4\pi^2}\over 8 \pi G}\right)\right]
	\end{align}
         where $H^{(1)}_a, H^{(2)}_a$ are the Hankel functions of the first and second kind respectively and  denoting the normalisation ${\cal N}$ in eq.(\ref{val}) by ${\cal N}=|{\cal N}| e^{i \theta}$ we have 
         \begin{align}
         \label{valp}
         e^{i \alpha}= -e^{i (\theta-\frac{\pi}{4})}
         \end{align}
         
         For $\alpha=0$ we find a rather nice result, 
         \begin{align}
	\label{asminia}
	\Psi({\hat l}, \phi_B)= -e^{S_0} \frac{|{\cal N}|}{2\sqrt{G}} \left( {{\hat l} \phi_B^2\over {\hat l}^2 - 4\pi^2} \right) J_2\left({\phi_B\sqrt{{\hat l}^2- 4\pi^2}\over 8 \pi G}\right)
	\end{align}
	This wave function is well defined at the turning point where ${\hat l}^2=4 \pi^2$ and it vanishes when ${\hat l}\rightarrow 0$ or $\phi\rightarrow 0$. 
	However, as mentioned above, we are not  able to fix the phase $e^{i\theta}$ with full certainty  and therefore  cannot establish whether this is the wave function which arises from the path integral quantisation described above. 
	 For other values of $\alpha$ the wave function diverges at the turning point. 
	 
	It is also worth commenting on our result above in relation to some of the recent literature. 
		The expanding branch of the HH wave function was discussed in \cite{Maldacena:2019cbz} and \cite{Iliesiu:2020zld}  . In \cite{Iliesiu:2020zld} the wave function for finite ${\hat l}, \phi$, as opposed to the asymptotic limit, was  obtained and a wave function of the form eq.(\ref{asmini}) was also considered. 
	In \cite{Vilenkin:2021awm}  these wave functions were also discussed  and by quantising the closely related Kantowski-Sachs model in the mini-superspace approximation  it was found that the HH wave function was in fact  of another form. 
	We leave a further study of these discrepancies and different possibilities for the future.

	{\bf Two Boundaries :}
	For two boundaries the path integral is carried out using the ``double trumpet" geometry of $(0,2)$ signature and analytically continuing it, as discussed in section 6 of \cite{Moitra:2021uiv}. This gives\footnote{There is an additional factor given by $\frac{\sqrt{\det'P^\dagger P}}{\det(\nabla^2+2)}$ in eq.\eqref{psiad} which can be b-dependent. However, there is an ambiguity in this b-dependence due to the way the determinants are regulated, as discussed in detail in \cite{Moitra:2021uiv}. To get agreement with the results in the first order formalism, we take this factor to be unity here and in the rest of the paper. }, 
	\begin{align}
		\label{psiad}
		\Psi [\hat{l}_R,\phi_{B_R},\hat{l}_L,\phi_{B_L}]= \int b \,db\, Z_{f}^\pm (\phi_{B_L}, {\hat l}_L,b) Z_{f}^\pm (\phi_{B_R},{\hat l}_R,b) 
	\end{align}

	Each ``flaring" boundary of the double trumpet gives rise to a $Z_f$ factor which is a function of the respective boundary values $\phi_B, {\hat l}$.
	The two boundaries are denoted by $L,R$ here.
	The two different analytic continuations, eq.(\ref{firsta}), eq.(\ref{seconda}) give rise to the two possible factors $Z_f^\pm$ respectively at each boundary, with 
	 $Z_{f}^\pm( \phi_B,{\hat l},b)$ being given by 
	\begin{align}
		\label{zv}
		Z_f^\pm(\phi_B, {\hat l},b)={1\over \sqrt{\pm 16 i\pi^2 GJl} } e^{\pm i \left({\phi_B {\hat l} \over 8 \pi G}  + { b^2 	\over 16 \pi G Jl}\right)}
	\end{align}
	and 
	\begin{align}
	{1\over J l}= {\phi_B\over {\hat l}}\label{jllhat}
	\end{align}. 
	{Note that in our conventions here $\sqrt{\pm i}= e^{\pm i \pi/4}$}. 
	

	The RHS in eq.(\ref{psiad}) arises after integrating over boundary diffeomorphisms at the two boundaries. The two sets of diffeomorphisms decouple from each  other, both sets are governed by a Schwarzian action,  and {the path integral} involves the  measure we mentioned above which arises in the one-boundary case, The variable $b$ in eq.\eqref{psiad}  is a modulus which arises from the metric degrees of freedom, we see that the $Z_f$ factors also depend on $b$.  In fact both the $b$ modulus and the boundary diffeomorphisms correspond to metric degrees of freedom which arise from zero modes of the operator $P^\dagger P$, defined in section 2.1 of \cite{Moitra:2021uiv}. The measures for integrating over them arises from the ultra local measure over the space of metric deformations (the Weil-Petersson metric),
	\begin{align}
		\label{ull}
		\langle \delta g_{ab}, \delta g_{cd}\rangle = \int d^2x \sqrt{g} g^{ac} g^{bd} \delta g_{ab}\delta g_{cd}
	\end{align}
	This has a close  parallel to  what happens in the first order formalism where these degrees of freedom arise from flat connections and a measure on  flat connections give rise to the {same  measure as obtained in the second order formalism}, for summing over the diffeomorphisms and the $b$ modulus. 
	
	
	In the discussion below we will be particularly interested in the case where one boundary say $L$ is continued to the ``past" using eq.(\ref{firsta}), while the second say $R$ is continued to the ``future", using eq.(\ref{seconda}). This will give rise to a tunnelling amplitude describing an initially contracting universe which will tunnel to an expanding one at late times. 
	In this context it is worth  noting that in  the double trumpet  integral for a fixed value of the modulus $b$ the dS$_2$ geometry after continuation from  -AdS$_2$  is given by the metric 
	\begin{align}
		\label{maf}
		ds^2=(r^2-1) d\theta^2 - {dr^2\over (r^2-1)},\quad \theta\sim \theta+b
	\end{align}
	where $\theta\sim \theta + b$. We therefore see that  for any given value of $b$ we   have an orbifold of the type discussed in section \ref{dssemicla} above, and  the full quantum  path integral   involves a sum over all values of the orbifold parameter $b$. Once this sum is done, in the full quantum theory, one can avoid the big crunch singularity we encountered in the semi-classical theory in section \ref{dssemicla}
	and for suitable boundary conditions on the matter at least,    get a finite transition amplitude to go from a contracting universe in the past to an expanding universe in the future as we will see below in section \ref{dswithmatter}. 
	
	Let us note that  the geometry for the tunnelling amplitude we are considering has two boundaries, and is therefore suppressied compared to the one boundary  HH wave function  by a factor  of $e^{-S_0}$.  
	
	{\bf Multiple Boundaries and Higher Genus:}
	The second order formalism has not been fully fleshed out beyond the two cases mentioned above. However there is good reason to expect, based on  the first order formalism,  \cite{Saad:2019lba,Cotler:2019nbi,Maldacena:2019cbz}, that the result for the path integral should be   given by, 
	\begin{align}
		\label{genI}
		I(n,g)= e^{S_0 (2-2g -n)} \int \prod^n_{i=1} [ b_i db_i Z_{f}^{\pm}(\phi_{B_i},\hat{l}_i, b_i) ] V_g(b_1, b_2, \cdots b_n)
	\end{align}
	Here $Z_f^\pm(\phi_{B_i}, l_i, b_i)$ is the factor given in eq.\eqref{zv} that would be associated with each ``flaring trumpet", depending on the analytic continuation carried out,  and $V_g(b_1,b_2, \cdots b_n)$ is 
	the 
	volume of moduli space, which follows from the Weil-Petersson metric, eq.(\ref{ull}),  for bordered Riemann surfaces of genus $g$ with $n$ boundaries of geodesic lengths $b_i, i=1,\cdots n$
	
	In the second order formalism the boundary diffeomorphisms and moduli would  both arise from the zero modes of $P^\dagger P$ \cite{Moitra:2021uiv}. We note that Fenchel-Nielsen type coordinates $(b_i,\tau_i)$ in hyperbolic space can be extended  to describe  spaces with the ``flaring trumpet" asymptotic boundaries  considered here as well, with one extra pair of moduli $(b,\tau)$ corresponding to  each asymptotic boundary. 
	A symplectic form can then be defined in moduli space  using these {coordinates \cite{wolperteleform,wolpertonweil,Wolpert1981AnEF} given by}, 
	$\Omega= \sum_i db_i \wedge d\tau_i$, and it is reasonable to expect, based on the double trumpet case worked out in \cite{Moitra:2021uiv} as well,  that  the  associated top  form $\Omega^k\over k!$ then   agrees with  the volume form in moduli space arising from the Weil-Petersson metric\footnote{We thank Mahan Mj. for patiently explaining some of these points to us.}, \cite{GOLDMAN1984200}, leading  to eq.(\ref{genI}).

	

	\subsection{ The Hologram for the Multi-Boundary dS case}
	\label{holjtds}
	
	In an important  analysis \cite{Saad:2019lba} showed that the path integral in JT theory in AdS$_2$, for any number of boundaries, can be related to the correlation functions in a Random Matrix theory in the double scaled limit and this correspondence holds to all orders in the genus expansion. 
	 
	In \cite{Maldacena:2019cbz} and \cite{Cotler:2019nbi}  it was pointed out that the analysis of \cite{Saad:2019lba}  could be extended to the dS case and the same Random matrix model in the double scaled limit could also provide a hologram for dS$_2$. This is a very interesting proposal which needs to be studied  further. We will only elaborate on a few points  here. 
	
	First, note that the JT path integral in dS space gives rise to an extra phase factor $e^{\pm i \phi_B {\hat l }\over 8 \pi G}$. 
	This factor is removed in the AdS case by adding a boundary term in the action, and the correspondence with the RMT follows thereafter. 
	To extend the discussion  to the dS case we therefore  have to remove this phase factor on the gravity side  for each boundary and then relate the result  to the matrix theory. 
	
	Since  these phase factors are independent of the $b_i$ moduli appearing in  eq.(\ref{genI}) they can be  taken out of the moduli space integral.
	Defining
	\begin{align}
		\label{hgenI}
		{\hat Z}_f^\pm(\phi_{B_i},l_i,b_i)=  e^{\mp i \phi_{B_i} {\hat l}_i\over 8 \pi G} Z_f^\pm (\phi_{B_i},\hat{l}_i,b_i)
	\end{align}
	we then consider in the gravity theory  the integral 
	\begin{align}
		\label{hatgenI}
		{\hat I}(n,g)= e^{S_0(2-2g-n)} \int \prod_{i=1}^n b_i db_i {\hat Z}_f^\pm(\phi_{B_i},l_i, b_i) V_g(b_1,\dots b_n)
	\end{align}
	Note that $l_i$ is related to $\hat{ l}_i$ as in eq.\eqref{jllhat}.
	The arguments in \cite{Maldacena:2019cbz} and \cite{Cotler:2019nbi} then imply that ${\hat I}$ is related to appropriate  correlation functions in the hologram. 
	
	
	It is worth reviewing some of the details. Consider the integral 
	\begin{align}
		\label{sumg}
		{\hat T}(l_i, {\tilde l}_j,g)=e^{(2-2g-n_- -n_+)S_0}\int \prod_{i=1}^{n_-} \prod_{j=1}^{n_+} b_i db_i  {\tilde b}_j d{\tilde b}_j {\hat Z}_f^+(\phi_{B_i},l_i, b_i) {\hat Z}_f^-(\phi_{B_j},{\tilde l}_j, {\tilde b}_j) V_g(b_1, \cdots b_{n_-}, {\tilde b}_1, \cdots {\tilde b}_{n_+})
	\end{align}
	which gives the genus $g$ contribution for the transition amplitude to go from $n_-$ contracting universes to $n_+$ expanding ones. 
	Summing over all genus then gives 
	\begin{align}
		\label{sumg}
		{\hat T}(l_i, {\tilde l}_j)=\sum_{g}{\hat T}(l_i, {\tilde l}_j, g) 
	\end{align}
	The argument is that this transition amplitude is equal to a  correlation function of the Random matrix $H$
	\begin{align}
		\label{cor}
		{\hat T}(l_i, {\tilde l}_j)=\langle \Tr	 e^{-i l_1 H}  \cdots \Tr e^{-i l_{n_-} H} \Tr e^{i {\tilde l}_{1} H} \cdots \Tr e^{i {\tilde l}_{n_+} H}\rangle 
	\end{align}
	where on the RHS  we are computing the expectation value in  a particular RMT in the double scaling limit (corresponding to eq.(58) in \cite{Saad:2019lba}).
	For the case when $n_-=0$ in eq.(\ref{sumg}) we are dealing the  with expanding universe branch of the Hartle-Hawking wave function and we obtain a relation between this branch $ \Psi_{\text{HH}}^{\text{exp}}$ and the matrix theory. More correctly we define, in analogy with ${\hat Z}_f^-$ in eq.(\ref{hgenI}), the ``stripped off"  expanding branch wave function
	\be
	\label{sthh}
	{\hat \Psi}_{\text{HH}}^{\text{exp}}({\tilde l}_i)=e^{+{i\phi_{B_i} \hat {\tilde l}_i \over 8 \pi G}} \Psi_{\text{HH}}^{\text{exp}}({\tilde l}_i).
	\ee
	Then we get from eq.(\ref{sumg}) the relation
	\be
	\label{brwf}
	{\hat \Psi}^{\text{exp}}_{\text{HH}}( {\tilde l}_i)= \langle \Tr e^{i {\tilde l}_{1} H} \cdots \Tr e^{i {\tilde l}_{n_+} H}\rangle
	\ee
	In a similar way we can relate the contracting branch of the HH wave function after removing the phase factor,
	\begin{align}
	\label{contb}
	{\hat \Psi}_{\text{HH}}^{\text{con}}({ l}_i)=e^{-{i\phi_{B_i} \hat { l}_i \over 8 \pi G}} \Psi_{\text{HH}}^{\text{con}}({ l}_i)
	\end{align}
	to the matrix model by 
	\begin{align}
	\label{brwf}
	{\hat \Psi}^{\text{con}}_{\text{HH}}( { l}_i)= \langle \Tr e^{-i { l}_{1} H} \cdots \Tr e^{-i { l}_{n_-} H}\rangle
	\end{align}
	
	The correspondence in eq.(\ref{cor})  arises from noting that the RHS after an integral transform gives the correlation functions for the Resolvant of $H$, which can be expanded  in a genus expansion. 
	The same integral transform of the LHS gives rise to an expression for the genus $g$ contribution coming from eq.(\ref{sumg}) involving the volume of moduli space of genus $g$ bordered Reimann surfaces. These two expression are known to be equal from the work of \cite{Eynard:2007fi} and \cite{Mirzakhani:2006fta}.
	The integral transform of the LHS is 
	\begin{align}
		\label{inta}
		\prod_{i=1}^{n_-} \int_0^{{- i \infty}} d l_i (i z_i e^{-i l_i z_i^2}) \prod_{j=1}^{n_+} \int_0^{{i \infty}} d{\tilde l}_j    (-i {\tilde z}_j) e^{i {\tilde l}_j {\tilde z}_j^2}  \hat{T}(l_i, {\tilde l}_j )
	\end{align}
	which gives
	\begin{align}
		\label{valw}
		W_g(z_i ,{\tilde z}_j)= \frac{1}{2^{n_-+n_+}}\int \prod_{i=1}^{n_-} b_i db_i \prod_{j=1}^{n_+}	 {\tilde b}_j d{\tilde b}_j e^{-{\sqrt {2 \gamma}} \sum_{i,j} (b_iz_i+{\tilde b}_j{\tilde z}_j)}V_g(b_1, \cdots b_{n_-}, {\tilde b}_1, \cdots {\tilde b}_{n_+})
	\end{align}
	Here $\gamma = \frac{1}{8\pi G J}$. 
	On the RHS of eq.\eqref{cor} we get 
	\begin{align}
		\label{prodv}
		(-1)^{n_-+n_+}\prod_{i,j} z_i  {\tilde z_j} \langle  R(-z_1^2) \cdots R(-z_{n_-}^2) R(-{\tilde z}_1^2)\cdots R(-{\tilde z}_{n_+}^2) \rangle 
	\end{align}
	where $R(-z^2)$ the Resolvant is given by 
	\begin{align}
		\label{resz}
		R(z^2)=\Tr \left({1\over z^2{-}H}\right)
	\end{align}
	On setting $\gamma= \half$, it follows from \cite{Eynard:2007fi} that eq.(\ref{prodv}) and eq.(\ref{valw}) are equal order by order in the genus expansion. 
	
	One can consider a further extension of this holographic correspondence by replacing the double scaled RMT by the SYK theory, in the large $N$ limit, see also,  \cite{Cotler:2019nbi}. The low-energy spectrum of the SYK  theory agrees with the spectrum  in the RMT but at higher energy the SYK theory is different. 
	An important comment in this context  is that the genus counting parameter $S_0$ of the JT theory, which is the analogue of the entropy of dS  space, (in fact on dimensional reduction from $4$ dim. $S_0$ is the entropy of the extremal dS black hole geometry, upto a factor of $2$, as was noted  in appendix \ref{pennhlim})  goes like the number of fermions, $N$,   with a   constant of proportionality that is calculable in the SYK model. 
	Also, in the SYK case the hologram comes equipped with  a well defined Hilbert space of  $N/2$ qubits on which the $N$ Majorana fermions act. The operator $H$ is the ``momentum operator" which generates translations along the spacelike boundary. Thus while there is a well-defined Hilbert space and an operator which generates translations along the boundary theory, there  is no notion of time in the hologram.
	
	
	There are several  important issues that remain to be understood. 
	Most importantly, in our minds,  an inner product needs to be defined to obtain probabilities from the wave function or transition amplitudes (this is of course different from the inner product on the space of the $N/2$ qubits mentioned above). 
	We have not been able to define such an inner product yet. The fact that  transitions  which can change the number of universes are allowed, suggests that one should work in the third quantised theory, consisting of the multi-verse with an  arbitrary numbers of universes, to define a norm which is conserved. We hope to return to this question in further work. 
	Also, let us note that the SYK model is just one example of a whole class of theories which exhibit the same pattern of symmetry breaking at low energies - resulting in  reparametrisation modes governed by a Schwarzian action. There is also the option of  considering a version of these SYK theories with random couplings or a particular realisation for these couplings. This whole class are candidate holograms for  dS$_2$, and constitute  a  large set of possibilities. Perhaps additional consistency conditions, including the existence of a well defined norm, might cut down this set.
	
	Let us  mention that it was also suggested recently in \cite{Susskind:2021esx} that the SYK model in a particular limit could be a hologram for dS space.
	But in this case the hologram being considered is on the cosmological horizon, instead of at asymptotic inifinity, as in our discussion, and also the limit being considered is  at high temperature, whereas in our discussion we expect a match approximately with the JT gravity limit when $T \ll J$. 

	\section{Addition of matter}
	\label{dswithmatter}
	We will now turn to the discussion in the presence of conformal matter. We  focus on the two boundary case with the double trumpet topology. 
	Mostly we will consider  massless scalar fields, our  analysis can be extended to other kinds of conformal matter in a straightforward manner. 
	The one boundary case, which gives the HH wave function, was discussed in \cite{Maldacena:2019cbz,Moitra:2021uiv} .
	
	It was argued in \cite{Moitra:2021uiv} that for Euclidean AdS$_2$, in the two boundary case,   the $b$ modulus integral, discussed in section \ref{moddivs},  diverges when $b\rightarrow 0$ in the presence of matter. This divergence can be thought of as arising due to the Casimir effect which gives rise to a negative ground state energy. 
	The metric for the double trumpet belongs to the conformal class, 
	\begin{align}
		\label{ccl}
		ds^2=d\theta^2+dr_*^2
	\end{align}
	where $\theta\simeq \theta+ b$, $r_*\in [0,\pi]$. To study the $b\rightarrow 0$ limit we can do a rescaling and take $\theta$ to have periodicity $2 \pi$ while $r_*$ now has range $[0,{2 \pi^2\over b}]$. If $H$ is the Hamiltonian generating translations along $r_*$, 
	the matter integral evaluates  the transition amplitude $\langle S_I | e^{-2 \pi^2 H\over  b} |S_F\rangle $, where $|S_I\rangle, |S_F\rangle$
	denote the initial and final states at the initial and final values of $r_*$. 
	As long as their overlap with the ground state $|0\rangle $ of $H$ is non-zero the ground state will give the leading contribution to the transition amplitude, when $b\rightarrow 0$,   giving rise to an exponential dependence, 
	\be
	\label{Tf}
	\langle S_I| e^{-2 \pi^2 H\over  b} |S_F\rangle \simeq \langle S_I |0\rangle \langle  0 |S_F\rangle  e^{-2 \pi^2 E_0 \over b}
	\ee
	If $E_0$, the ground state energy, is negative this diverges when $b\rightarrow 0$. In fact, it is well known that for a real scalar field satisfying periodic boundary conditions along the $ \theta$ direction (with periodicity $2\pi)$, $E_0=-{1\over 12}$ leading to a divergence which goes like
	$e^{\pi^2\over 6 b}$. 
	When we go to dS space by continuing the AdS$_2$ result, the divergence persists. 
	
	Here we will discuss two possible ways to control this divergence. First, in subsection \ref{twistbc} we take the scalars to be in a twisted sector where they do not satisfy periodic boundary conditions along the $\theta$ direction. For a range of values of the twist parameter the ground state 	energy is now positive and the path integral is well behaved. Continuing to dS space we show that the final  state of the universe, produced by tunnelling from   an initial contracting phase,
	is different from the state described by the HH wave function, analysed in \cite{Maldacena:2019cbz}.The tunnelling transition amplitude is suppressed compared to the  the HH wave function by a factor of $e^{-S_0}$.  This toy calculation suggests  the interesting possibility  for the universe to have been born by a tunnelling event from a prior dS or Robertson Walker phase and for 
	quantum perturbations, which  in this model are analogous to those  giving rise to the CMB perturbations in $4$ dimensions, then   allow us to distinguish  between the tunnelling wave function and other possibilities like the HH wave function.    
	
	Second, in subsection \ref{matcorrel} we consider standard periodic boundary conditions for the scalars but now study correlation functions which include cross-boundary correlations. We find that sometimes the calculation for such correlations can also be free from the divergence mentioned above. 
	
	Before going any further let us alert the reader to one notational inconsistency which we will indulge in here. In section \ref{onetwob} we referred to the two boundaries in the double trumpet as the left and right boundaries, L,R. The superscripts $\pm$ in $Z_f$, eq.(\ref{psiad}) referred to the two different analytic continuations, eq.(\ref{firsta}), eq.(\ref{seconda}). In this section, we will almost exclusively focus   on a transition amplitude which is obtained by carrying out the continuation eq.(\ref{seconda}) on  the R boundary and eq.(\ref{firsta}) on the L boundary. And at the risk of some confusion, we  then refer to the L,R  boundaries here as  the $-,+$ boundaries respectively, so that the $-$ boundary corresponds to the past with a contracting universe, and the $+$ boundary to the future with an expanding one. The notation of this section will also be used in rest of this section  and in the appendices \ref{amainadsdt},\ref{fdsdt}. {Another important point is that the analytic continuations discussed in section \ref{transamp}, in particular eq.\eqref{firsta},\eqref{seconda} are based on a local coordinate system at each  asymptotic boundary. However, for the case of double trumpet there exists a single coordinate chart that covers the entire geometry and is given by 
\begin{align}
	ds^2=(r^2+1)d\theta^2+\frac{dr^2}{r^2+1}\label{adsdtm}
\end{align}	
where the asymptotic boundaries are now located at $r=r_+\gg 1,r= r_-\ll -1$. The corresponding analytic continuations to obtain a past to a future transition in terms of this $r$ coordinate is then given by 
\begin{align}
	&r_+\rightarrow - i r_+ \nonumber\\
	 &r_- \rightarrow i r_-,\label{radsconts}
\end{align} 	 
	The analytic continuation for two expanding universes to be produced from ``nothing" is
	\begin{align}
		&r_+\rightarrow - i r_+ \nonumber\\
		&r_- \rightarrow- i r_-,\label{radscont02}
	\end{align}
}
	
	Before going to the twisted boundary conditions let us warm up with the periodic boundary case and work out a formula for the path integral with two boundaries as a function of the boundary values taken by the scalar field. This will already bring out some of the important differences between the wave function obtained from tunnelling and the HH wave function. 
	
	The metric for AdS$_2$ asymptotically takes the form 
	\begin{align}
		\label{asads}
		ds^2 \simeq  r^2 d\theta^2 + {dr^2\over r^2}
	\end{align}
	For the $b$ modulus having value $b$, $\theta\in [0,b]$. 
	We expand the scalar field in modes 
	\begin{align}
		\label{ms}
		\varphi =\sum_{k=-\infty}^\infty e^{i \tilde{k} \theta} \varphi_k(r)
	\end{align}
	where $k$ takes values over integers and 
	\begin{align}
		\tilde{k}=\frac{2\pi k}{b}\label{ktildef}
	\end{align}
	At the  boundary 
	\begin{align}
		\label{bca}
		r\rightarrow -\infty,  &\,\, \varphi_k(r)\rightarrow {\varphi}_k^- \nonumber\\
		r\rightarrow \infty, &\,\, \varphi_k(r)\rightarrow {\varphi}_k^+. 
	\end{align}
	The full path integral then gives\footnote{Note that there is in general an additional factor, due to various determinants, which is b-dependent as was also metioned near eq.\eqref{psiad}. Again, to get agreement with the first order formalism without matter we set this factor to unity.} 
	
	\begin{align}
		{	I=\int b db [\mathcal{D}\theta_-] [\mathcal{D}\theta_+]  e^{-\frac{\phi_{B_+}\epsilon_+}{8\pi G}\int du\, {\text{Sch}\{\tanh({\theta_+(u)\over 2}) \} } -\frac{\phi_{B_-}\epsilon_-}{8\pi G}\int du\, {\text{Sch}\{\tanh({\theta_-(u)\over 2}) \} } }e^{-S_{M,b} } Z_M}
		\label{piinth}
	\end{align}
where $\text{Sch}$ denotes the Schwarzian action. 
 Expanding $\theta_{\pm}(u)$ as
 \begin{align}
 	\theta_\pm (u)=\frac{b}{2\pi}u+\epsilon_\pm(u)\label{timintheep}
 \end{align}
and noting that $u\in[0,2\pi]$,
the path integral becomes
	\begin{align}
		\label{fullpath}
	{	I=\int b db [\mathcal{D}\epsilon_-] [\mathcal{D}\epsilon_+]  e^{{-b^2\over 16 \pi GJ }\left({1\over l_+}+{1\over l_-}\right)} e^{-\left({\text{S}\{\epsilon_- \}\over  l_- } + {\text{S}\{\epsilon_+ \}\over l_+}\right) } e^{\bigl( -S_{M,b} + \cdots \bigr)} Z_M}
	\end{align}
	where $\mathcal{D}\epsilon_{\pm}$ corresponds to the integral over the time reparametrization modes at the left and right boundaries. The measure for the sum over these modes is the symplectic measure discussed in \cite{Stanford:2017thb,Moitra:2021uiv} and $\text{S}\{\theta\}$ denotes the action for these modes given by 
	\begin{align}
		S\{\theta\}=\frac{1}{8GJ}\int_0^{2\pi} du \pqty{\theta'(u)^2+\frac{4\pi^2}{b^2}\theta''(u)^2}\label{timac}
	\end{align}
	with $u$ being the proper time along the respective boundary.

	Here we have carried out the path integral over the scalar. $S_M$ above is given in terms of the boundary values of the  scalar by, see appendix \ref{mdsdt} for an analogous calculation in dS, 
	\begin{align}
		S_{M,b}=\frac{b}{2}\sum_{k=- \infty}^{\infty}\tilde{k}\left(({\varphi}^+_{-k}{\varphi}^+_{k}+{\varphi}^-_{-k}{\varphi}^-_{k})\coth\tilde{k}\pi-2{\varphi}^+_{-k}{\varphi}^-_{k}\csch\tilde{k}\pi\right)\label{dsdt02mact}
	\end{align}
	The ellipses in the last exponent in eq.(\ref{fullpath})  denote the couplings of these boundary values to the time reparametrisation modes at the two boundaries which we have denoted by $\theta_-, \theta_+$.  $ l_- ,l_+$ are the renormalised lengths of the two boundaries and 
	\begin{align}
		\label{defzm}
		Z_M[b]= {e^{-b/{24}}\over \eta({i b \over 2 \pi})}
	\end{align}
	is the matter determinant. 
	The couplings of the boundary values of the scalars to the reparametrisation modes results in corrections to the matter correlations which are suppressed in $G$, the gravitational coupling. Neglecting these couplings and integrating out the reparametrisation modes then gives
	\begin{align}
		\label{nact}
		I=\int b db \,e^{{-b^2\over 16 \pi GJ }\left({1\over  l_+ }+{1\over l_-}\right)} {1\over 16 \pi^2 GJ \sqrt{ l_+ l_-}} Z_M[b]e^{-S_{M,b}}
	\end{align}
	
	We can continue to dS space in different ways as discussed in the previous section. If we  consider the transition amplitude
	obtained by the continuations eq.\eqref{radsconts}
	 we get 
		\begin{align}
		\label{actdsa}
		I=e^{-{i \phi_{B_+}{\hat l}_+\over 8 \pi G}}\, e^{i \phi_{B_-}{\hat l}_-\over 8 \pi G} \int b db \, e^{{-ib^2\over 16 \pi GJ }\left({1\over l_+}-{1\over  l_- }\right)}
		{1\over 16 \pi^2 GJ \sqrt{ l_+ l_-}} Z_M[b]e^{-S_{Mb}}
	\end{align}
	Here, besides introducing the relevant factors of $i$ we have also introduced the two phase factors in front, which were discussed extensively in the previous section.  To summarise, eq.(\ref{actdsa}) gives the transition amplitude to go from an initial contracting dS universe with dilaton and length $\phi_{B_-}, {\hat l}_-$ to a final universe with values $\phi_{B_+}, {\hat l}_+$. The initial and final asymptotic values of the scalar are ${\varphi}^-_k, {\varphi}^+_k$ respectively. Note also that ${\phi_B\over {\hat l}}={1\over J l}$, at both boundaries, eq.\eqref{dbasymcond}. 
	
	Now consider the case where we start with an initial state for the matter field of the form, 
	\begin{align}
		\langle {\varphi}^-|S_I\rangle =\exp(-2\pi\sum_{m=-\infty}^{\infty}{E_{m}}{\varphi}^-_{-m}{\varphi}^-_{m})\label{defsi}
	\end{align}
	where the state $| \varphi^-\rangle$ is an eigenstate of the asymptotic value of the field operator  ${\hat \varphi}$ in the far past. When 
	\begin{align}
	\label{asfp}
	r\rightarrow -\infty, \hat{\varphi}\rightarrow \sum_m \hat{\varphi}_m^- e^{i m \theta}
	\end{align}
	and $| \varphi^-\rangle$ satisfies the condition, 
	\begin{align}
	\label{assfm}
{	\hat{\varphi}^-_m| \varphi^-\rangle= \varphi^-_m |\varphi^-\rangle}
	\end{align}
	An initial state $|S_I\rangle $ with such  a Gaussian wave function  is a reasonable one to consider, e.g. the ground state in the vacuum with respect to the coordinates, $z^\pm $, defined in eq.\eqref{milnezpm}. The details of the computation of eq.\eqref{defsi} for this ground state  are shown in  appendix \ref{initstate}, for which we get $E_m=|m|$.
	 
	 We can now compute the final state  wavefunction's  dependence on $\varphi^{+}$ by integrating over $\varphi^-$ for the initial state eq.(\ref{defsi}) to get 
	\begin{align}
		\Psi[{\varphi}^+]=e^{-\frac{i \phi_{B_+}  l_+}{8\pi G}+\frac{i \phi_{B_-} l_- }{8\pi G}}\int b db e^{-\frac{ib^2}{16\pi GJ}\pqty{\frac{1}{ l_+ }-\frac{1}{ l_- }}}  {Z_M[b] \over {16\pi^2 G J\sqrt{ l_-  l_+ }}}\int [D{\varphi}^-]\exp(-S_{M,b})\langle \varphi^-|S_I\rangle\label{wfpfinte}
	\end{align}
	 On carrying out the ${\varphi}^-$ integral this gives
	 \begin{align}
		\Psi[\varphi^+]
		=&e^{-\frac{i \phi_{B_+}  l_+}{8\pi G}+\frac{i \phi_{B_-} l_-  }{8\pi G}}\int b db e^{-\frac{ib^2}{16\pi G J}\pqty{\frac{1}{ l_+ }-\frac{1}{ l_- }}}\frac{Z_M[b]}{16\pi^2 G J\sqrt{ l_-  l_+ }}\left(\frac{2\pi^2}{b+4\pi^2 E_0}\right)^{\frac{1}{2}}e^{-2\pi b E_0(\varphi_0^+ )^2\over b+4\pi^2 E_0 }\nonumber\\
						&\prod_{m> 0}\frac{\pi}{4\pi {E_m}+b\tilde{m}\coth\tilde{m}\pi}
		\exp{-\varphi^+_m \varphi^+_{-m}\left(b\,\tilde{m}\coth\tilde{m}\pi-{b^2\tilde{m}^2\csch^2\tilde{m}\pi\over 4\pi{E}_m+b\tilde{m}\coth\tilde{m}\pi }\right)}
		\label{futwfval}.
	\end{align}
We note that the $b$ modulus integral above diverges. In the next section we consider twisted boundary conditions for which it will converge.
	 The reason why we have all the details of  eq.(\ref{futwfval}) is that the dependence on $\varphi^+$ in the twisted case will  not be very different from the equation above which is somewhat simpler and allows us  to extract most of the important physics as we will see in subsection \ref{socons}.

	\subsection{Twisted Boundary Condition}
	\label{twistbc}
	We now turn to considering twisted boundary conditions. We take a complex scalar field satisfying the boundary condition
	\begin{align}
		\label{twistedbc}
		\varphi(\theta+b)=e^{2\pi i \alpha} \varphi(\theta),\quad  -{1\over 2}\le \alpha \le {1\over 2}
	\end{align}
	A standard calculation now shows that the Casimir energy for the ground state of $H$ - the translation operator along $r_*$, eq.\eqref{ccl}- is given 
	by
	\begin{align}
		\label{cen}
		E=\frac{1}{12}-\frac{1}{4}(2\abs{\alpha}-1)^2
	\end{align}
	For 
	\begin{align}
	\label{ral}
	{1\over 2}-{1\over 2 \sqrt{3}}<|\alpha|\leq {1\over 2}
	\end{align}
	we see that $E_0>0$ and the divergence when $b\rightarrow 0$ discussed above will be absent. 
	
	In this case, as discussed in appendix \ref{gbcforscalars}  the scalar determinant is given by 
	\begin{align}
		\det(-\nabla^2)={e^{{b\over 12}}}q^{\frac{\alpha^2}{2}}	\frac{i\vartheta_{11}(v,\tau)}{2\sin(\pi\alpha)\eta(\tau)}\label{matgenbcdet}
	\end{align}
	where 
	\begin{align}
		& \vartheta_{11}(v,\tau)=-2\sin(\pi v)q^{\frac{1}{8}}\prod_{m=1}^{\infty}(1-q^m)(1-zq^m)(1-z^{-1}q^m),\nonumber\\
		&\eta(\tau)=q^{\frac{1}{24}}\prod_{m=1}^{\infty}(1-q^m)\nonumber\\
		&\tau=\frac{2\pi i}{b},\,\, v=\frac{2\pi i\alpha}{b},\,\, q=e^{2\pi i \tau},\,\,z=e^{2\pi i v}
		\label{gbctv}
	\end{align}
	with $Z_M[b]={1\over \det( -\nabla^2)}$. With the scalar field operator taking the form
	\begin{align}
	r\rightarrow \infty,  {\hat \varphi} & \rightarrow  \sum_{m=-\infty}^\infty {\hat \varphi}^+_m e^{i(\tilde{m}+\tilde{\alpha}) {\theta }} \nonumber\\
	r\rightarrow -\infty, {\hat \varphi} & \rightarrow \sum_{m=-\infty}^\infty {\hat \varphi}^-_m e^{i(\tilde{m}+\tilde{\alpha}) {\theta}} \label{formassn}
	\end{align}
{the on-shell action is obtained to be, see \ref{mdsdt}}
\begin{align}
	S_M={b}\sum_{k=-\infty }^{\infty}(\tilde{k}+\tilde{\alpha})\left((({\varphi}^+_{k})^\dagger{\varphi}^+_{k}+({\varphi}^-_{k})^\dagger{\varphi}^-_{k})\coth(\tilde{k}+\tilde{\alpha})\pi-(({\varphi}^+_{k})^\dagger{\varphi}^-_{k}+({\varphi}^-_{k})^\dagger{\varphi}^+_{k})\csch(\tilde{k}+\tilde{\alpha})\pi\right)\label{onacttwi}
\end{align}
	We choose an initial state of the form
	\begin{align}
	\label{Inc}
	|S_I\rangle= \exp(-2\pi \sum_{m=-\infty}^{\infty} E_m(\varphi_{m}^-)^\dagger  \varphi_m^-)
	\end{align}
	
	
	This gives a final state wave function 	
	
	\begin{align}
		\Psi[\varphi_m^+, \varphi_n^+]
		=&e^{-\frac{i \phi_{B_+}  \hat{l}_+}{8\pi G}+\frac{i \phi_{B_-} \hat{l}_-  }{8\pi G}}\int b db e^{-\frac{ib^2}{16\pi G J}\pqty{\frac{1}{l_+}-\frac{1}{ l_- }}}\frac{Z_M[b]}{16\pi^2 G J\sqrt{ l_- l_+}}\nonumber\\
		&\prod_{{m}=-\infty}^{\infty}\frac{\pi}{2\pi {E_m}+b(\tilde{m}+\tilde{\alpha})\coth(\tilde{m}+\tilde{\alpha})\pi}\nonumber\\
		&\exp{-\varphi^+_m(\varphi^+_{m})^\dagger \pqty{b(\tilde{m}+\tilde{\alpha})\coth(\tilde{m}+\tilde{\alpha})\pi-\frac{b^2(\tilde{m}+\tilde{\alpha})^2\csch^2(\tilde{m}+\tilde{\alpha})\pi}{2\pi{E}_m+b(\tilde{m}+\tilde{\alpha})\coth(\tilde{m}+\tilde{\alpha})\pi}}} 
 \label{ffa}.
	\end{align}

	Note that the $b$ integral above is now finite but complicated to carry out, and depending on the relative importance of the various terms the support for  $b$ in the integral can come from different  regions of moduli space. 
	
{ Let us end this subsection with one more comment. The contribution to the wave function with two disconnected boundaries, i.e. with two disks,  would  be enhanced compared to the double trumpet geometry by a factor of $e^{2S_0}$. However once twisted boundary conditions for the matter fields are imposed the disk amplitude, i.e. the amplitude to tunnel out of nothing,  vanishes. Thus the tunnelling amplitude would be the leading contribution. }
	

	\subsubsection{Some Consequences}
	\label{socons}
	Let us now discuss some of the physical consequences of the wave function eq.(\ref{ffa}) obtained above.
	For the HH wave function as was noted in \cite{Maldacena:2019cbz,Moitra:2021uiv}, the dependence of the wave function on the  boundary values of the scalar (for a real scalar) is given by 
	\begin{align}
	\label{reHH}
	\Psi\sim \exp{-2\pi\sum_{m>0}\varphi^+_{-m}\varphi^+_m m }
	\end{align}
	This is to be compared with the dependence on the boundary values for the real scalar  with periodic boundary conditions in eq.\eqref{futwfval} and in eq.(\ref{ffa}) for a complex scalar with twisted periodic boundary conditions. Note that eq.(\ref{ffa})  has double the number of modes,
	$\varphi^+_m, {\tilde \varphi}^+_n$ since we are dealing with a complex scalar. 
	
	Comparing eq.(\ref{futwfval}) and eq.(\ref{ffa}) it is easy to see that the effect of the twisted boundary condition, as far as the dependence on boundary values $\varphi^+$ are concerned, drop out when we take modes of mode number $m,n\gg 1$, since then $m+\alpha\simeq m$. This is simply because modes which are sensitive to the boundary conditions must have a wavelength of   order the size of the universe and  for modes which are much smaller in size the dependence on boundary condition, parametrised by  $\alpha$,  becomes unimportant. 
	
	More interesting is the deviation compared to the HH case in the width of the Gaussian in  $\varphi^+$  which appears in  the wave function. The dependence in the HH case corresponds to a scale invariant spectrum, with 
	\begin{align}
	\label{exs}
	\langle {\varphi}^+_m {\varphi}^+_{-m}\rangle ={ \frac{1}{4\pi \abs{m}}}
	\end{align}
	In contrast we see from eq.(\ref{futwfval}) and eq.(\ref{ffa}) that the two point function now has departures from scale invariance. 
	It is easy to see that the departure persists for mode numbers, $m\le  b$. 
	How big these departures end up being therefore  depend on what region of the $b$ modulus integral contributes dominantly. This could depend on the details of what the total matter content is, etc. 
	If  the region  $b\ll 1$ contributes, then only the smallest mode numbers, corresponding to wavelengths of order the size of the universe,  will be sensitive to the departure. 
	However if the region $b\gg 1$ contributes in a significant way, the departures will persist upto much larger mode numbers, i.e. upto wave lengths much smaller than the size of the universe.  
	
	These departures from scale invariance  lead to a bigger value for the width of the Gaussian than in the HH case and thus to smaller power at the wavelengths where the departures occur. The departures also depend on the parameters $E_m$ which 
	are determined by   the initial state, thus the departures from scale invariance would tell us about the nature of the initial state from which the tunnelling occurs.
	If in a variant of our model the final dS phase can end and match at late times to a more conventional FRW type cosmology, then the shorter wavelengths will re-enter the horizon earlier than the very long ones and there would be a chance of observing them, and thus by detecting the departure from scale invariance observe  the nature of the wave function of the universe and the initial state. 
	 
	{ Suppressing the twist angle $\alpha$, we see from eq.(\ref{futwfval}) that the width of the scalar Gaussian is determined by the function  $f[m,b,E_m]$ given by 
		\begin{align}
			f[m, b, E_m]=\left({m}\coth\tilde{m}\pi-{{m}^2\csch^2\tilde{m}\pi\over 2{E}_m+{m}\coth\tilde{m}\pi }\right)\label{fmbem}
		\end{align} Thus the ratio $m/f[m,b,E_m]$ is a measure of the departure of the scalar power spectrum from scale invariance. We give a plot of  this ratio in   Fig \ref{cmb} with $E_m=m$, which illustrates that the departure from scale invariance persists uptill  bigger momentum $m$, as $b$ increases. 
	\begin{figure}[h!]
		\centering
		\includegraphics[scale=0.4]{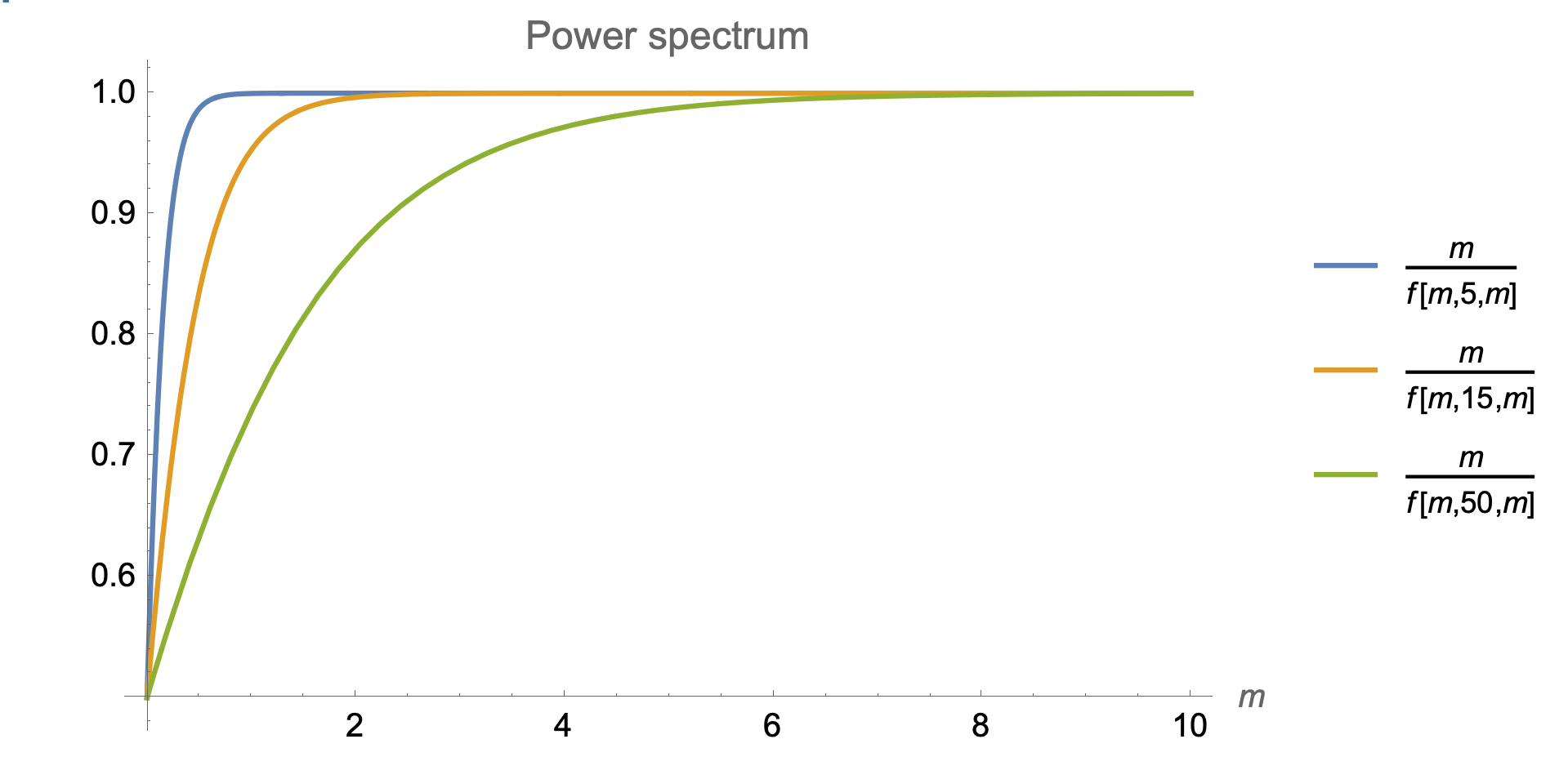}
		\caption{The plot for  the ratio ${m\over f[m,b,m]}$ shows that the departure from scale invariance in the scalar two-point function persists for modes with momentum $m<\sim b$. }
		\label{cmb}
	\end{figure}
}
	
	 The  suggestive lesson which can be drawn for the early universe from this toy model is then as follows: 
	 there could be  alternatives to the HH wave function in which the universe tunnels from a prior dS or FRW phase and the spectrum of perturbations could carry signatures of this tunnelling wave function which depends on the initial state. 
	
	We end this subsection  with two comments. 
	{ First, the breaking of scale invariance in the matter correlators in eq.(\ref{fmbem}) is connected to the dependence of the matter correlator on the $b$ modulus. The disk geometry has an SL(2,R) isometry and this  is reflected in the matter correlators which arise from the HH wave function being scale invariant. In contrast, the double trumpet geometry only has an U(1) isometry, corresponding to translations along the $\theta$ direction, eq.\eqref{edtrt}. The remaining two isometries are broken by the identification $\theta \simeq \theta+b$, and this breaking  then allows for the  
	lack of scale invariance in  $f[m,b,E_m]$. When $b\rightarrow 0$,  the two boundaries, in effect,  move far apart with  a distance going like $1/b$, and  the past boundary become unimportant; this is why $f[m,b,E_m]$ approaches the HH result, eq.(\ref{exs}) when $m/b\rightarrow \infty$. 
In this way  we see that the geometry   of the double trumpet instanton   is directly responsible for the violations of scale invariance in the matter correlators. 
  }

Second, while using twisted boundary conditions allowed us to avoid the divergence in the $b$ modulus integral, similar divergences are expected, even with the twisted boundary conditions,  at  higher genus or  with larger numbers of  boundaries in the  path integrals. To get a finite result at all orders, we therefore have to consider embedding the JT theory in a more complete UV theory, analogous to SYK model with matter, as we  discuss further  in section \ref{conclusion}.

	\subsection{Matter Correlators}
	\label{matcorrel}
	Here we return to the case AdS$_2$ and take matter fields to have  periodic boundary conditions. The action is given by eq.(\ref{actdsa}) with the matter action $S_{M,b}$ being given by eq.\eqref{dsdt02mact}. 
	
	Writing the matter action in position space variables on the boundaries gives, 
	\begin{align}
		S_{M,b}=&\frac{1}{2}
		\left(\frac{2\pi}{b}\right)^2\left[\int_+\int_+d\theta d\tilde{\theta}\hat{\varphi}^+(\theta)\hat{\varphi}^+(\tilde{\theta})G(\theta,\tilde{\theta})+\int_-\int_-d\theta d\tilde{\theta}\hat{\varphi}^-(\theta)\hat{\varphi}^-(\tilde{\theta})G(\theta,\tilde{\theta})\right]\nonumber\\
		&-\left(\frac{2\pi}{b}\right)^2\left[\int_+\int_-d\theta d\tilde{\theta}\hat{\varphi}^+(\theta)\hat{\varphi}^-(\tilde{\theta})H(\theta,\tilde{\theta})\right]\label{dssmredef}
	\end{align}
	where $G(\theta,\tilde{\theta}),H(\theta,\tilde{\theta})$ are defined as
	\begin{align}
		\frac{4\pi^2}{b}G(\theta,\tilde{\theta})=\sum_{n\in\mathbb{Z}-\{0\}} \tilde{n}\coth(\tilde{n}\pi) e^{i\tilde{n}(\theta-\tilde{\theta})}\nonumber\\
		\frac{4\pi^2}{b}H(\theta,\tilde{\theta})=\sum_{n\in\mathbb{Z}-\{0\}} \tilde{n}\csch(\tilde{n}\pi) e^{i\tilde{n}(\theta-\tilde{\theta})}\label{G1G2def}
	\end{align}
	with 
	\begin{align}
		\label{defnt}
		{\tilde n}= \frac{2\pi n}{b}
	\end{align}
	In general the $\theta$ variables along each  boundary differ from the proper time $u_1,u_2$ along them due to reparametrisation modes being turned on, but neglecting these modes here, since their coupling will make a subdominant contribution to the correlators we are calculating,  we can take 
	\begin{align}
		\theta(u)=\frac{b}{2\pi}u
		\label{timrepthetaa}
	\end{align}

	The Kernel functions $G,H$ defined in eq.\eqref{G1G2def} can be written in terms of Weierstrass functions after a rewriting of them in terms of $\csc$ function series, some of the details of which are provided in appendix \ref{bosadsdtmat}, see p.434 of \cite{whittaker_watson_1996}, as follows
	\begin{align}
		&\frac{4\pi^2}{b}G(u_1,u_2)=\sum_{n\neq 0}\tilde{n}\coth(\tilde{n}\pi) e^{inu_{12}}=-\wp(\hat{u}_{12},\omega_1,\omega_2)+{c},\nonumber\\
		&\frac{4\pi^2}{b}H(u_1,u_2)=\sum_{n\neq 0}\tilde{n}\csch(\tilde{n}\pi) e^{inu_{12}}=-\wp(\hat{u}_{12}+\omega_2,\omega_1,\omega_2)+{c},
		\label{G1Weier}
	\end{align}
	where
	$\wp(z,\omega_1,\omega_2)$ is the Weierstrass P function with half periodicities $\omega_1$ and $\omega_2$ satisfying
	\begin{align}
		\wp(z,\omega_1,\omega_2)=\wp(z+2 \omega_1,\omega_1,\omega_2)=\wp(z+2\omega_2,\omega_1,\omega_2)\label{wpperiods}.
	\end{align}
	Also, in the formula above $u_{12}=u_1 -u_2$, and the coordinate $\hat{u}$, the half-periodicities $\omega_1,\omega_2$ and the constant $c$ are 
	\begin{align}
		\hat{u}=\sqrt{\frac{b}{\pi}}\frac{u}{2},\quad \omega_1=\frac{\pi}{2}\sqrt{\frac{b}{\pi}},\quad\omega_2=i\pi\sqrt{\frac{\pi}{b}}\nonumber\\
		c=-\frac{\pi}{b}\pqty{\frac{1}{3}-\sum_{n\in \mathbb{Z}-\{0\}}\csch^2\left(\frac{2n\pi^2}{b}\right)}\label{weierargdefs}
	\end{align}
	
	Now, we come to main point of this subsection.  Consider a two point function for boundary operator dual to the scalar ${\varphi}$ with one operator being on inserted the $1^{st}$ boundary and the other on the $2^{nd}$ boundary. 
	In this case only the term proportional to $H(u_1,u_2)$ will be relevant and we get 
	\begin{align}
		\label{excro}
		{\delta^2 I \over \delta {\varphi}^{(1)}(u_1) \delta{\varphi}^{(2)}(u_2)} \bigg\vert_{ {\varphi}^{(1)}=0, {\varphi}^{(2)} =0}\equiv \langle O(u_1) O(u_2)\rangle ={1\over 16 \pi GJ \sqrt{  l_+  l_-}}\int_{0}^\infty b db 
		e^{-{b^2\over l_{\text{eff}} G_{\text{e}}}} Z_M[b]  H( u_{12})
	\end{align}
	where 
	\begin{align}
		\label{defgeff}
		G_{\text{e}}= 16\pi G J ,\quad l_{\text{eff}}=\frac{ l_+ l_-}{ l_+ +l_-}
	\end{align}
	
	As an aside, if the reader is wondering physically how relative locations on the two boundaries are being compared, we note that for any value of the $b$ modulus there is a minimum length geodesic from any point on the $1^{st}$ boundary to some point on the $2^{nd}$ boundary  and this allows us to relate the locations on the two boundaries. 
	
	We will see now that the $b$ integral on the RHS is well behaved when $b\rightarrow 0$. This follows from eq.\eqref{defzm} that while $Z_M\rightarrow e^{\pi^2\over 6 b}$, $H( u_{12})\rightarrow e^{-\pi^2\over b}$, independent of $u_{12}$, as discussed in appendix \ref{modstab}. 
	The integral is also well behaved as $b \rightarrow \infty$, since  $Z_M\rightarrow \order{(1)}$, in that limit  while, 
	\begin{align}
		H(\Delta u)=\begin{cases}
			-k_2 b^2 +\dots,\quad &b\rightarrow \infty,\,\, \Delta u=0,\\
			-k_2 b^2\exp(-k_3 b)+\dots,\quad &b\rightarrow \infty,\,\, \Delta u\neq 0
		\end{cases}
		\label{hfasymm}
	\end{align}
	where $k_2,k_3$ are $O(1)$, which can be obtained from the full expression eq.\eqref{hucschdeff},\eqref{huincscderiv}.
	As a result the factor $e^{{-b^2\over l_{\text{eff}} G_{\text{e} }}}$ in eq.(\ref{excro})  dominates the behaviour and ensures convergence. 
	
	For $N$ matter fields if there are $n$ cross boundary contractions the integral will be well defined as long as 
	\begin{align}
		\label{nb}
		n>{N\over 6}
	\end{align}
	
	Some saddle points which can arise in evaluating such correlators when $N\rightarrow \infty$ are discussed in appendix \ref{modstab}. 
	
	Let us note that there will of course always continue to be some correlators which are divergent. The   path integral without any operator insertions is the simplest example,  or more generally if the number of cross-boundary contractions is small in number not meeting  eq.\eqref{nb}.
	
	{ We end with three comments. First, in the paper, \cite{Stanford:2020wkf}  similar cross-boundary correlators were considered for massive matter and it was argued that the modular integral would then be finite.
	Second, let  us also  comment on  the dS case. Doing the analytic continuation to go from -AdS$_2$ to dS$_2$ leaves the matter action $S_M$
	the same as eq.\eqref{dsdt02mact} above, with the  qualification that $\hat{\varphi}^+, \hat{\varphi}^-$ now refer to the asymptotic values the scalar takes at the two dS boundaries. Consider the transition amplitude discussed in eq.(\ref{actdsa}). For this case, the calculation above of the two point function, with one matter operator inserted at each boundary, tells us about how the wave function of the state in the future will change in response to a change in the
	wave function of the initial state in the far past. This is connected to what we saw in the previous subsection, namely that the matter correlators do depend on the initial state, e.g. the parameter $E_m$, eq.(\ref{Inc}).}

		Third, though not related to moduli stabilization, in appendix \ref{otocadsdt}, we discuss  the calculation of the four point out-of-time order (OTOC) correlator, which is useful to diagnose the chaotic nature of a  system. Interestingly, we find that the OTOC for the  double trumpet geometry does not exhibit  exponential growth at large time, as opposed to the  case of the disk geometry which is well known to have   exponential growth with the Lyapunov exponent saturating the chaos bound. Instead,  in the double trumpet the OTOC  has an  oscillating exponential at large times eq.\eqref{4pt1bzero},\eqref{otoc1bdyval},\eqref{otoc2bdyval}. The underlying reason for this can be traced to the absence of an SL(2,R) worth of zero modes for the time reparametrisation modes in the double trumpet which instead only has  an  U(1) worth of zero modes. We elaborate more on this in appendix \ref{otocadsdt}.

	\section{ Comments on The Double Trumpet Spectral Form Factor}
	\label{diskdtsff}
	
	The spectral form factor is defined to be
	\begin{align}
		\label{sff}
		\text{SFF}(\beta, T)= \langle  Z(\beta+ i T) Z(\beta-i T)\rangle = \langle \text{Tr} e^{-(\beta-i T) H} \text{Tr} e^{-(\beta+i T) H}\rangle .
	\end{align}
	It has proved to be an important correlator to study, for understanding chaos and more generally the behaviour of complex systems. 
	Here we will consider its behaviour for the theory with conformal matter and specifically will be interested in the contribution made by 
	the Double trumpet topology  to it. 
	
	Before we turn to a detailed analysis let us briefly review  some  basic features. 
	The specific heat of the theory  arises from the disk topology. The presence of matter fields does not change the $\beta$  dependence of the partition function, so that the specific heat of the JT theory with matter is the same as that without matter and given by 
	\begin{align}
		\label{spheat}
		C={\pi\over {2}G \beta J}+\frac{3}{2}
	\end{align}
	Thermodynamics is a good approximation when $C\gg 1$, i.e.
	\begin{align}
		\label{condb}
		\beta\ll {1\over G J}
	\end{align}
	In terms of energy, $E= \pi /4G J\beta^2 +3/2\beta$,  the condition is 
	\begin{align}
		\label{enc}
		E\gg  G J
	\end{align}
	
	For the case without matter the leading contribution to the SFF which arises from the topology of two disconnected disks gives, 
	\begin{align}
		\label{dsff}
		S_{\text{Disk, SFF}}= e^{2 S_0}{e^{{\pi \beta\over 2 G J(\beta^2 + T^2)}}\over 16\pi  G_e^3 (\beta^2+ T^2)^{3/2}}
	\end{align}
where $G_e$ is defined in eq.\eqref{defgeff}.
	For the Double trumpet we get 
	\begin{align}
		\label{dtsff}
		S_{\text{DT,SFF}}={ \sqrt{\beta^2+T^2}\over 4 \beta}
	\end{align}
	For large times 
	\begin{align}
		\label{lt}
		T\gg \beta
	\end{align}
	we get, from eq.(\ref{dsff}), eq.(\ref{dtsff})
	\begin{align}
		S_{\text{Disk, SFF}} & \simeq   e^{2 S_0} {e^{ {\pi \beta\over 2 G J T^2}}\over 16\pi G_e^3 T^3}\rightarrow {e^{2 S_0}\over 16\pi G_e^3 T^3}\label{ltdisksff} \\
		S_{\text{DT,SFF}} & \simeq {T\over 4 \beta}\label{ltdtsff}
	\end{align}
	We see that for $T\sim e^{S_0/2} \left({\beta\over G_e^3}\right)^{1/4}$ the double trumpet topology begins to dominate over the disconnected disk ones and one  enters what is referred to as  the ``ramp" region, \cite{Saad:2018bqo}. 
	
	Now let us turn to the case with matter. From eq.\eqref{nact} it is easy to see that after the matter path integral is done we get 
	\begin{align}
		Z_{DT}(\beta-i T ,\beta+ i T )&=\int_0^\infty b db Z_M[b] \frac{\exp{-\frac{b^2}{G_e  }\pqty{\frac{2 \beta }{\beta^2+T^2} }}}{\sqrt{G_e^2(\beta^2+T^2)}}\label{dtbpimat}
	\end{align}
	Here $Z_M[b]$ is the matter determinant, eq.\eqref{defzm},  it is $\beta, T$ independent, and only depends on the $b$ modulus.
	This is easy to see since the matter does not couple to the dilaton and by dimensional analysis the lengths of the two boundaries only appear  together with the scale $J$ characterising the fall off of the dilaton towards the boundaries, eq.\eqref{dbasymcond}.  
	It then follows that the $b$ integral in eq.(\ref{dtbpimat}) diverges when $b\rightarrow 0$, since in this limit $Z_M\rightarrow e^{\pi^2\over 6 b}$, eq.\eqref{defzm}. 
	
	One way to try and get a meaningful answer  is   to consider the microcanonical, instead of canonical, ensemble in the SFF. More specifically we carry out an inverse  Laplace transform of the SFF, eq.\eqref{sff}, going from $\beta$ to  energy $E$,
	\begin{align}
		Z_{DT}(E,T)=\int_C d\beta \,e^{2\beta E}Z_{DT}(\beta-iT,\beta+i T)\label{dtmicdef}
	\end{align}
	and  examine if $Z_{DT}(E,T)$ is well behaved. The contour for the $\beta$ integral above is the Bromwich contour, $\beta\in (\gamma-i\infty,\gamma+i\infty)$, where $\gamma$ is real and  chosen so that the contour is to the right, i.e., at a larger value along the horizontal-axis, than  all the singularities of the integrand in the complex $\beta$ plane and  it is parallel to the imaginary axis. We can then write the integral variable $\beta=\gamma+ i x$ where, $ x\in [\infty,\infty]$. Note that we will  also take the energy $E$ to be big enough and satisfying eq.(\ref{enc}).	
	
	To flesh this out more, notice that 
	by exchanging the order of the $b$ modulus and $\beta$ integrals on the RHS of eq.(\ref{dtmicdef}) we get  
	\begin{align}
		\label{idd}
		Z_{DT}(E,T)= \int b db  Z_M[b]\int_{\gamma-i \infty}^{\gamma+i \infty} d\beta e^{2 \beta E} \frac{\exp{-\frac{b^2}{G_e  }\pqty{\frac{2 \beta }{\beta^2+T^2} }}}{\sqrt{G_e^2(\beta^2+T^2)}}
	\end{align}
	The idea is that perhaps $Z_{DT}(E,T)$ as defined in eq.(\ref{idd}), after reversing the order of the $b, \beta$ integrals, will be  well defined. 
	Although at first sight this looks promising, here we argue after a more careful analysis, that it is not true. Even with the order reversed the integrals on the RHS of eq.(\ref{idd}) give a diverging answer with the divergence coming still  from the $b\rightarrow 0$ region of the $b$ modulus integral.  
	
	The reason to be hopeful at first sight \cite{Saad:2019pqd} is as follows. Taking the limit $T\gg \beta$ in eq.(\ref{idd}), one might then like to approximate the exponential term, $e^{ -{b^2\over  G_e}\left({2\beta\over \beta^2+T^2}\right)}\simeq  e^{ - {2\beta b^2\over  G_e T^2}}$
	This gives, 
	\begin{align}
		Z_{DT}(E,T)&\simeq \int b\,db Z_M[b]\int_{-\infty}^\infty \frac{dx }{G_e T} e^{2(\gamma+ i x)\pqty{E-\frac{b^2}{G_eT^2}}}\label{he1}\\
		&\simeq \int db \frac{bZ_M[b]}{2G_e T}\delta \pqty{E-\frac{b^2}{G_e T^2}}\label{he2}\\
		&\simeq \frac{1}{2}T Z_M[b_0]\label{nvdtmicroval}
	\end{align}
	where  
	\begin{align}
		\label{valb0}
		b_0^2=E G_e T^2
	\end{align}
	So it seems that the divergence at $b\rightarrow 0$ has indeed been successfully cut-off by the  $\delta$ function which only has support at $b=b_0$. 
	
	However this argument is too quick. As a first pass at being  more careful  let us express the  $\delta$ function above as a distribution 
	\begin{align}
		\label{deld}
		\delta \left(E-{b^2\over G_e T^2}\right)=\lim_{ \alpha\rightarrow 0}   {1\over 2\sqrt{\pi \alpha}} e^{-{\left(E-{b^2\over G_e T^2}\right)^2 \over 4\alpha}},
	\end{align}
	then eq.(\ref{he2}) gives
	\begin{align}
		\label{midf}
		Z_{DT}(E,T)= \lim_{ \alpha \rightarrow 0}{1\over 2\sqrt{\pi\alpha}} \int_{b_c}^\infty b db Z_M[b] e^{-{\left(E-{b^2\over G_e T^2}\right)^2 \over \alpha}}
	\end{align}
	where  we have introduced a lower cut-off in the $b$ integral as well. Now since $Z_M[b]\rightarrow e^{\pi^2\over 6 b}$ when $ b\rightarrow 0$ we see that if we take $b_c\rightarrow 0$ first and then take $\alpha\rightarrow 0$, the result will diverge, whereas if we take the limit the other way around it will be finite.
	This alerts us to the fact  us that we need to be more careful in evaluating eq.(\ref{idd}). 
	
	\subsection{A More Careful Analysis}
	
	To proceed more carefully, we break up the $b$ integral in eq.\eqref{idd} into two parts. From $[0,b_c]$ and $[b_c,\infty]$, where $b_c$ is sufficiently small. We will make more precise how small $b_c$ needs to be shortly. This gives, 
	\begin{align}
		\label{bu}
		Z_{DT}(E,T)=\int_0^{b_c}  db \,\frac{b Z_M[b]}{{G_e }}\hat{Z}(b) + \int_{b_c}^\infty db \,\frac{b Z_M[b]}{{G_e }}\hat{Z}(b) 
	\end{align}
	where with $x={\beta\over T}$ we have
	\begin{align}
		\hat{Z}(b) =& \int_C dx {e^{S(x)}\over\sqrt{1+x^2} }\nonumber\\
		S(x) = & 2ET x\pqty{1-\frac{b^2}{{b}_0^2}\frac{1}{1+x^2}}\label{jhbh}
	\end{align}
	Here $C$ in the Bromwich contour and $b_0$ is given by eq.\eqref{valb0}.
	
	Now consider the case where $0<b<b_c$. To evaluate it we expand the exponent 
	\begin{align}
		\label{expexp}
		e^{S(x)}=e^{2ETx}\biggl(1-{b^2\over { b_0^2}}{1\over 1+x^2}+\cdots\biggr)
	\end{align}
	This gives\footnote{A  quick  way to obtain the result eq.(\ref{bpertzhat}) is to use the ``InverseLaplaceTransform" function in Mathematica.}
	\begin{align}
		\hat{Z}(b)&=\int_{\gamma-i\infty}^{\gamma+i\infty} dx \frac{e^{2 ET x}}{\sqrt{1+x^2}}\pqty{1-\frac{2ETx}{1+x^2}\frac{b^2}{{b}_0^2}+\order{(b^4)}}\nonumber\\
		&=J_0(2ET)\pqty{1-\pqty{\frac{2ETb}{{b}_0}}^2+\order{(b^4)}}
		\label{bpertzhat}
	\end{align}
	Here $J_0$ is the Bessel function of the first kind with index $0$. 
	We see that subsequent terms in the expansion are suppressed in
	\begin{align}
		\frac{2ETb}{{b}_0}\ll 1\Rightarrow b\ll {1\over \sqrt{E G_e^{-1}} }\label{bpertres}
	\end{align}
	We  take $b_c$ in eq.\eqref{bu} to satisfy this condition, 
	\begin{align}
		\label{condbc}
		b_c< {1\over \sqrt{E G_e^{-1}} }
	\end{align}
	and this is the  precise condition referred to above for its smallness. 
	
	In particular we have learnt that in  the region $0<b<b_c$, ${\hat Z}(b)$ makes a leading order contribution independent of $b$. On the other hand $Z_M$ diverges as $e^{\pi^2\over 6 b}$. It therefore follows that the contribution to $Z_{DT}(E,T)$ from this region diverges. 
	
	On the  other hand  we will argue below and in appendix \ref{sffapp} that when $b>b_c$ the contribution of the second term in eq.(\ref{bu}) is finite. 
	Putting these together then leads to the conclusion that working in the microcanonical ensemble does not help after all and $Z_{DT}(E,T)$, like its  Laplace transform $Z_{DT}(\beta,T)$, also  diverges. 
	In fact once a lower cut-off $b_c$ is  non-vanishing- the   argument  in eq.(\ref{he2})-eq.(\ref{midf}) would already suggest  
	that the second term in eq.(\ref{bu}) is well behaved and finite. A careful analysis outlined in appendix \ref{sffapp} indeed establishes that  this is true. 
	
	The main points of the argument in the appendix \ref{sffapp} are as follows. 
	We divide the integration region into three parts:
	\begin{align}
		&b> b_0,\label{r1}\\	
		&{b-b_0\over b_0}\ll  1,\label{r2}\\
		&b^2 EG_e^{-1}\geq 1 \,\,\& \,\, b<b_0\label{r3}
	\end{align}
	Note that the lower limit in eq.(\ref{r3}) can be made of order $b_c$,  which is    the upper limit in eq.(\ref{bpertres}).
	We then  argue that the saddle point approximation involved in evaluating ${\hat Z}(b)$ eq.\eqref{jhbh} is a good one when  $b$ lies in 1$^{\text{st}}$ and 3$^{\text{rd}}$ regions, eq.(\ref{r1}), eq.(\ref{r3}). From this we obtain an estimate of ${\hat Z}(b)$ which leads to the conclusion that the $b$ integral in these regions is finite and in fact in region 3, eq.(\ref{r3}), its contribution mainly comes from the region near the  lower limit, which is of order,   $b_c$. 
	In the middle region eq.(\ref{r2}) the $x$ integral involved in ${\hat Z}(b)$ has a   rapidly oscillating phase factor  which suppresses it and we use this to conclude  again the finiteness of the result. These arguments establish that the dominant contribution to eq.(\ref{bu}) comes from the first term on the RHS which we have argued diverges. 
	
	We will have more to say about how to control this divergence in a more complete embedding of JT coupled with matter in section \ref{conclusion}.

	\section{Discussion}
	\label{conclusion}
	We end with a discussion of some open questions. 
	
	We saw above that once matter is added, the  partition function in AdS, for surfaces with two or more boundaries or at higher genus, and correspondingly wave function/transition amplitudes in dS space,  can be divergent, due to contributions arising from the boundaries of moduli space. While this divergence is absent in the double trumpet, once the matter satisfies suitable twisted boundary conditions,  it is expected to reappear in higher genus contributions,  because states in the vacuum sector  of the theory, as opposed to the twisted sector,  can propagate along the additional handles.  We studied conformal matter above but such a divergence is also expected with massive matter,  of  mass $m$. This is  because in the limit  when the modular parameter $b$ defined in eq.\eqref{edtrt} vanishes, $b \rightarrow 0$, the divergence arises from wavelengths much smaller than $1/m$. 
	
	Let us now discuss a proposal for how such divergences might be avoided once the JT theory with matter is embedded in a more complete SYK theory which also contains, correspondingly,  additional matter. 
	
	\subsection{SYK +  Additional Matter}
	We had mentioned in section \ref{holjtds} while discussing the pure JT theory without matter, a somewhat revised  proposal for holography with the RMT being replaced by SYK theory, see also \cite{Cotler:2019nbi}. 	This could serve as the hologram in both the AdS or dS cases along the lines discussed in section \ref{holjtds}. Here we will discuss a further revision of this proposal which  includes additional matter coupled to the SYK theory, and argue that it could serve as a hologram for the gravity theory with additional massless or light matter.

	For completeness, before proceeding, let us  briefly note that the SYK theory is given by the action (here for simplicity we only consider the $q=4$ model):
	\begin{align}
		\label{actsyk}
		S_{SYK}= \int \sum_i dt \psi^i \partial_t \psi^i +\sum_{i_1<i_2<i_3<i_4} J_{i_1 i_2 i_3 i_4} \psi^{i_1}\psi^{i_2}  \psi^{i_3}\psi^{i_4}
	\end{align}
	In eq.(\ref{actsyk})  $\psi^i, i= 1\cdots N$ are $N$ flavours of Majorana fermions, and the coupling $J_{i_1 i_2  i_3 i_4}$ is a random coupling drawn from a Gaussian ensemble with two-point correlator
	\begin{align}
		\label{twopt}
		\langle J_{i_1 i_2,i_3, i_4} J_{j_1 j_2 j_3 j_4}\rangle =  {6J^2 \over N^{3}} \delta_{i_1j_1}\delta_{i_2j_2}\delta_{i_3j_3}\delta_{i_4j_4}
	\end{align} 
	 $J$ is the  only energy scale in   the theory, and the large $N$ limit is obtained  by taking $N\rightarrow \infty$ while keeping $J$ fixed. 
	 The system can be solved in this  limit using saddle point methods. At low energies, $E\ll J$, in the conformal limit, the  saddle point value of the two-point flavour singlet Green's function is given by 
	 \begin{align}
	 \label{tpg}
	 G_{c}={1\over N} \sum_i \langle  \psi^i(t) \psi^i(t')\rangle= {g_\psi\over |t-t'|^{1/2}},\quad g_{\psi}^4=\frac{1}{4\pi J^2}
	 \end{align}
	At low temperatures, $T$, or energies, $E$,  with, $T/J, E/J\ll 1$ the fluctuations about the saddle point are  given by  the time reparametrisation modes.
	At higher $E,T$, extra degrees of freedom come into play. {This means that pure JT gravity will agree with  the SYK theory when the length of boundaries becomes small,
	$1/J l\ll O(1)$, where $l$ is the renormalised length eq.\eqref{dbasymcond}, but will  differ when $1/Jl\ge O(1)$}. It would be interesting to figure out on the gravity side what the additional high energy degrees of freedom are when the dual hologram is taken to be the SYK model.
	
	Now consider adding  an additional scalar  field $\varphi$  to the SYK theory with the action,  
	\begin{align}
		\label{scc}
		S_{scalar}=  \int dt \pqty{{{\dot \varphi}^2 \over 2} {+ \sum_{i<j<k<l }g_{ijkl}\psi^i\psi^j\psi^k\psi^l\varphi}}
	\end{align}
	where the coupling $g_{ijkl}$ is  taken to be random and drawn from a Gaussian ensemble
	\be
	\label{gge}
	\langle g_{ijkl} g_{ijkl}\rangle = {g^2\over N^4}  \\({\rm no \,\,sum}). 
	\ee
	
	At low enough energies the time derivative term in eq.(\ref{scc}) can be neglected. 
	Integrating out the fermions gives  a description in terms of the flavour singlet bi-local   $G_\psi, \Sigma_\psi$ fields. The resulting coupling to 
	$\varphi$ is 
	\begin{align}
		\label{rscc}
		S_{lsc}=\frac{g^2}{48 N^4} \int dt dt'  G_{\psi}(t,t')^4 \varphi(t) \varphi(t')
	\end{align}
	It is easy to see that the saddle point equations can be consistently solved by taking $\varphi$ at the saddle to vanish.
	For good measure, note that we have taken the coupling $g^2$ to be O(1) rather than O(N) in eq.(\ref{rscc}) to ensure that the coupling $g$  is  sufficiently weak and the 
	saddle point for the fermions  continues to be that discussed in the theory without $\varphi$. With $G_\psi$ replaced by $G_c$, eq.(\ref{tpg}) we then get 
	eq.(\ref{rscc}) to become
	\begin{align}
		\label{newc}
		S_{lsc}=\frac{ g^2g_\psi^4}{48N^4} \int dt \int dt' {\varphi(t) \varphi(t') \over |t-t'|^{2}}
	\end{align}
	This shows that $\varphi$ behaves like the source of a dimension $1$ operator. 
	
	In general a bulk scalar of mass $m^2$ (in $R_{AdS}$ units) would give rise to an operator of dimension
	\begin{align}
		\label{dimop}
		\Delta= {1\over 2} +\sqrt{{1\over 4}+m^2}
	\end{align}
	we see therefore that $\varphi$ above corresponds to a massless scalar in the bulk. This was to be expected since its  coupling to the fermions was marginal.

	We note in passing that a Yukawa-like coupling of $\varphi$ to the fermions, which would be the simplest one to consider, 
	\begin{align}
		\label{yuk}
		S_{yuk}=\int dt g_{ij} \psi^i \psi^j \varphi,
	\end{align}
	after averaging over the random coupling $g_{ij}$, would give rise to a bulk scalar of mass $m^2={-1/4}$. This is the lowest mass allowed in AdS space. 
	
	The coupling of time reparametrisations to $\varphi$ follow from the action eq.(\ref{newc}) and it is easy to see that the resulting dynamics of the the reparametrisation modes coupled to $\varphi$ would then exactly agree with what is obtained in  the bulk by coupling a  bulk massless scalar to JT gravity theory with $\varphi(t)$ playing the role of the boundary value for the bulk scalar. 
	
	In this way we see how degrees of freedom corresponding to extra massless or light matter (with mass $m^2 R_{AdS}^2 \le 1$ )on the gravity side can be added to the SYK theory. Our proposal  is  then  to  consider this SYK theory with the additional matter  as a UV completion of the JT theory in AdS with additional massless or light matter. 
	
	The SYK theory with the extra matter is of course finite with no divergences in multi-boundary correlation functions. 
 This suggests that the extra degrees of freedom in the SYK theory must regulate the divergences we discussed above which appear in the JT theory \cite{Maldacena:2018lmt,Maldacena:2019ufo}. 
 Understanding how this works more explicitly is an interesting question which we leave for the future. 
 
 The discussion above can also be extended to dS space. As discussed in section \ref{holjtds}, one could also consider a version of holography for pure JT theory in dS space with  the SYK theory rather than RMT being the hologram. This could then  be extended with additional matter on both sides along the lines above.  
 
 In parallel with the  AdS discussion above, here one expects that the  resulting transition amplitudes or wave functions obtained from the SYK+matter theory would be closely related to JT dS gravity with matter, when the renormalised lengths of all boundaries are large,  meeting the condition ${1\over J l}\ll 1$. 
	Note also that the correspondence in the dS case extends to both transition amplitudes of the type considered in section \ref{holjtds}  and also to those where the boundary value for matter fields are turned on,  giving rise to the wave function or transition amplitudes as a function of these boundary values. This more general case would correspond on the SYK side to calculating correlations of $\Tr\, e^{\pm i H {l}} $ in the presence of the corresponding sources for the matter fields.

	\subsection{Some Additional Comments}
	Let us end with three more comments. 
	First, as also noted in section \ref{holjtds},  the proposals for a hologram in the dS$_2$ case are incomplete in  one  important respect. We have not defined a suitable norm on the space of wave functions. This will probably need to be done in the third quantised theory, since the number of universes can  change through quantum tunnelling.
	
	Second, there are several different models besides the SYK model which give rise to the same low-energy  theory of boundary reparametrisation modes and  all of them can be candidates for dS holograms as well. It will be interesting to explore if additional consistency conditions can cut down this bewilderingly large set of possibilities. 
	
	An important distinction is between models which involve random averages over couplings and those which involve single realisations of these couplings. The (connected) correlation functions for the partition function in a single realisation case would vanish, and this suggests that the related transition amplitudes or multi-boundary wave functions in the dS case would also vanish. The presence or absence of the  multiverse in these systems would therefore seem to  depend on whether we are dealing with a boundary theory that has a random average over couplings or one whose coupling constants are fixed, say at typical values.   It is worth emphasising that the random average theories might be sensible ones to consider, especially in the dS context, as candidate holograms. 
	
	Finally, one of our motivations behind this work was to explore in the  precise manner afforded to us in two dimensions,  Coleman's important idea that wormholes could determine the values that  coupling constants take in  nature \cite{COLEMAN1988643}. In studying this idea one wants a model where there are local degrees of freedom present, along with their associated coupling constants, and this was one of our motivations for  considering gravity  in the  presence of extra matter. Hopefully, having dealt with  some of the issues outlined above  in a satisfactory manner,   we can also complete  a careful study  of Coleman's ideas in the context of two dimensional  gravity.

		We leave an exploration of these fascinating ideas for the future.

		\acknowledgments
	\label{acknw}
	We thank the TIFR String theory group, in particular, A. Gadde, G. Mandal, S. Minwalla,  O. Parrikar and P. Shrivastava for very helpful discussions. We also acknowledge useful   discussion with  D. Skliros and  A. Vilenkin. 	We acknowledge support of the Department of Atomic Energy, Government of India, under Project Identification No. RTI 4002 and support from the Infosys Foundation in form of the Endowment for the Study of the Quantum Structure of Spacetime. S. P. T. acknowledges support from a J. C. Bose Fellowship, Department of Science and Technology, Government of India.  Most of all, we are grateful to the people of India for generously supporting research in String Theory.

	\newpage
	
	\appendix
	
	\section{Coordinate transformations}
	\label{cordtrans}

	In this appendix we will list out various coordinates systems and the corresponding transformations between them that are used throughout the main text. We will also show important Penrose diagrams that illustrate the regions in the full spacetime covered by these coordinate systems. First let us consider the de Sitter spacetime. 
	
	\subsection{de Sitter}
	\label{ds2cr}
	Let us first note that dS$_2$ can be thought of as an embedding in $R^{1,2}$ Minkowski spacetime as
	\begin{align}
		&ds^2=-dx_0^2+dx_1^2+dx_2^2,\nonumber\\
		&-x_0^2+x_1^2+x_2^2=1\label{dsasembed}
	\end{align}
	where we have taken the dS radius to be unity. The global coordinate parametrization of de Sitter is obtained by taking 
	\begin{align}
		x_0=\sinh \hat{\tau},
		\,\,x_1=\cosh \hat{\tau}\,\cos\hat{\theta},\,\,	x_2=\cosh \hat{\tau}\,\sin\hat{\theta}\label{globalembed}
	\end{align}
	Using this to find the metric in terms of $\hat{\tau},\hat{\theta}$ will give 
	\begin{equation}
		ds^2=-d\hat{\tau}^2+\cosh^2\hat{\tau}\,d\hat{\theta}^2\label{dslerho}
	\end{equation}
	It is easy to see from eq.\eqref{globalembed} that  with the range of $\hat{\tau},\hat{\theta}$ 
	\begin{align}
		\hat{\tau}\in (-\infty,\infty),\quad \hat{\theta}\in [0,2\pi].\label{dstauhthh}
	\end{align}
	the coordinates $x_i$ have the range $x_i\in[-\infty,\infty]$ satisfying the constraint in eq.\eqref{dsasembed} and so cover whole dS spacetime. Casting the line element eq.\eqref{dslerho} in the conformal form by defining the coordinate $\hat{r}_{*}$ as 
	\begin{align}
		d\hat{r}_{*}&=\frac{d\hat{\tau}}{\cosh(\hat{\tau})}\nonumber\\
		\Rightarrow \hat{r}_*&=2\arctan(\tanh(\hat{\tau}\over 2)) \label{dsrsrho}
	\end{align}
	the metric in the $\hat{r}_*,\hat{\theta}$ coordinates becomes
	\begin{equation}
		ds^2=\frac{d\hat{\theta}^2-d\hat{r}_{*}^2}{\cos^2\hat{r}_*}\label{dslers}
	\end{equation}
	We will find it useful to shift the $\hat{r}_*$ coordinate by $\frac{\pi}{2}$,i.e.,
	\begin{align}
		\hat{r}_*\rightarrow \hat{r}_*+\frac{\pi}{2}\label{rsshift}
	\end{align}
	so that the range of the coordinate $\hat{r}_*$ and the new line element are
	\begin{align}
		ds^2=\frac{d\hat{\theta}^2-d\hat{r}_*^2}{\sin^2\hat{r}_*},\quad \hat{r}_*\in [0,\pi],\hat{\theta}\in[0,2 \pi]\label{dsrsshift}
	\end{align} 
	as in eq.\eqref{dsrsshift}, it is easy to see that the Penrose diagram in this case would be as shown below.
	
	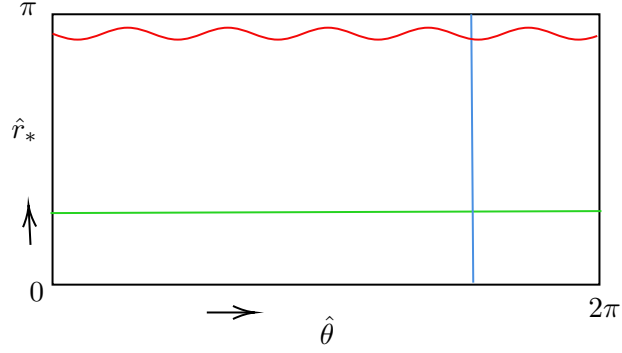
\begin{figure}[h!]
		\tikzset{every picture/.style={line width=0.75pt}} 
		\centering
		\begin{tikzpicture}[x=0.75pt,y=0.75pt,yscale=-1,xscale=1]
			
			\draw   (100,111) -- (373,111) -- (373,247) -- (100,247) -- cycle ;
			\draw    (89,227) -- (88.1,209) ;
			\draw [shift={(88,207)}, rotate = 87.14] [color={rgb, 255:red, 0; green, 0; blue, 0 }  ][line width=0.75]    (10.93,-3.29) .. controls (6.95,-1.4) and (3.31,-0.3) .. (0,0) .. controls (3.31,0.3) and (6.95,1.4) .. (10.93,3.29)   ;
			\draw    (177,261) -- (199,261) ;
			\draw [shift={(201,261)}, rotate = 180] [color={rgb, 255:red, 0; green, 0; blue, 0 }  ][line width=0.75]    (10.93,-3.29) .. controls (6.95,-1.4) and (3.31,-0.3) .. (0,0) .. controls (3.31,0.3) and (6.95,1.4) .. (10.93,3.29)   ;
			\draw [color={rgb, 255:red, 39; green, 211; blue, 33 }  ,draw opacity=1 ]   (99,211) -- (374,210) ;
			\draw [color={rgb, 255:red, 74; green, 144; blue, 226 }  ,draw opacity=1 ]   (309,111) -- (310,246) ;
			\draw  [color={rgb, 255:red, 243; green, 9; blue, 9 }  ,draw opacity=1 ] (100,120.75) .. controls (104.08,122.29) and (107.98,123.75) .. (112.5,123.75) .. controls (117.02,123.75) and (120.92,122.29) .. (125,120.75) .. controls (129.08,119.21) and (132.98,117.75) .. (137.5,117.75) .. controls (142.02,117.75) and (145.92,119.21) .. (150,120.75) .. controls (154.08,122.29) and (157.98,123.75) .. (162.5,123.75) .. controls (167.02,123.75) and (170.92,122.29) .. (175,120.75) .. controls (179.08,119.21) and (182.98,117.75) .. (187.5,117.75) .. controls (192.02,117.75) and (195.92,119.21) .. (200,120.75) .. controls (204.08,122.29) and (207.98,123.75) .. (212.5,123.75) .. controls (217.02,123.75) and (220.92,122.29) .. (225,120.75) .. controls (229.08,119.21) and (232.98,117.75) .. (237.5,117.75) .. controls (242.02,117.75) and (245.92,119.21) .. (250,120.75) .. controls (254.08,122.29) and (257.98,123.75) .. (262.5,123.75) .. controls (267.02,123.75) and (270.92,122.29) .. (275,120.75) .. controls (279.08,119.21) and (282.98,117.75) .. (287.5,117.75) .. controls (292.02,117.75) and (295.92,119.21) .. (300,120.75) .. controls (304.08,122.29) and (307.98,123.75) .. (312.5,123.75) .. controls (317.02,123.75) and (320.92,122.29) .. (325,120.75) .. controls (329.08,119.21) and (332.98,117.75) .. (337.5,117.75) .. controls (342.02,117.75) and (345.92,119.21) .. (350,120.75) .. controls (354.08,122.29) and (357.98,123.75) .. (362.5,123.75) .. controls (365.86,123.75) and (368.88,122.94) .. (371.88,121.9) ;
			
			\draw (78,161.4) node [anchor=north west][inner sep=0.75pt]    {$\hat{r}_{*}$};
			\draw (232,261.4) node [anchor=north west][inner sep=0.75pt]    {$\hat{\theta }$};
			\draw (87,244.4) node [anchor=north west][inner sep=0.75pt]    {$0$};
			\draw (366,252.4) node [anchor=north west][inner sep=0.75pt]    {$2\pi $};
			\draw (82,103.4) node [anchor=north west][inner sep=0.75pt]    {$\pi $};
		\end{tikzpicture}
		\label{dsglpenr}
		\caption{The coloured horizontal line corresponds to fixed $\hat{r}_*$ and coloured vertical line corresponds to fixed $\hat{\theta}$ and the wavy line indicates the singularity where $\phi=-1$.}
	\end{figure}	
	Various other useful coordinate systems to describe the global dS are the following. The transformation
	\begin{align}
		r=\cot\hat{r}_*=\sinh\hat{\tau}\label{dsrsr}
	\end{align}
	gives 
	\begin{align}
		ds^2=(1+r^2)d\hat{\theta}^2-\frac{dr^2}{r^2+1}\label{dsrthcs}
	\end{align}
		%
	Defining the null coordinates
	\begin{equation}
		\zeta^{\pm}={\hat{\theta}\pm \hat{r}_{*}}\label{dsrsze}
	\end{equation}
	the line element in these coordinates becomes,
	\begin{equation}
		ds^2=\frac{d\zeta^{+}d\zeta^{-}}{\sin^2(\frac{\zeta^{+}-\zeta^{-}}{2})}\label{dsleze}
	\end{equation}
	The coordinate transformation 
	\begin{equation}
		x^+= \tan({\zeta^+\over 2}) , x^-=\tan ({\zeta^-\over 2})\label{dspcgl}
	\end{equation}
	The line element  eq.\eqref{dsleze} becomes
	\begin{equation}
		ds^2= \frac{4 \, dx^+ dx^-}{(x^+- x^-)^2}\label{dspclc}
	\end{equation}
	Written in terms of the conventional Poincare coordinates $\eta,z$ given by 
	\begin{align}
		x^\pm =z\pm \eta\label{dszetapo}
	\end{align}
	in terms of which the line element becomes
	\begin{align}
		ds^2=\frac{dz^2-d\eta^2}{\eta^2}\label{dspnmet}
	\end{align}
	The metric in eq.\eqref{dspnmet} can be obtained by the following parametrization of the embedding coordinates $x_i$ in terms of $\eta, z$ 
	\begin{align}
		x_0=\frac{1}{2}\pqty{\frac{1}{\eta}-\eta}+\frac{z^2}{2\eta},\,\,x_1=\frac{1}{2}\pqty{\frac{1}{\eta}+\eta}-\frac{z^2}{2\eta},\,\,
		x_2=\frac{z}{\eta}\label{xipoinrel}
	\end{align}
	Instead of using the above to construct the Penrose diagram, 
	we use the relations between Poincare and global coordinates to understand the region of full dS covered by these coordinates. From eq.\eqref{dspcgl}, we have
	\begin{align}
		z\pm\eta=\tan(\frac{\hat{\theta}\pm \hat{r}_*}{2}),\,\label{zetainrsths}
	\end{align}
	{{It is easy to see from the above relations that}}
	\begin{align}
	z=\frac{\sin\hat{\theta}}{\cos\hat{\theta}+\cos\hat{r}_*},\,	\eta=\frac{\sin\hat{r}_*}{\cos\hat{\theta}+\cos\hat{r}_*}\label{zexpxmpoin}
	\end{align}
	Using the above information, it is easy to see that the patch of the full dS spacetime covered by the Poincare coordinates can be represented as below,
	
	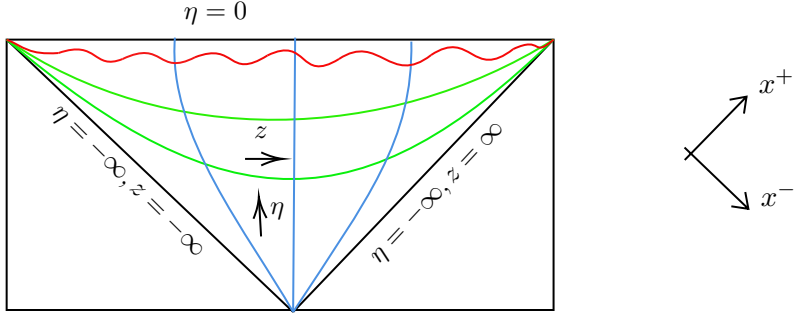
\begin{figure}[h!]
		\centering

		\tikzset{every picture/.style={line width=0.75pt}} 
		
		\begin{tikzpicture}[x=0.75pt,y=0.75pt,yscale=-1,xscale=1]
			
			\draw   (128,41) -- (401,41) -- (401,177) -- (128,177) -- cycle ;
			\draw    (128,41) -- (271,178) ;
			\draw    (401,41) -- (271,178) ;
			\draw [color={rgb, 255:red, 41; green, 239; blue, 5 }  ,draw opacity=1 ]   (128,41) .. controls (206.88,95.75) and (317.88,93.75) .. (401,41) ;
			\draw [color={rgb, 255:red, 9; green, 239; blue, 13 }  ,draw opacity=1 ]   (128,41) .. controls (238,135) and (306,134) .. (401,41) ;
			\draw [color={rgb, 255:red, 74; green, 144; blue, 226 }  ,draw opacity=1 ]   (212,40) .. controls (208,89) and (249,139) .. (271,178) ;
			\draw [color={rgb, 255:red, 74; green, 144; blue, 226 }  ,draw opacity=1 ]   (330,42) .. controls (331,92) and (308,121) .. (271,178) ;
			\draw [color={rgb, 255:red, 74; green, 144; blue, 226 }  ,draw opacity=1 ]   (272,40) -- (271,178) ;
			\draw    (255,140) -- (253.99,122) ;
			\draw [shift={(253.88,120)}, rotate = 86.8] [color={rgb, 255:red, 0; green, 0; blue, 0 }  ][line width=0.75]    (10.93,-3.29) .. controls (6.95,-1.4) and (3.31,-0.3) .. (0,0) .. controls (3.31,0.3) and (6.95,1.4) .. (10.93,3.29)   ;
			\draw    (247,101) -- (264,101) ;
			\draw [shift={(266,101)}, rotate = 180] [color={rgb, 255:red, 0; green, 0; blue, 0 }  ][line width=0.75]    (10.93,-3.29) .. controls (6.95,-1.4) and (3.31,-0.3) .. (0,0) .. controls (3.31,0.3) and (6.95,1.4) .. (10.93,3.29)   ;
			\draw [color={rgb, 255:red, 241; green, 9; blue, 9 }  ,draw opacity=1 ]   (128,41) .. controls (136.88,44.75) and (138.88,48.75) .. (153.88,47.75) ;
			\draw [color={rgb, 255:red, 248; green, 15; blue, 15 }  ,draw opacity=1 ]   (153.88,47.75) .. controls (173.88,40.75) and (170.88,56.75) .. (185.88,50.75) ;
			\draw [color={rgb, 255:red, 243; green, 23; blue, 23 }  ,draw opacity=1 ]   (185.88,50.75) .. controls (206.88,40.75) and (202.88,59.75) .. (222.01,49.88) ;
			\draw [color={rgb, 255:red, 241; green, 9; blue, 9 }  ,draw opacity=1 ]   (222.01,49.88) .. controls (242.88,41.75) and (237.88,61.75) .. (258.13,49) ;
			\draw [color={rgb, 255:red, 248; green, 9; blue, 9 }  ,draw opacity=1 ]   (258.13,49) .. controls (278.88,38.75) and (273.51,63.37) .. (293.88,50.75) ;
			\draw [color={rgb, 255:red, 241; green, 11; blue, 11 }  ,draw opacity=1 ]   (293.88,50.75) .. controls (314.11,37.33) and (319.28,64.89) .. (339.88,48.75) ;
			\draw [color={rgb, 255:red, 246; green, 9; blue, 9 }  ,draw opacity=1 ]   (339.88,48.75) .. controls (360.11,35.33) and (360.28,61.89) .. (380.88,45.75) ;
			\draw [color={rgb, 255:red, 239; green, 5; blue, 5 }  ,draw opacity=1 ]   (380.88,45.75) .. controls (395.88,38.75) and (389.88,50.75) .. (401,41) ;
			\draw  (466.38,95.27) -- (498.24,126.12)(497.32,69.69) -- (466.48,101.54) (496.69,117.66) -- (498.24,126.12) -- (489.73,124.85) (488.86,71.24) -- (497.32,69.69) -- (496.04,78.2)  ;
			
			\draw (258,121.4) node [anchor=north west][inner sep=0.75pt]  [font=\small]  {$\eta $};
			\draw (250,83.4) node [anchor=north west][inner sep=0.75pt]  [font=\small]  {$z$};
			\draw (215,20.4) node [anchor=north west][inner sep=0.75pt]  [font=\small]  {$\eta =0$};
			\draw (155.93,70.21) node [anchor=north west][inner sep=0.75pt]  [font=\small,rotate=-44.98]  {$\eta =-\infty ,z=-\infty $};
			\draw (305.89,148.44) node [anchor=north west][inner sep=0.75pt]  [font=\small,rotate=-312.84]  {$\eta =-\infty ,z=\infty $};
			\draw (122,131) node [anchor=north west][inner sep=0.75pt]   [align=left] {};
			\draw (502,54.4) node [anchor=north west][inner sep=0.75pt]    {$x^{+}$};
			\draw (503,111.4) node [anchor=north west][inner sep=0.75pt]    {$x^{-}$};

		\end{tikzpicture}
		\label{dsponpenrose}
		\caption{The green curves correspond to curves of constant $\eta$ and blue curves corresponds to cuves of constant $z$ and red curve corresponds to $\phi=-1$ singularity. }
	\end{figure}

	The Milne patch is given by the metric
	\begin{align}
		ds^2=-\frac{d\tilde{r}^2}{\tilde{r}^2-1}+(\tilde{r}^2-1)\,d\theta^2,\qquad \tilde{r}>{1}\label{rthemilne}
	\end{align}
	Defining the coordinate $\tau$ as
	\begin{align}
		\tilde{r}=\cosh{\tau}\label{taurmil}
	\end{align}
	the line element becomes
	\begin{align}
		ds^2=-d{\tau}^2+\sinh^2{\tau}\,d\theta^2,\quad \tau\in(-\infty,\infty)\label{tauthmil}
	\end{align}
	To rewrite the metric eq.\eqref{rthemilne} in conformally flat form, we define $r_*$ coordinate as
	\begin{align}
		\tilde{r}=-\coth \,r_*\label{rthmil}
	\end{align}
	in terms of which the line element becomes
	\begin{align}
		ds^2= \frac{d\theta^2-dr_{*}^2}{\sinh^2  r_*},\qquad r_*\in(-\infty,0)\label{milnersmet}
	\end{align}
	Defining the null coordinates 
	\begin{align}
		\zeta^\pm=\theta\pm {r}_*\label{milnezetaco}
	\end{align}
	the line element in terms of $\zeta^\pm$ becomes
	\begin{align}
		ds^2=\frac{ \,d\zeta^+ d\zeta^-}{\sinh^2(\frac{\zeta^+ - \zeta^-}{2})}\label{milnezetamet}
	\end{align}
	Further defining the coordinates $x^\pm$ as 
	\begin{align}
		x^\pm=\tanh(\frac{\,\zeta^\pm}{2})\label{milnepoinco}
	\end{align}
	the line element becomes
	\begin{align}
		ds^2=\frac{4\, dx^+ dx^-}{(x^+ - x^-)^2}\label{milnepoinmet}
	\end{align}
	When we consider the theory with identification of $\theta$ coordinate, discussed in section \ref{dssemicla}, one particular set of null coordinates $z^\pm$ that will be useful, are defined as
	\begin{align}
		z^\pm=\exp(\frac{\pi^2}{b}\pm i\frac{2\pi}{b}\zeta^\pm)\label{milnezpm}
	\end{align}
	
	Let us now analyze the Penrose diagram. 
	We see that the above coordinate system breaks down at $\tilde{r}=\pm 1$. Indeed, the cosmological and black hole horizons are located at $\tilde{r}=\pm 1$ where the geometry is smooth even though coordinates break down. So these locations are just coordinate singularities. To find the maximal extension of the spacetime, we devise a pair of Kruskal-like coordinates.  As a first step, we define the coordinates $u,v$ as
	\begin{align}
		v=\theta+\int {d\tilde{r}\over \tilde{r}^2-1}=\theta+\frac{1}{2}\ln \abs{\tilde{r}-1\over \tilde{r}+1}=\theta+r_*\nonumber\\
		u=\theta-\int {d\tilde{r}\over \tilde{r}^2-1}=\theta-\frac{1}{2}\ln \abs{\tilde{r}-1\over \tilde{r}+1}=\theta-r_*\label{uv}
	\end{align}
	 Define the Kruskal coordinates $X^+_K,X^-_K $ in the future Milne wedge,  shown as region $\text{I}$ in figure  below where $\tilde{r}>1$ as
	\begin{align}
		X^-_K=-e^{-u}=-e^{-\theta+r_*}, \, X^+_K =e^{v}=e^{\theta+r_*}.\label{UV1st}
	\end{align}
	
	\tikzset{every picture/.style={line width=0.75pt}} 
	\begin{figure}[h!]
		\centering
		\vspace{-0.3cm}

		\tikzset{every picture/.style={line width=0.75pt}} 
		
		\begin{tikzpicture}[x=0.75pt,y=0.75pt,yscale=-1,xscale=1]
			
			\draw    (223.69,65.57) -- (370.44,222.2) ;
			\draw    (224.79,224) -- (369.34,63.77) ;
			\draw    (151.98,146.89) -- (223.69,65.57) ;
			\draw    (151.98,146.89) -- (224.79,224) ;
			\draw    (369.34,63.77) -- (438,143.73) ;
			\draw    (438,143.73) -- (370.44,222.2) ;
			\draw    (223.69,65.57) -- (369.34,63.77) ;
			\draw    (370.44,222.2) -- (224.79,224) ;
			
			\draw (272,47.4) node [anchor=north west][inner sep=0.75pt]  [font=\scriptsize]  {$X^+_KX^-_K =-1$};
			\draw (277,227.4) node [anchor=north west][inner sep=0.75pt]  [font=\scriptsize]  {$X^+_K X^-_K =-1$};
			\draw (282,170.4) node [anchor=north west][inner sep=0.75pt]  [font=\scriptsize]  {$ \begin{array}{l}
					X^+_K < 0\\
					X^-_K > 0
				\end{array}$};
			\draw (338,129.4) node [anchor=north west][inner sep=0.75pt]  [font=\scriptsize]  {$ \begin{array}{l}
					X^+_K > 0\\
					X^-_K  >0
				\end{array}$};
			\draw (275,94.4) node [anchor=north west][inner sep=0.75pt]  [font=\scriptsize]  {$ \begin{array}{l}
					X^+_K  >0\\
					X^-_K  <0
				\end{array}$};
			\draw (234,130.4) node [anchor=north west][inner sep=0.75pt]  [font=\scriptsize]  {$ \begin{array}{l}
					X^+_K  <0\\
					X^-_K < 0
				\end{array}$};
			\draw (161.38,117.24) node [anchor=north west][inner sep=0.75pt]  [font=\scriptsize,rotate=-312.27]  {$\theta=-\infty $};
			\draw (231.38,195.24) node [anchor=north west][inner sep=0.75pt]  [font=\scriptsize,rotate=-312.27]  {$\theta=\infty $};
			\draw (336.69,169.03) node [anchor=north west][inner sep=0.75pt]  [font=\scriptsize,rotate=-45.61]  {$\theta=-\infty $};
			\draw (180.54,159.75) node [anchor=north west][inner sep=0.75pt]  [font=\scriptsize,rotate=-45.6]  {$\theta=\infty $};
			\draw (401.54,85.75) node [anchor=north west][inner sep=0.75pt]  [font=\scriptsize,rotate=-45.6]  {$\theta=\infty $};
			\draw (331.38,89.24) node [anchor=north west][inner sep=0.75pt]  [font=\scriptsize,rotate=-312.27]  {$\theta=\infty $};
			\draw (240.48,66.93) node [anchor=north west][inner sep=0.75pt]  [font=\scriptsize,rotate=-44.97]  {$\theta=-\infty $};
			\draw (292,71) node [anchor=north west][inner sep=0.75pt]   [align=left] {I};
			\draw (383,141) node [anchor=north west][inner sep=0.75pt]   [align=left] {III};
			\draw (290,201) node [anchor=north west][inner sep=0.75pt]   [align=left] {II};
			\draw (201,135) node [anchor=north west][inner sep=0.75pt]   [align=left] {IV};
		\draw  (534.45,103.51) -- (567.6,135.99)(566.16,77.77) -- (534.61,109.98) (566.1,127.52) -- (567.6,135.99) -- (559.1,134.66) (557.69,79.27) -- (566.16,77.77) -- (564.83,86.27)  ;
		
		\draw (569,60.4) node [anchor=north west][inner sep=0.75pt]    {$X_{K}^{+}$};
		\draw (574,127.4) node [anchor=north west][inner sep=0.75pt]    {$X_{K}^{-}$};

		\end{tikzpicture}
		
		\label{eefig3}
	\end{figure}
	These $X^+_K , X^-_K $ coordinates can be extended to the regions $\text{II, III, IV}$ by just taking the appropriate signs 
	\begin{align}
		X^-_K =\pm e^{-\theta+r_*},\, X^+_K =\pm e^{\theta+r_*}\label{fUVinfull}
	\end{align}
	From the above, we see that 
	\begin{align}
		\tilde{r}^2-1=\frac{-4 X^+_K X^-_K }{(1+X^+_K X^-_K )^2}.\label{rinUV}
	\end{align}
	Moreover, we also have that 
	\begin{align}
		&\tilde{r}\rightarrow 0\implies X^+_K X^-_K \rightarrow 1 ,\quad\tilde{r}\rightarrow 1\implies X^+_K X^-_K \rightarrow 0,  \quad \tilde{r}\rightarrow -1\implies X^+_K X^-_K \rightarrow \infty
		\label{UVlimits}
	\end{align}
	From the above, we see that the $X^+_K ,X^-_K $ coordinates break down near the black hole horizon corresponding to $\tilde{r}=-1$. So, we need to define a new pair of Kruskal-like coordinates $\tilde{X}^+_K ,\tilde{X}^-_K$ as
	\begin{align}
		\tilde{X}^+_K=-e^u,\,\tilde{X}^-_K=-e^{-v}\label{UtVtset}
	\end{align}
	This procedure can be repeated infinitely and so we get an infinite chain for the Penrose diagram.

	\tikzset{every picture/.style={line width=0.75pt}} 
	\begin{figure}[h!]

		\tikzset{every picture/.style={line width=0.75pt}} 
		
		\begin{tikzpicture}[x=0.75pt,y=0.75pt,yscale=-1,xscale=1]
			
			\draw   (212.79,55.93) -- (324.52,55.71) -- (323.88,185.84) -- (212.16,186.06) -- cycle ;
			\draw   (100.21,59.12) .. controls (103.67,61.66) and (106.98,64.08) .. (110.84,64.07) .. controls (114.7,64.06) and (118.04,61.63) .. (121.52,59.08) .. controls (125.01,56.53) and (128.35,54.1) .. (132.21,54.09) .. controls (136.07,54.08) and (139.38,56.5) .. (142.84,59.04) .. controls (146.31,61.58) and (149.62,64) .. (153.48,63.99) .. controls (157.34,63.98) and (160.68,61.55) .. (164.16,59) .. controls (167.65,56.45) and (170.99,54.02) .. (174.85,54.01) .. controls (178.71,54) and (182.02,56.42) .. (185.48,58.96) .. controls (188.95,61.5) and (192.26,63.91) .. (196.12,63.91) .. controls (199.98,63.9) and (203.32,61.47) .. (206.8,58.92) .. controls (209.09,57.24) and (211.31,55.63) .. (213.65,54.71) ;
			\draw   (99.58,188.75) .. controls (103.04,191.54) and (106.35,194.2) .. (110.21,194.19) .. controls (114.07,194.18) and (117.41,191.51) .. (120.9,188.71) .. controls (124.39,185.9) and (127.72,183.23) .. (131.58,183.22) .. controls (135.44,183.21) and (138.75,185.87) .. (142.22,188.66) .. controls (145.68,191.46) and (148.99,194.12) .. (152.85,194.11) .. controls (156.71,194.1) and (160.05,191.43) .. (163.54,188.62) .. controls (167.03,185.82) and (170.36,183.14) .. (174.22,183.14) .. controls (178.08,183.13) and (181.39,185.79) .. (184.86,188.58) .. controls (188.32,191.38) and (191.63,194.04) .. (195.49,194.03) .. controls (199.35,194.02) and (202.69,191.35) .. (206.18,188.54) .. controls (208.18,186.93) and (210.14,185.36) .. (212.17,184.32) ;
			\draw    (212.79,55.93) -- (323.88,185.84) ;
			\draw    (212.16,186.05) -- (324.52,55.72) ;
			\draw    (100.21,59.12) -- (212.16,186.05) ;
			\draw    (212.79,55.93) -- (99.57,189.24) ;
			\draw    (100.21,59.12) -- (99.57,189.24) ;
			\draw   (437.14,51.52) -- (548.87,51.3) -- (548.23,181.43) -- (436.5,181.65) -- cycle ;
			\draw   (324.55,54.71) .. controls (328.01,57.25) and (331.33,59.67) .. (335.19,59.66) .. controls (339.04,59.65) and (342.38,57.22) .. (345.87,54.67) .. controls (349.36,52.12) and (352.7,49.69) .. (356.55,49.68) .. controls (360.41,49.67) and (363.73,52.09) .. (367.19,54.63) .. controls (370.65,57.17) and (373.97,59.59) .. (377.83,59.58) .. controls (381.68,59.57) and (385.02,57.14) .. (388.51,54.59) .. controls (392,52.04) and (395.34,49.61) .. (399.19,49.6) .. controls (403.05,49.59) and (406.37,52.01) .. (409.83,54.55) .. controls (413.29,57.09) and (416.61,59.5) .. (420.47,59.5) .. controls (424.32,59.49) and (427.66,57.06) .. (431.15,54.51) .. controls (433.44,52.84) and (435.65,51.22) .. (438,50.3) ;
			\draw   (323.92,184.34) .. controls (327.38,187.13) and (330.7,189.79) .. (334.55,189.78) .. controls (338.41,189.78) and (341.75,187.1) .. (345.24,184.3) .. controls (348.73,181.49) and (352.07,178.82) .. (355.93,178.81) .. controls (359.79,178.8) and (363.1,181.46) .. (366.56,184.26) .. controls (370.02,187.05) and (373.34,189.71) .. (377.19,189.7) .. controls (381.05,189.69) and (384.39,187.02) .. (387.88,184.21) .. controls (391.37,181.41) and (394.71,178.74) .. (398.57,178.73) .. controls (402.43,178.72) and (405.74,181.38) .. (409.2,184.17) .. controls (412.66,186.97) and (415.98,189.63) .. (419.83,189.62) .. controls (423.69,189.61) and (427.03,186.94) .. (430.52,184.13) .. controls (432.53,182.52) and (434.48,180.95) .. (436.52,179.91) ;
			\draw    (437.14,51.52) -- (548.23,181.43) ;
			\draw    (436.51,181.64) -- (548.86,51.31) ;
			\draw    (324.55,54.71) -- (436.51,181.64) ;
			\draw    (437.14,51.52) -- (323.92,184.83) ;
			\draw    (324.55,54.71) -- (323.92,184.83) ;
			
			\draw (557,99) node [anchor=north west][inner sep=0.75pt]  [font=\Large] [align=left] {. . . .};
			\draw (49,102) node [anchor=north west][inner sep=0.75pt]  [font=\Large] [align=left] {. . . .};
		\end{tikzpicture}
		\label{penrosefig}
	\end{figure}
	For completeness, we note down the relation between the Kruskal coordinates eq.\eqref{fUVinfull} and the Poincare coordinates eq.\eqref{milnepoinco}, say in the region II, to be
	\begin{align}
	x^+=\frac{X^+_K+1}{X^+_K-1},\quad x^-=\frac{1-X^-_K}{X^-_K+1} 	\label{kporel}
	\end{align}
	The analog of the coordinate system eq.\eqref{rthemilne} in the static patch corresponding to the region $\text{III}$ in Fig. \ref{eefig3} is given by 
	\begin{align}
		ds^2=\frac{d\tilde{r}^2}{1-\tilde{r}^2}-(1-\tilde{r}^2)\,d\theta^2,\qquad \abs{\tilde{r}}<{1}\label{rthstatic}
	\end{align}
	So, in this region we have a time-like Killing vector $\del_\theta$. To go to the conformally flat form, we can define the coordinate $r_*$ as
	\begin{align}
		\tilde{r}=-\tanh r_*\label{statictori}
	\end{align}
	in terms of which the metric becomes
	\begin{align}
		ds^2=\frac{dr_*^2-d\theta^2}{\cosh^2r_*}\label{dsstametor}
	\end{align}
	To further go to the Poincare coordinate system, we can do the coordinate transformation 
	\begin{align}
		x^+ =- \coth(\frac{\theta+r_*}{2}),\,\,x^-=-\tanh(\frac{\theta-r_*}{2}) \label{xpxminstatic}
	\end{align}
	following which the metric becomes
	\begin{align}
		ds^2=\frac{4 dx^+ dx^-}{(x^+ - x^-)^2}\label{dsstmpoin}
	\end{align}

	\subsection{Euclidean AdS double trumpet}
	\label{eads2dtg}
	The line element for this geometry is given by 
	\begin{equation}
		ds^2=(r^2+1)d\theta^2+\frac{dr^2}{(r^2+1)},\quad \theta\sim \theta+b\label{edtrt}
	\end{equation}
	The two boundaries correspond to the limits $r\rightarrow \infty$ and $r\rightarrow-\infty$. Performing the coordinate transformations
	\begin{equation}
		r=\sinh(\rho)\label{edtrrho}
	\end{equation}
	we find that the metric is given by
	\begin{equation}
		ds^2=d\rho^2+\cosh^2(\rho)d\theta^2\label{edtlerho}
	\end{equation}
	It has to be noted that the $\theta$ direction is periodic with period $2\pi$.
	Defining $r_{*}$ coordinate as 
	\begin{equation}
		r=\cot(r_*)\label{edtrsrho}
	\end{equation}
	In term of the $r_*$ coordinates the metric becomes
	\begin{equation}
		ds^2=\frac{dr_*^2+d\theta^2}{\sin^2(r_*)}\label{edletrs}
	\end{equation}
	This can be written in complex coordinates as 
	\begin{equation}
		\zeta={r_*+i \theta},\bar{\zeta}={r_*-i \theta}\label{edtrsze}
	\end{equation}
	the metric becomes
	\begin{equation}
		ds^2=\frac{d\zeta\, d\bar{\zeta}}{\sin^2(\frac{\zeta+\bar{\zeta}}{2})}\label{edtleze}
	\end{equation}
	To get it to the Poincare form, consider the further coordinate transformation
	\begin{equation}
		x=i\tan(\frac{\zeta}{2}),\,\bar{x}=-i\tan(\bar{\frac{\zeta}{2}})\label{edtpcze}
	\end{equation}
	and hence the metric becomes
	\begin{equation}
		ds^2=-\frac{4 dx d\bar{x}}{(x-\bar{x})^2}\label{edtlepc}
	\end{equation}

	\section{Orbifold theory}
	\label{orbfoapp}
	\subsection{Classical matter}
	\label{obcm}
	In this subsection, we shall mention some formulae to compute the stress tensor in the orbifold theory discussed in section \ref{dssemicla}. Consider a matter stress tensor $T_{++}$ of the form
	\begin{align}
		T_{++}=\mu_0^+ \delta(x^+ - x^+_0)\label{stinor}
	\end{align}
where 
\begin{align}
	x^\pm_0=\tanh(\frac{\theta_0\pm r_{*0}}{2})\label{x0def}
\end{align}
	To compute the backreaction in the orbifold theory, it is convenient to use the procedure of method of images and work in the un-orbifolded theory. Let the full stress tensor including the images be given by 
	\begin{align}
		T_{++}^{I}=\sum_{n}\mu_n^+ \delta(x^+ - x^+_n )\label{strunorb}
	\end{align}
where 
\begin{align}
	x^\pm_n=\tanh({\theta_0+n b\pm r_{*0}\over 2})\label{xndef}
\end{align}
Now, we need to find the compute the strength of the images $\mu_n, n\neq 0$. To do this, we note that the full stress tensor with images in eq.\eqref{strunorb} should be invariant under the shift of the coordinate $\theta$ as $\theta\rightarrow \theta+b$, where $x^\pm $ is related to $\theta$ by eq.\eqref{milnepoinco}.Viewing the shift $\theta\rightarrow \theta+b$ as a coordinate transformation under which $x^\pm \rightarrow \hat{x}^\pm $ where $\hat{x}^{\pm}$ are given by 
\begin{align}
	\hat{x}^\pm =\tanh(\frac{\theta+b\pm r_*}{2})\label{xhco}
\end{align}
and hence, using the transformation rule for the stress tensor as a rank two covariant tensor, and noting that the full stress tensor remains unchanged under this coordinate transformation, we arrive at
\begin{align}
	\sum_m \mu_m^+ \delta(\hat{x}^+ - {x}^+_m)=\frac{(1-(x^+)^2)}{(1-(\hat{x}^+)^2)}\sum_n \mu_n^+ \delta(x^+ - {x}^+_n)\label{strimeq}
\end{align}
From the above, we get
\begin{align}
	\mu_{n+1}^+=\mu_n^+\frac{(1-(x^+_n)^2)^2}{(1-(x^+_{n+1})^2)^2}\label{imrecur}
\end{align}
And so, we get
\begin{align}
	\mu_n^+=\mu_0^+\frac{(1-(x^+_0)^2)^2}{(1-(x^+_{n})^2)^2}\label{muninmu0}
\end{align}
Now, we need to solve for the dilaton given the stress tensor eq.\eqref{strunorb}. Since, the equations of motion involving the dilaton in eq.\eqref{pm},\eqref{ppmm} are linear in dilaton and the stress tensor, we can solve for each of the image sources and obtain the full solution by superposition of the response from each image source. Starting from a dilaton solution of the form eq.\eqref{dilinhk},\eqref{stc}, and noting the solution eq.\eqref{laso} for a source of the form eq.\eqref{strt}, we get the full solution for the stress tensor eq.\eqref{strunorb} to be given by 
\begin{align}
	k(x^+)=\sum_n \mu_n^+ (x^+ - x^+_n)^2\Theta(x^+ - x^+_n )\label{kfuim}
\end{align}
We can now construct the solution for the dilaton by substituting this function in eq.\eqref{dilinhk}. Consider $x^+_j<x^+<x^+_{j+1}$ for some $j$. We find the dilation solution to be given by 
\begin{align}
	\phi=\frac{\tilde{a}+\tilde{b}(x^+ + x^-)+\tilde{c}x^+ x^-}{x^+ - x^-}\label{fulimso}
\end{align}
where 
\begin{align}
	\tilde{a}=1+\sum_{n\leq j}\mu_n^+ (x^+_n)^2, \quad \tilde{b}=-\sum_{n\leq j}\mu_n^+ x^+_n,\quad\tilde{c}=-\mu^- +\sum_{n\leq j}\mu_n^+\label{atbtctde}
\end{align}.
We see from eq.\eqref{muninmu0} and eq.\eqref{xndef} that for $b$ large, the sources become larger as $n$ increases and so the backreaction becomes stronger and stronger.

	\subsection{Stress tensor transformation}
	\label{sttrn}
	
	In this appendix we give some formulae for the stress tensor of a massless free scalar field transforms corresponding to a choice of the coordinates with respect to which  the vaccuum state of the matter fields is defined \cite{Moitra:2019xoj,Christensen:1977jc}. 
	Consider two coordinate systems denoted by $(X^+,X^-)$ and $(x^+,x^-)$ with metric,
	\begin{align}
		\mathrm{d}s^2 & =  - F(X^+,X^-)  \mathrm{d} X^+ \mathrm{d} X^- \label{metrics},\\
		\mathrm{d} s^2 & =  -f(x^+,x^-) \mathrm{d}x^+ \mathrm{d}x^- \label{metricsb}.
	\end{align}
	Then, the stress tensor components for the system in the vaccuum with respect to coordinates $X^\pm$ but computed in the coordinate system $x^\pm$ is given by
	\begin{equation}
		\langle X | T_{x^+ x^+} | X \rangle = \langle x | T_{x^+ x^+} | x \rangle -\frac{1}{24\pi} \text{Sch}(X^+,x^+)\label{strtran}.
	\end{equation}
	where 
	\begin{equation}
		\langle x|T_{x^+x^+}|x\rangle = -\frac{1}{12\pi}f^{\frac{1}{2}}\del^2_{x^+}f^{-\frac{1}{2}}\label{txvac}.
	\end{equation}
	There is an analogous formula with $(X^+,x^+)$ being replaced by $(X^-,x^-)$.

	Now, in the case of orbifold theory considered in section \ref{dssemicla}, for the matter fields in the vaccum with respect to the coordinates eq.\eqref{milnezpm},  
	\begin{align}
		\langle z^+| T_{z^+z^+}|z^+\rangle =\frac{1}{48\pi (z^+)^2}\pqty{1+\frac{b^2}{4\pi^2}}\label{Tzzval}
	\end{align}
	Using the coordinate transformations eq.\eqref{milnezetaco}, we get the stress tensor components as
	\begin{align}
		\langle T_{\zeta^\pm,\zeta^\pm}\rangle \equiv \frac{\mathcal{A}}{4}=-\frac{1}{48\pi}\pqty{1+\frac{4\pi^2}{b^2}}<0\label{tzetappmm}
	\end{align}
	where $\zeta^\pm$ are defined in eq.\eqref{milnezetaco}.
	Further using the coordinate transformation to go to the Poincare coordinates,\eqref{milnepoinco}, we see that the stress tensor becomes
	\begin{align}
		\langle T_{\pm,\pm}\rangle \equiv \langle T_{x^\pm,x^\pm}\rangle =\frac{\mathcal{A}}{(1-(x^\pm)^2)^2}\label{poinstress}
	\end{align}

	\section{dS$_2$ from dS$_4$}
	\label{pennhlim}
	
	In this subsection, we shall show the steps illustrating how $\text{dS}_2\times \text{S}^2$ arises in the near-Nariai limit of the $dS_4$ black hole. The 4D black hole solution to the action 
	\begin{align}
		I_{4D}=\frac{1}{16\pi G_4}\int d^4 x \sqrt{-g}(R-2)\label{4daction}
	\end{align}
	is given by
	\begin{align}
		ds^2=-f(r)dt^2+{dr^2\over f(r)}+\Phi^2d\Omega_2^2\quad
		\text{where }\quad 
		f(r)=1-r^2-\frac{\mu}{r},\quad \Phi=r\label{4dmet}
	\end{align}
	Requring $f(r)=0$ gives a cubic equation in $r$ whose solutions correspond to horizons. Let them be denoted by $r_c, r_b, r_a$ where $r_c>r_b>r_a$. It is easy to see that these satisfy the relations
	\begin{align}
		r_c+r_b+r_a=0, \quad r_c r_b r_a=-\mu \label{rcrbrarels}
	\end{align}
	The cosmological horizon is given by $r=r_c$ and the black hole horizon by $r=r_b$. The temperature, as computed by requiring that the Euclidean circle shrinks smoothly, say as $r\rightarrow r_c$ is given by 
	\begin{align}
		T_{r_c}={f'(r_c)\over 4\pi}={1-3 r_c^2\over 4\pi r_c}\label{costemp}
	\end{align}
	In the near-Nariai limit $r_c\rightarrow r_b$. Let the extremal value of the horizon be $r_h$.  The value of $r_h$ can be easily calculated from eq.\eqref{rcrbrarels} and is found to be 
	\begin{align}
		r_h=\left(\frac{\mu}{2}\right)^{\frac{1}{3	}}=\frac{1}{\sqrt{3}}\label{rhmuval}
	\end{align}
	
	In this limit, $f(r)$ has a double zero given by 
	\begin{align}
		f(r)\simeq -3(r-r_h)^2\label{nariaf}
	\end{align}
	Slightly away from this limit, parametrizing $r_c, r_b$ as 
	\begin{align}
		r_c=r_h+\delta, r_b=r_h-\delta , \quad \delta\ll r_h\label{nnarrcrb}
	\end{align}
	we get the form of $f(r)$ as 
	\begin{align}
		f(r)=-3((r-r_h)^2-\delta^2)\label{finnnar}
	\end{align}
	Let $\tilde{r}={r-r_h\over \delta}$ and so  $f(r)= -3 \delta^2(\tilde{r}^2-1)$. Taking $t\rightarrow \frac{t}{3\delta}$ the line element becomes
	\begin{align}
		ds^2=\frac{1}{3}\pqty{-\frac{d\tilde{r}^2}{\tilde{r}^2-1}+(\tilde{r}^2-1)dt^2}+\Phi_0^2(1+\phi)d\Omega_2^2\label{nds2le}
	\end{align}
	where 
	\begin{align}
		\Phi_0=r_h,\,\phi\simeq {{{2}}(r-r_h)\over r_h}
		\label{phidef}
	\end{align}
	We see from eq.\eqref{nds2le}, the geometry becomes $\text{dS}_2\times \text{S}^2$.
	The full 4D entropy would be given by the sum of contributions from the cosmological and black horizons and reads
	\begin{align}
		S^{4D}=\frac{\pi}{G_4}\left(\Phi_0^2\bigg\vert_{\text{BH}}+\Phi_0^2\bigg\vert_{\text{CH}}\right)\rightarrow \frac{2\pi \Phi_0^2}{G_4}=\frac{\pi\mu^3}{4G_4}\label{ex4dent}
	\end{align}
	where the limit at the end in the above equation corresponds to the extremal limit with BH and  CH denoting black hole and cosmological horizons respectively.
	Doing a dimensional reduction of the action in eq.\eqref{4daction} with the ansatz
	\begin{align}
		ds^2=\tilde{g}_{\alpha\beta}dx^\alpha dx^\beta+\Phi_0^2(1+\phi)d\Omega_2^2\label{dimredas}
	\end{align}
	would give the 2D JT action in eq.\eqref{jtact} with the identification
	\begin{align}
		\frac{1}{G}=\frac{4\pi}{G_4}\label{Gg4}
	\end{align}
	The genus counting parameter in JT gravity that corresponds to the topological term has the value
	\begin{align}
		S_0={4\pi \Phi_0^2\over 4 G_4}=\frac{\pi \mu^3}{8G_4}\label{s0ext}
	\end{align}
	which is half the extremal entropy of the full 4D black hole solution eq.\eqref{ex4dent}.
	
		The spacetime is defined to have a singularity when the dilaton, $\phi$, becomes sufficiently negative that the radius of the transverse sphere in $4D$ shrinks to zero. This happens at 
	\begin{align}
		\phi=-1\
		\label{singul}
	\end{align}

	\section{Remarks on Canonical quantisation}
	\label{canquant}

	In this section, we shall discuss some aspects of canonical quantisation of the JT theory in dS both in the second and first order formalisms. 
	
	{\bf Second Order Formalism: } The quantisation of the theory in the second order formalism was carried out in \cite{Hennauxjt} and \cite{Louis-Martinez:1993bge}. Here we will mainly review a few points which allow us to connect to the main text especially section \ref{transamp},
	and also bring out some key features of the quantisation. 
	
	Writing the metric as
	\begin{align}
		ds^2=-dt^2+e^{2\rho}d\theta^2,\quad \theta\sim \theta+2\pi\label{thlmet}
	\end{align}
	The action for JT gravity  in the minisuperspace approximation in which $\phi,\rho$ depend only on time $t$, becomes
	\begin{align}
		S_{JT}=-\frac{1}{4 G}\int dt \, (\dot{\phi}\dot{\rho}+\phi )e^{\rho}\label{minsupact}
	\end{align}
	The canonical conjugate variable $\pi_\phi,\pi_\rho$ are given by
	\begin{align}
		\pi_\rho=-\frac{\dot{\phi}\,e^\rho}{4 G}, \,	\pi_\phi=-\frac{\dot{\rho}\,e^\rho}{4 G}\label{conjmom}
	\end{align}
	Written in terms of the length of the boundary $\hat{l}=2\pi e^\rho$, we get
	\begin{align}
		\pi_{\hat{l}}=\frac{\pi_{\rho}}{\hat{l}}=-\frac{\dot{\phi}\,e^\rho}{4G\hat{l}}\label{pilin}
	\end{align}
	We see that $\dot{\phi}>0$ leads to $\pi_{\hat{l}}<0$ as discussed in section \ref{transamp}.
	
	The Hamiltonian constraint for the wavefunction in the quantum theory then becomes, \cite{Maldacena:2019cbz},
	\begin{align}
		((8\pi G)^2\del_{\hat{ l}}\del_\phi+\phi\hat{ l})\Psi[\hat{l},\phi]=0\label{wdwinlp}
	\end{align}
		
		A more careful quantisation, as discussed in \cite{Hennauxjt}, see also, \cite{Iliesiu:2020zld} which keeps track of factor ordering ambiguities shows  that 
	it is ${\tilde \Psi}({\hat l},\phi)= {\Psi({\hat l},\phi)\over {\hat l}}$ which actually satisfies the WDW equation. 
	Keeping this in mind it is easy to see that an exact solution of the WDW equation is given by 	
	\begin{align}
	\label{exf}
	\Psi({\hat l},\phi)=  {{\hat l} \phi^2\over {\hat l}^2-4 \pi^2} \left[A H^{(2)}_2\left({\phi \sqrt{{\hat l}^2-4\pi^2} \over 8 \pi G}\right)+ B H^{(1)}_2\left({\phi \sqrt{{\hat l}^2-4\pi^2}\over 8 \pi G}\right)\right]
	\end{align}
	where $A,B$ are two constants and $H_2^{(1,2)}$ are the Hankel functions with index 2 of first or second kind. For appropriate $A,B$ this agrees with eq.(\ref{asmini}) in section \ref{onetwob}. 
	
	The WDW equation is second order and involves two functional derivatives. 
	One of the main points  in the quantisation discussed by Henneaux \cite{Hennauxjt}  is to simplilfy  the  constraints in the classical theory   by partial solving  and replacing them with constraints which involve only one functional derivative. And then going to the quantum theory where these simplified constraints are imposed.  The step where the constraints are partially solved gives, see also \cite{Iliesiu:2020zld}, 
	\begin{align}
	\Pi_\rho & =  \pm \sqrt{(\partial_\theta \phi)^2 + {(M + W(\phi))}e^{2\rho}}\rightarrow  \pm e^\rho\sqrt{M+W(\phi)} \label{miniaa}\\
	\Pi_\phi & =  \pm {{1\over 2} U  e^{2 \rho} + \partial_\theta^2\phi -\partial_\theta \phi \partial_\theta\rho\over \sqrt{(\partial_\theta \phi)^2 +( M + W(\phi))e^{2\rho}}}
	\rightarrow \pm{{1\over 2} U e^{\rho}\over \sqrt{M+W(\phi)}}\label{minibb}\end{align}
	Here $\partial_\phi W=U$, and $U=\Lambda \phi$, with $\Lambda=2$ in our notation, eq.\eqref{jtact}, in units where $8\pi G=1$. $M$ is an integration constant, and the last expression on the RHS in both equations are the values in the mini superspace approximation.
	 Comparing with the mini superspace approximation, eq.(\ref{conjmom})  we see that we have to  choose the $-$ sign in eq.(\ref{miniaa}), eq.(\ref{minibb}), to describe the expanding branch and the $+$ sign to describe the contracting branch. 
	
	However to describe the possibility of a tunnelling solution which goes from a contracting universe to an expanding one, we cannot make a particular choice and one must instead therefore return  to the original WDW equation which involves second order functional derivatives and try to solve it.    This is more complicated and provides us with  the motivation for considering the canonical quantisation in the first order formalism which we describe next.

	{\bf First Order Formalism}
	
	The quantisation of the  theory in the first order formalism was considered by Strobl,  \cite{Strobl:1993yn} and we briefly review it next. 
	In particular we will not be working in the mini-superspace approximation, and instead  consider the theory  with all its degrees of freedom. It will then turn out that imposing the constraints allow us to set all the spatially varying modes to vanish, showing that the mini-superspace approximation is in fact exact in this theory. 
	
	The action is 
	\begin{align}
		S&=\int   \,\,\mathcal{L}\nonumber\\
		\mathcal{L}&=\pi_A(dX^A+\frac{1}{2}f^{A}_{BC}X^B\wedge X^C)\label{dsfirstorder}
	\end{align}
	where  $X^A$ is the vector of 1-forms made of Vierbeins, $e^0,e^1$ and spin connection, $\omega$, and $f^A_{BC}$ are the structure constants of the de Sitter symmetry group $SO(2,1)$ parametrized by the generators $T_A$, where
	\begin{align}
		X^A&=(e^0,e^1,\omega)\nonumber\\
		\pi_A&=(\pi_0,\pi_1,\pi_\omega)\nonumber\\
		T_A&=\frac{1}{2}(\sigma_1,i\sigma_2,\sigma_3)\label{xsandts}
	\end{align}
	with $\sigma_i$ being the Pauli matrices. Note that $\pi_\omega$ is the dilaton $\phi$. 
	 In order to describe tunnelling transitions we take  the bulk manifold to have  a  cylindrical  topology with the spacelike direction being identified,
	 $\theta \simeq \theta+2 \pi$.	
	
	Expanding the Lagrangian eq.\eqref{dsfirstorder} in terms of components  reads
	\begin{align}
		\mathcal{L}=\pi_A\pqty{\del_t X^A_\theta-\del_\theta X^A_t+\half f^{A}_{BC}(X^B_t X^C_\theta-X^B_\theta X^C_t)}\label{lagincomp}
	\end{align}
We see that the momenta  conjugate to $X^A_\theta$ are $\pi_A$, and the momenta conjugate to $X^A_t$ vanish.  The $X^A_t$ variables therefore serve as Lagrange multipliers imposing the Gauss law constraints which are given by 
\begin{align}
		G_A\equiv\frac{\delta\mathcal{L}}{\delta	 X^A_t}=\del_\theta\pi_A+ f^B_{AC}\pi_B X^C_\theta=0\label{constraint}
	\end{align}

To quantise the theory we take the dynamical variables to be $\pi_A, X^A_\theta$ and impose the canonical commutation relations 
 on the pair $(\pi_A,X^A_\theta)$ 
	\begin{align}
		\{X^A_\theta(t,\theta_1),\pi_B(t,\theta_2)\}=\delta^A_B \delta(\theta_1-\theta_2)\label{canonposi}
	\end{align}
	In addition we impose the constraints $G_A\simeq0$ on the states. 
	
	The constraints eq.\eqref{constraint}  satisfy an  $SO(2,1)$ algebra 
	\begin{align}
		\{G_A(t,\theta_1),G_B(t,\theta_2)\}=f^C_{AB}\delta(\theta_1-\theta_2)G_C(t,\theta_1)\label{GAbrackets}
	\end{align}
	and in particular  close among themselves. The constraints are therefore    not Abelian, i.e. do not commute with each other. One strategy followed by Strobl\cite{Strobl:1993yn}, which is the one we discuss below,  is to carry out a canonical transformation   on the $\pi_A, X^A_\theta $ variables after which the constraints become Abelian.  This is done for the non-zero modes with respect to the spatial direction $\theta$. 
	
	One set of such variables  are 
	\begin{align}
		\pqty{\frac{G_+}{\pi_+},-\frac{G_\omega}{\pi_+},Q;\pi_\omega,\pi_+,P^+}\label{canoncord}
	\end{align}
	with 
	\begin{align}
		\del_\theta Q=\pi_+ G_-+\pi_- G_+\label{delzqincons}
	\end{align} and 
	\begin{align}
		\pi_\pm&=\pi_0\pm\pi_1,\quad X^\pm_\theta=\half(e^0_\theta\pm e^1_\theta)\nonumber\\
		P^+&=-\frac{X^-_\theta}{\pi^+},\quad Q=\pi_+\pi_-+\pi_\omega^2\label{PQpipmdef}
	\end{align}
	Straightforward algebra then shows that the only non-zero commutation relations at equal time  are
	\begin{align}
		\{Q(\theta_1),P^+(\theta_2)\}&=\delta(\theta_1-\theta_2),\nonumber\\
		\left\{\frac{G_+}{\pi_+}(\theta_1),\pi_\omega(\theta_2)\right\}&=-\delta(\theta_1-\theta_2)\nonumber\\
		\left\{-\frac{G_\omega}{\pi_+}(\theta_1),\pi_+(\theta_2)\right\}&=-\delta(\theta_1-\theta_2)\label{abconmute}
	\end{align}
	Having Abelianised the constraints we can now simply set the constraints to vanish,
	\begin{align}
	\label{conp}
	{G_+\over \pi+}={G_-\over \phi_+}= Q=0
	\end{align}
	and remove all the non-zero mode degrees of freedom. 
	

	All that remains then is the  sector involving the zero modes  of $\theta$. Thus we see clearly that the system has no local degrees of freedom, and that the mini-superspace approximation is in fact exact. 
	
	  For the zero mode sector we continue to work with the original variables, i.e. with the $\theta$ independent modes of 
	$\pi_A, X^A_\theta$, satisfying the commutation relations $\{X^A_\theta(\theta_1), \pi_B(\theta_2)\}=  \delta^A_B \delta(\theta_1-\theta_2)$
	 
	The constraints in this sector become, 
	\begin{align}
		G_+=2X^-_\theta \pi_\omega - X^3_\theta\pi_+=0,\nonumber\\
		G_-=2 X^+_\theta\pi_\omega - X^3_\theta\pi_-=0\nonumber\\
		G_\omega=X^+_\theta\pi_+ - X^-_\theta \pi_- =0\label{gacons}
	\end{align}
	Only two of these three constraints are independent. 
	
	To proceed, we now fix gauge in the zero mode sector by setting
	\begin{align}
		X^0_\theta=0\label{zerogauge}
	\end{align}
	It then follows from the $G_\omega$ constraint that
	\begin{align}
		\pi_+ +\pi_-=\pi_0=0\label{pizerogag}
	\end{align}
	From the other independent equation,  we get
	\begin{align}
		X^1_\theta \pi_\omega+ X^3_\theta\pi_1=0\label{nontrgau}
	\end{align}
In the quantum theory we then have the variables $X^1_\theta, X^3_\theta$ and their conjugate momenta $\pi_1, \pi_\omega$.
Let us remind the reader  that $\pi_\omega$ is actually  the dilaton, $\phi$, and $X^1_\theta=e^1_\theta$, which we denote below as $e^1$.
 The spatial component of the metric $g_{\theta\theta}=(e^{1})^2$. 
 And the length along the spatial direction is then given by 
 \begin{align}
 \label{length}
 \hat{l}^2=(2\pi e^1)^2
 \end{align}
 
In the quantum theory with $X^3_\theta=i \partial_\phi$, $\pi_1=-i \partial_{e^1}$ we then get from eq.(\ref{nontrgau})
	\begin{align}
		(\partial_\phi \partial_{e^1} + \phi e^1) \Psi=0\label{wfncondti}
	\end{align}
	We see that this  agrees with the equation we obtained above in the second order formalism, eq.\eqref{wdwinlp}, (noting that we have set $8 \pi G=1$ here) with one important caveat.
	$\hat{l}$ - the length of the spatial circle - must be positive but $e_1$ - the value of the vierbein - can be either positive or negative.

	 Indeed, for the case of transition amplitude from the past to future, the spacetime has the metric and the dilaton given by 
	\begin{align}
		ds^2=-dt^2+\sinh^2t\,d\theta^2,\quad \phi=A\cosh t\label{metindscomp}
	\end{align}
	Taking the vierbein  to be $e^1=\sinh t$,  we see that it continuously extrapolates from the far past to the future, being  negative  in the far past, as  $t\rightarrow -\infty$ and positive in the far future $t\rightarrow \infty$. Thus $e^1$ could serve as a good ``clock" for describing such transitions. 
See also the discussion in section \ref{transamp} in this context. 

In summary, we see that  the variables 	of the first order formalism could provide a more convenient description for the study of transition amplitudes. 
We have not completed this study, including  an analysis of how to solve the problem of time, define a good norm, and the relation to the quantisation in the second order formalism. We leave this   for the future.


	\section{Matter in AdS double trumpet}
	\label{amainadsdt}

	\subsection{General boundary conditions for scalar fields}
	\label{gbcforscalars}
	In this appendix we shall outline the calculations for the determinant of the scalar fields for the general boundary conditions mentioned in eq.\eqref{twistedbc}.
	
	We shall carefully evaluate the determinant of scalar laplacian, $\text{det}(-{\nabla}^2)$. We will consider massless scalar in the background of the  AdS double trumpet topology with the metric written in conformally flat coordinate system, eq.\eqref{edletrs}, as
	\begin{align}
		ds^2={dr_*^2+d\theta^2\over \sin^2{r_*}},\quad \quad r_*\in\left[0,{\pi}\right],\quad \theta\in [0,b]\label{dbconfmet}
	\end{align}
	We can compute the dependence on $b$ by noting that the metric above is conformally flat and so we can use the conformal anomaly to evaluate the contribution due to the conformal factor and then compute the contribution from the flat metric separately.
	The $b$ dependence coming from the conformal factor can be evaluated using the conformal anomaly since the theory of a massless scalar field is a conformal field theory. The relation between determinants of conformally related metrics ${g}_{ab}=e^{2\sigma} \bar{ g}_{ab}$ is given by 
	\begin{align}
		{\text{det}(-{\nabla}^2)\over \text{det}(-\bar{\nabla}^2)}=&\exp{-\frac{1}{6\pi}\left[\half\int d^2x \sqrt{\bar{g}}{(\bar{g}^{ab}\del_a\sigma\del_b\sigma+\bar{R}\sigma)}+\int_\del d\bar{s}\bar{K}\sigma\right]
		}\label{detndco}
	\end{align}
	where quantities denoted by bars are calculated with respect to the metric $\bar{ g}$. The computation for the conformal factor is explained in detail in \cite{Moitra:2021uiv} in  appendix I  and we shall just present the results here which is
	\begin{align}
		{\text{det}(-{\nabla}^2)\over \text{det}(-\bar{\nabla}^2)}=e^{\frac{b}{12}}\label{confactscdet}
	\end{align}
	Now, we shall compute the contribution from the flat part of the metric in eq.\eqref{dbconfmet}. 
	To implement these boundary conditions, we expand the scalar fields in a mode expansion as
	\begin{align}
		\varphi(r_*,\theta)=\sum_{m}e^{i(\tilde{m}+\tilde{\alpha})\theta}\varphi_{m}(r_*)\label{gbcmexp}
	\end{align}
	The eigenvalue equation for the scalar field then becomes
	\begin{align}
		\del_{r_*}^2\varphi_m-(\tilde{m}+\tilde{\alpha})^2\varphi_m=-\lambda \varphi_m\label{gbcevaleq}
	\end{align}
	where 
	\begin{align}
		\tilde{m}=\frac{2\pi m}{b},\quad \tilde{\alpha}=\frac{2\pi \alpha}{b}\label{mbabdef}
	\end{align}
	The general solution to the above equations is given by 
	\begin{align}
		\varphi_m=A_me^{i\sqrt{\lambda -(\tilde{m}+\tilde{\alpha})^2}r_*}+B_m e^{-i\sqrt{\lambda -(\tilde{m}+\tilde{\alpha})^2}r_*}\label{gbcssol}
	\end{align}
	Imposing the Dirichlet boundary conditions at the boundary $r_*=0$ then relates $A_m$ and $B_m$ as
	\begin{align}
		A_m=-B_m\label{gbcsondb}
	\end{align}
	following which the solution becomes
	\begin{align}
		\varphi_m=2i A_m \sin(r_*\sqrt{\lambda-(\tilde{m}+\tilde{\alpha})^2})\label{gbcscaldsol}
	\end{align}
	Further imposing Dirichlet boundary condition on this solution at $r_*=\pi$ then gives the eigenvalues to be
	\begin{align}
		\lambda_{m,n}=n^2+(\tilde{m}+\tilde{\alpha})^2,\quad n\geq 1, m\in \mathbb{Z}\label{gbcevals}
	\end{align}
	The determinant is then given by 
	\begin{align}
		\det(-\bar{\nabla}^2)=\prod_{n=1}^{\infty}\prod_{m=-\infty}^{\infty}\lambda_{m,n}\label{gbcdet}
	\end{align}
	We shall compute the product above by using the $\zeta$-function regularization. Following are some useful formulae that we shall use,
	\begin{align}
		\sum_{n=1}^{\infty}\log(n-n_0)&=-\ln{\Gamma(1-n_0)\over \sqrt{2\pi}}\nonumber\\
		\prod_{n=1}^{\infty}(n^2+\tilde{\alpha}^2)&=\frac{2}{\tilde{\alpha}}\sinh(\pi\tilde{\alpha})\nonumber\\
		\sum_{n=1}^{\infty}(n-\alpha)&=\frac{1}{24}-\frac{1}{8}(2\alpha-1)^2\label{polf}
	\end{align}
	First, let us regulate the product over $n$ in eq.\eqref{gbcdet}. Using the second formula in eq.\eqref{polf}, we get
	\begin{align}
		\det(-\bar{\nabla}^2)&=\prod_{m=-\infty}^{\infty}\frac{b}{\pi(m+\alpha)}\sinh (\frac{2\pi^2}{b}(m+\alpha))\nonumber\\
		&=\frac{b}{\pi\alpha}\sinh(\pi\tilde{\alpha})\prod_{m=1}^{\infty}\frac{b^2e^{\frac{2\pi^2}{b}((m+\alpha)+(m-\alpha))}}{4\pi^2(m+\alpha)(m-\alpha)}\pqty{1-e^{-\frac{4\pi^2}{b}(m+\alpha)}}\pqty{1-e^{-\frac{4\pi^2}{b}(m-\alpha)}}\nonumber\\
		&=\frac{\sinh(\pi\tilde{\alpha})\Gamma(1+\alpha)\Gamma(1-\alpha)q^{\frac{\alpha^2}{2}+\frac{1}{12}}}{\pi \alpha}\prod_{m=1}^{\infty}(1-zq^m)(1-z^{-1}q^m)\nonumber\\
		&=\frac{iq^{\frac{\alpha^2}{2}}\vartheta_{11}(v,\tau)}{2\sin(\pi\alpha)\eta(\tau)}\label{gbcscaldet}
	\end{align}
where $\vartheta_{11}(v,\tau)$ is the theta function with characteristics and $\eta(\tau)$ is the Dedekind Eta function.
	\begin{align}
			\vartheta_{11}(v,\tau)&=-2 \sin(\pi v)q^{1\over 8}\prod_{m=1}^{\infty}(1-q^m)(1-z q^m)(1-z^{-1}q^m)\nonumber\\
		\eta(\tau)&=q^{\frac{1}{24}}\prod_{m=1}^{\infty}(1-q^m)\nonumber\\
		q&=e^{2\pi i \tau},\quad z=e^{2\pi i v}\nonumber\\
		\tau&=\frac{2\pi i}{b},\quad v=\frac{2\pi i\alpha}{b}\label{zetaformulae}
	\end{align}
	which then implies that $q=e^{-\frac{4\pi^2}{b}},z=e^{-\frac{4\pi^2\alpha}{b}}$. 
	The full determinant including the contribution from the conformal factor is given by 
	\begin{align}
		\det(-{\nabla^2})=e^{\frac{b}{12}}\frac{i q^{\frac{\alpha^2}{2}}\vartheta_{11}(v,\tau)}{2\sin(\pi\alpha)\eta(\tau)}\label{fuldetindt}
	\end{align}
	We can now look at the various limiting cases. First consider the limit $b\rightarrow0$. It is easy to see that in this limit $q\rightarrow 0$ and so we have
	\begin{align}
		\eta(\tau)\rightarrow q^\frac{1}{24},\vartheta_{11}(v,\tau)\rightarrow -e^{{\pi\abs{\tilde{\alpha}}}}q^{\frac{1}{8}}
		\Rightarrow \det(-\bar{\nabla}^2)\rightarrow \frac{ie^{{2\pi^2\over b}E}}{2\sin(\pi\alpha)}\label{gbcbzerolim}
	\end{align} 
	where $E$ is given in eq.\eqref{cen}.
	So the matter contribution in the partition function can be made finite by choosing $\alpha$ that satisfies eq.\eqref{ral}.
	Now let us analyze the limit of $b\rightarrow \infty$. To estimate the value of the determinant in this limit, we have to use the modular transformation properties of the $\vartheta_{11}$ and $\eta$ functions which read
	\begin{align}
		\vartheta_{11}(v,\tau)&={i}{(-i\tau)^{-\frac{1}{2}}}e^{-\frac{i\pi v^2}{\tau}}\vartheta_{11}\left(\frac{v}{\tau},-\frac{1}{\tau}\right)\nonumber\\
		\eta(\tau)&=\frac{\eta(-\frac{1}{\tau})}{(-i\tau)^\frac{1}{2}}\label{gbcetamodtr}
	\end{align}
	Using these transformation properties, we find that the determinant is given by 
	\begin{align}
		\det(-\bar{\nabla}^2)=-\frac{e^{-\frac{i\pi v^2}{\tau}}}{2\sin(\pi\alpha)}\frac{q^{\frac{\alpha^2}{2}}\vartheta_{11}(\frac{v}{\tau},-\frac{1}{\tau})}{\eta(-\frac{1}{\tau})}\label{gbcdetmod}
	\end{align}
	In the limit of $b\rightarrow\infty$, we find
	\begin{align}
		\det(-\bar{\nabla}^2)\sim e^{-\frac{b}{12}}\Rightarrow\det(-{\nabla}^2)\sim\order(1)\label{gbcsdetbinf}
	\end{align}
	Now, we show an alternative way to derive the casimir energy in eq.\eqref{cen}. The alternate way is to canonically quantise the scalar matter field and compute the zero point energy directly. For the complex scalar field with the action
	\begin{align}
		S=\int d^2x\, \del_a\varphi^\dagger \del^a\varphi\label{comsc}
	\end{align}
	The equations of motion then read $\nabla^2\varphi=0$ the solution for which can be written in the mode-expanded form as
	\begin{align}
		\varphi=\frac{1}{\sqrt{2}}\sum_{n=-\infty}^{\infty} \pqty{\frac{a_n}{\sqrt{2\pi\omega_n}} e^{-i\omega_n (t+\theta)}+\frac{\tilde{a}_n}{\sqrt{2\pi\tilde{\omega}_n}} e^{-i\tilde{\omega}_n (t-\theta)}}\label{moexsctw}
	\end{align}
	The periodicity condition that
	\begin{align}
		\varphi(\theta+2\pi)=e^{2\pi i \alpha}\varphi(\theta)\label{twpcap}
	\end{align}
	determines the quantisation condition on $\omega,\tilde{\omega}$ as
	\begin{align}
		\omega_n=n-\alpha,\,\tilde{\omega}_n=n+\alpha\label{omtwom}
	\end{align}
	Without loss of generality, we take $\alpha>0$. 
	The canonical conjugate momenta are given by $\Pi_\varphi=\dot{\varphi}^\dagger, \Pi_{\varphi^\dagger}=\dot{\varphi}$. We then impose the canonical commutation relations $[\varphi(\theta),\Pi_\varphi(\theta')]=i\delta(\theta-\theta')$. The Hamiltonian is given by
	\begin{align}
		H=\int (\dot{\varphi}^\dagger \dot{\varphi}+\del_\theta\varphi^\dagger \del_\theta \varphi)\label{hatwex}
	\end{align}
	which after inserting the mode expansions becomes
	\begin{align}
		H=&\sum_{n=-\infty}^{\infty} \omega_n a^\dagger_n a_n +\sum_{n=-\infty}^{\infty} \tilde{\omega}_n \tilde{a}^\dagger_n \tilde{a}_n\label{hatwbc}
	\end{align}
	To compute the normal ordering constant, we note from the expansion eq.\eqref{moexsctw} that the annihilation operators for the right movers correspond to $a_n$ for $n\geq 1$ and for the left movers correspond to $\tilde{a}_n$ for $n\geq 0$. Thus, the zero point energy that is obtained by the normal ordering of the operator expansion above for the Hamiltonian gives
	\begin{align}
		E= \sum_{n=1}^{\infty}\omega_n+\sum_{n=0}^{\infty}\tilde{\omega}_n=\sum_{n=1}(n-\alpha)+\sum_{n=0}(n+\alpha)\label{zpe}
	\end{align}
	which gives the correct Casimir energy mentioned in eq.\eqref{cen} upon using eq.\eqref{polf}.

	\subsection{Bosonic fields in AdS double trumpet}
	\label{bosadsdtmat}
	In this appendix, we will elaborate on the details of the calculations related to the two-point kernels for the matter action in AdS double trumpet given in eq.\eqref{G1G2def}. We will show the steps leading to eq.\eqref{G1Weier}. First consider the function $G(u)$ given by 
	\begin{align}
		\frac{4\pi^2}{b}G(u)=\sum_{n=-\infty}^{\infty}\tilde{n}\coth(\tilde{n}\pi)e^{inu}\label{guincoth}
	\end{align}
	The derivation goes as follows.
	\begin{align}
		\frac{4\pi^2}{b}G(u)=&-\frac{2\pi i }{b}\del_u\pqty{\sum_{n=-\infty, n\neq 0}^{\infty}\coth(\tilde{n}\pi)e^{inu}}\nonumber\\
		=&-\frac{2\pi i }{b}\del_u\pqty{\sum_{n=1}^{\infty}\coth(\tilde{n}\pi)e^{inu}}+(u\rightarrow -u)\label{gderexp}
	\end{align}
	So we shall focus on the computation of the sum above for $n\ge 1$. We can futher split this sum as follows
	\begin{align}
		\sum_{n=1}^{\infty}\coth(\tilde{n}\pi)e^{inu}\equiv t_1+t_2=\sum_{n=1}^{\infty}e^{inu}+\sum_{n=1}^{\infty}\frac{2e^{inu-\tilde{n}\pi}}{e^{\tilde{n}\pi}-e^{-\tilde{n}\pi}}\label{nge1sumcoth}
	\end{align}
	where $t_1$ and $t_2$ denote the first and second sums after the last equality above. The sum denoted by $t_1$ is easier to do for which we get
	\begin{align}
		t_1=\frac{e^{iu}}{1-e^{iu}}\label{t1sum}
	\end{align}
	The sum in $t_2$ is slightly non-trivial which is done as follows
	\begin{align}
		t_2&=\sum_{n=1}^{\infty}\frac{2e^{inu-\tilde{n}\pi}}{e^{\tilde{n}\pi}-e^{-\tilde{n}\pi}}=2\sum_{n=1}^{\infty}\sum_{p=1}^{\infty}e^{n
			(iu-2\pi\tilde{p})}=2\sum_{p=1}^{\infty}\frac{1}{e^{(2\pi\tilde{p}-iu)}-1}\label{t2inpsum}
	\end{align}
	Combining these, we get
	\begin{align}
		\sum_{n=1}^{\infty}\coth(\tilde{n}\pi)e^{inu}&=\frac{e^{iu}}{1-e^{iu}}+2\sum_{p=1}^{\infty}\frac{1}{e^{(2\pi\tilde{p}-iu)}-1}\nonumber\\
		-\frac{2\pi i}{b}\del_u\pqty{\sum_{n=1}^{\infty}\coth(\tilde{n}\pi)e^{inu}}&=-\frac{\pi}{2b}\csc^2\left(\frac{u}{2}\right)-\frac{\pi}{b}\sum_{p=1}^{\infty}\csc^2\left(\frac{u}{2}+i\pi\tilde{p}\right)\label{firsing1}
	\end{align}
	From eq.\eqref{gderexp} we then get
	\begin{align}
		\frac{4\pi^2}{b}G(u)=-\frac{\pi}{b}\sum_{p=-\infty}^{\infty}\csc^2\left(\frac{u}{2}+i\pi\tilde{p}\right)\label{gcscsum}
	\end{align}
	It then immediately follows from p.434 of \cite{whittaker_watson_1996} that the function $G(u)$ is related to the Wieierstrass function as in eq.\eqref{G1Weier}
	The above series for $G(u)$ is well-suited to analyze the region $b\rightarrow 0$ but not for the region $b\rightarrow\infty$. A small manipulation can be done to rewrite this series in a different form as below. We will have to use the following infinite sum representations of the $\csc$ and $\csch$ functions which are given by 
	\begin{align}
		\csc^2(u)=-\sum_{m=-\infty}^{\infty}(u-m\pi)^{-2},\quad  	\csch^2(u)=\sum_{m=-\infty}^{\infty}(u-i m\pi)^{-2}\label{csccschsums}
	\end{align}
	Using the $\csc$ sum above, we have
	\begin{align}
		\frac{4\pi^2}{b}G(u)=&-\frac{\pi}{b}\sum_{p=-\infty}^{\infty}\csc^2\left(\frac{u}{2}+i\pi\tilde{p}\right)\nonumber\\
		=&-\frac{\pi}{b}\sum_{p=-\infty}^{\infty}\sum_{m=-\infty}^{\infty}\pqty{\frac{u}{2}+\frac{2\pi^2 i p}{b}-m\pi}^{-2}\nonumber\\
		=&-\frac{b}{4\pi}\sum_{p=-\infty}^{\infty}\sum_{m=-\infty}^{\infty}\pqty{\frac{bu}{4\pi}-\frac{m  b}{2}+	i\pi p}^{-2}\nonumber\\
		=&-\frac{b}{4\pi}\sum_{m=-\infty}^{\infty}\csch^2\left(\frac{bu}{4\pi}+\frac{m  b}{2}\right)\label{gsinhderiv}
	\end{align}
	where we used $\csch$ representation as an infinite sum in eq.\eqref{csccschsums} to obtain the final result. We shall now repeat the steps for the function $H(u)$ in eq.\eqref{G1Weier}
	\begin{align}
		\frac{4\pi^2}{b}H(u)=\sum_{n=-\infty,n\neq 0}^{\infty}\tilde{n}\csch(\tilde{n}\pi)e^{inu}\label{hufndef}
	\end{align}
	Following the same series of steps as in the case of $G(u)$ above to obtain eq.\eqref{gcscsum}, we get
	\begin{align}
		\frac{4\pi^2}{b}H(u)
		=&-\frac{2\pi i }{b}\del_u\pqty{\sum_{n=1}^{\infty}\csch(\tilde{n}\pi)e^{inu}}+(u\rightarrow -u)\nonumber\\
		=&-\frac{4\pi i }{b}\del_u \pqty{\sum_{n=1}^{\infty}\sum_{p=1}^{\infty}e^{n
				(iu-(2p-1)\frac{2\pi^2}{b})}}+(u\rightarrow -u)\nonumber\\
		=&-\frac{4\pi i }{b}\del_u \pqty{\sum_{p=1}^{\infty}\pqty{e^{-iu+(2p-1)\frac{2\pi^2}{b}}-1}^{-1}}+(u\rightarrow -u)\nonumber\\
		=&-\frac{\pi}{b}\sum_{p=-\infty}^{\infty}\csc^2\left(\frac{iu}{2}-(2p-1)\frac{\pi^2}{b}\right)\label{huincscderiv}
	\end{align}
	A similar manipulation as in eq.\eqref{gsinhderiv} will then give a $\csch$ function representation for the function $H(u)$ as
	\begin{align}
		\frac{4\pi^2}{b}H(u)=-\frac{b}{{{4}}\pi}\sum_{n=-\infty}^{\infty}\csch^2\left(\frac{bu}{4\pi}+\frac{i\pi}{2}-\frac{nb}{2}\right)\label{hucschdeff}
	\end{align}

	\subsection{OTOC calculations}
	\label{otocadsdt}
	In this appendix, we  calculate few out-of-time order correlation functions for the matter fields in AdS double trumpet geometry. For this appendix and the next one,  \ref{modstab}, we follow the convention used in section \ref{matcorrel}.
	We  introduce a time reparametrization as 
	\begin{align}
		\theta(u)=\frac{b}{2\pi}u+\epsilon(u)\label{timreptheta}
	\end{align}
	The reparametrization mode $\epsilon(u)$ at the boundaries is mode expanded as follows
	\begin{align}
		\epsilon(u)\big\vert_{r\rightarrow\pm \infty}=\epsilon^\pm(u)=\sum \epsilon^\pm_m e^{i{m}u}\label{epmodes}
	\end{align}
	We now expand various terms in the action in eq.\eqref{dssmredef} noting that the boundary values of the scalar field $\hat{\varphi}^\pm(\theta)$ remain fixed under these reparametrizations, i.e. $\hat{\varphi}^\pm(\theta)\rightarrow \hat{\varphi}^\pm(u)$. Thus, we get, the terms of $\order(\epsilon)$ as
	\begin{align}
		S_{M,b}=&S_{M,b}^{(0)}+S_{M,b}^{(1)}\nonumber\\
		S_{M,b}^{(1)}=&\frac{\pi}{b}\int_+\int_+du d\tilde{u}\,\hat{\varphi}^+(u)\hat{\varphi}^+(\tilde{u})((\epsilon^+{}'(u)+\epsilon^+{}'(\tilde{u}))G(u,\tilde{u})+\epsilon^+(u)\del_u G(u,\tilde{u})+\epsilon^+(\tilde{u})\del_{\tilde{u}} G(u,\tilde{u}))\nonumber\\
		+&\frac{\pi}{b}\int_-\int_-du d\tilde{u}\,\hat{\varphi}^-(u)\hat{\varphi}^-(\tilde{u})((\epsilon^-{}'(u)+\epsilon^-{}'(\tilde{u}))G(u,\tilde{u})+\epsilon^-(u)\del_u G(u,\tilde{u})+\epsilon^-(\tilde{u})\del_{\tilde{u}} G(u,\tilde{u}))\nonumber\\
		-&\frac{2\pi}{b}\int_+\int_-du d\tilde{u}\,\hat{\varphi}^+(u)\hat{\varphi}^-(\tilde{u})((\epsilon^+{}'(u)+\epsilon^-{}'(\tilde{u}))H(u,\tilde{u})+\epsilon^+(u)\del_u H(u,\tilde{u})+\epsilon^-(\tilde{u})\del_{\tilde{u}} H(u,\tilde{u}))\label{dstiminpos}
	\end{align}
	where $S_{M,b}^{(0)}$ is just given by the $\order{(\epsilon^0 )}$ term, eq.\eqref{dsdt02mact}.
	%
	We now calculate various 4-pt functions. For this we consider two different species of matter fields $V,W$. 
	Also, to study the 4-pt functions of matter correlators, we need the propagator for the time reparametrization mode $\epsilon(u)$. This can be derived from the action for these modes which is given by 
	\begin{align}
		S_{\epsilon}=-\frac{1}{8\pi G Jl_+}\int_{\del_{+}}du\,\, \text{Sch}\left[\tanh(\frac{\theta(u)}{2}),u\right]-\frac{1}{8\pi G Jl_- }\int_{\del_{-}}du\,\, \text{Sch}\left[\tanh(\frac{\theta(u)}{2}),u\right]\label{timrepdbact}
	\end{align}
	where $l_+,l_- , J$ are the quantities that characterizes the dilaton on the boundary and the length of the boundary in the asymptotic AdS limit as
	\begin{align}
		\phi_{\pm}\sim \frac{1}{J\epsilon},\quad \hat{l}_{\pm}\sim \frac{l_\pm}{\epsilon}\label{asymadslimit}
	\end{align}
	and $\text{Sch}(u)$ is the Schwarzian function given by 
	\begin{align}
		\text{Sch}\left[f(u),u\right]=\frac{f'''(u)}{f'(u)}-\frac{3}{2}\pqty{\frac{f''(u)}{f'(u)}}^2\label{schfundef}
	\end{align}
	Expanding the action in eq.\eqref{timrepdbact} to quadratic order using eq.\eqref{timreptheta},\eqref{epmodes}, we get
	\begin{align}
		S_{\epsilon}=\frac{\pi}{4 G J}\pqty{\frac{2\pi}{b}}^2\left[\sum_{m\neq 0}m^2\pqty{m^2+\pqty{\frac{b}{2\pi}}^2}\pqty{{\epsilon^+_m\epsilon^+_{-m}\over l_+}+{\epsilon^-_m\epsilon^-_{-m}\over l_- }}\right]\label{quadtimdbac}
	\end{align}
	The propagator for the $\epsilon$ modes is then given by 
	\begin{align}
		\langle \epsilon^\pm_m\epsilon^\pm_{-m}\rangle =\frac{4GJl_\pm}{\pi}\left(\frac{b}{2\pi}\right)^2\frac{1}{m^2\left(m^2+(\frac{b}{2\pi})^2\right)}\label{momtimprop}
	\end{align}
	In terms of position space the propagator becomes
	\begin{align}
		\langle\epsilon^\pm(u)\epsilon^\pm(0) \rangle =\frac{4GJl_{\pm}}{\pi}\left[\frac{(\abs{u}-\pi)^2}{2}-\frac{2\pi^2}{b}\csch(\frac{b}{2})\cosh(\frac{b}{2\pi}(\abs{u}-\pi))+\frac{4\pi^2}{b^2}-\frac{\pi^2}{6}\right]\label{postimprop}
	\end{align}
	One crucial thing to note here in the above propagator for the time reparametrization mode is the presence of hyperbolic $\cosh$ function unlike the case of disk where we would get the trignometric $\sin$ function. The reason for this can be traced to the momentum space two point function for $\epsilon_m$ in eq.\eqref{momtimprop} which is different from the case of disk which has
	\begin{align}
		&\langle \epsilon_m\epsilon_{-m}\rangle_{\text{disk}} \sim \frac{1}{m^2(m^2-1)}\nonumber\\
		\Rightarrow &\langle \epsilon(u)\epsilon(0)\rangle_{\text{disk}}=\left[-\frac{(\abs{u}-\pi)^2}{2}+(\abs{u}-\pi)\sin(\abs{u})\right]\label{diskmomtim}
	\end{align}
	Moreover to construct the position space propagator one has to exclude the $m=0,\pm 1$ modes which are the SL(2,R) zero modes on the disk whereas in the double trumpet we have to only exclude the $m=0$ mode corresponding to the U(1) isometry. 
	This difference results in the absence of an exponential growth of the 4pt OTOCs as we show below. 
	
	For two different matter fields $V,W$, there are various combinations of four point function with two $V$'s and two $W$'s  on either of the boundaries. We shall focus particularly on two types of correlators, one which has all the fields on the same boundary and another in which one field is on either boundary for each of the species,
	\begin{align}
		\frac{\langle  V^+(u_1)W^+(u_3)V^+(u_2) W^+(u_4) \rangle}{\langle W^+(u_3)W^+(u_4) \rangle \langle V^+(u_1)V^+(u_2)\rangle},\quad 
		\frac{\langle V^+(u_1)W^+(u_3)V^-(u_2) W^-(u_4) \rangle}{\langle W^+(u_3)W^-(u_4) \rangle \langle V^+(u_1)V^-(u_2)\rangle}\label{otocpos}
	\end{align}
	We shall compute these correlators with the ordering of the times as
\begin{align}
	u_4<u_2<u_3<u_1\label{ordering}
\end{align}
	To compactify the notation a bit, we shall use shorthand notation as mentioned below
	\begin{align}
		&u_{ij}=u_i - u_j, \,\del_i=\del_{u_i},\,\epsilon_i^\pm=\epsilon^\pm(u_i),\epsilon_{ij}^\pm=\langle\epsilon^\pm(u_i) \epsilon^\pm(u_j)\rangle ,\nonumber\\
		&G_{ij}=G(u_{ij}),\quad \tilde{G}_{ij}=\frac{\del_i G(u_{ij})}{G(u_{ij})}\nonumber\\
		&H_{ij}=H(u_{ij}),\quad \tilde{H}_{ij}=\frac{\del_i H(u_{ij})}{H(u_{ij})}\label{varsym}
	\end{align}
	Using the vertex functions in eq.\eqref{dstiminpos}, we have for the first of the correlators above
	\begin{align}
		&\frac{\langle  V^+(u_1)W^+(u_3)V^+(u_2) W^+(u_4) \rangle}{\langle W^+(u_3)W^+(u_4) \rangle \langle V^+(u_1)V^+(u_2)\rangle}\nonumber\\
		&=\left\langle \pqty{\del_1\epsilon_1^+ + \frac{\epsilon_1^+\del_1 G_{12}}{G_{12}} +\del_2\epsilon_2^+ + \frac{\epsilon_2^+\del_2 G_{12}}{G_{12}}}\pqty{\del_3\epsilon_3^+ + \frac{\epsilon_3^+\del_3 G_{34}}{G_{34}} +\del_4\epsilon_4^+ + \frac{\epsilon_4^+\del_4 G_{34}}{G_{34}}}\right \rangle \nonumber\\
		&=-\del_1^2\epsilon_{13}^+ -\del_1^2\epsilon_{14}^+ -\del_2^2\epsilon_{23}^+
		-\del_2^2\epsilon_{24}^+ +\tilde{G}_{12}\tilde{G}_{34}(\epsilon_{13}^+ -\epsilon_{14}^+- \epsilon_{23}^+ + \epsilon_{24}^+)\nonumber\\
		&\,\,\,\,\,+(\tilde{G}_{12}-\tilde{G}_{34})(\del_{2}\epsilon_{24}^+ - \del_{1}\epsilon_{13}^+)+(\tilde{G}_{12}+\tilde{G}_{34})(\del_{2}\epsilon_{23}^+ - \del_{1}\epsilon_{14}^+)\label{otoc1bdy}
	\end{align}
	The exact analysis for arbitrary value of $b$ is difficult as the functions $G(u),H(u)$ are complicated functions of $b$. We first consider the limiting case of $b\rightarrow 0$. In this limit, the function $G(u)$ becomes
	\begin{align}
		G(u)\simeq -\frac{1}{4\pi}\csc^2\pqty{\frac{u}{2}}\label{gucsc}
	\end{align}
	To diagnose the chaos behaviour, we consider the analytic continuation to Minkowski time and take the coordinates $u_i$ as
	\begin{align}
		u_1=\pi,u_2=0, u_3=\pi+i t, u_4= i t\label{minkucordss}
	\end{align}	
	Further, we take $l_{\pm}=2\pi$ and $t$ to be large so that $tb\gg 1$ . In this limit, we get
	\begin{align}
		\frac{\langle V^+(u_1)W^+(u_3)V^+(u_2) W^+(u_4) \rangle}{\langle W^+(u_3)W^+(u_4) \rangle \langle V^+(u_1)V^+(u_2)\rangle}={\alpha\over b^2}\sin^2\left(\frac{tb}{4\pi}\right)+\order{\left(\frac{1}{b}\right)}\label{4pt1bzero}
	\end{align}
	where $\alpha$  is a constant, the exact value of which is not important for the following discussion. 
	Thus, we see that the $t$ dependence is not an exponential but a phase factor. In the case of disk the $\sin$ function appearing above is replaced by $\sinh$ and so gives rise to an exponential growth. As mentioned earlier, although the function $G$ matches with the disk case, the propagator for time reparametrization differs from the case of disk in that it has a hyperbolic function, see eq.\eqref{postimprop}, whereas the disk propagator has a trignometric function, see eq.\eqref{diskmomtim}. More generally, this conclusion about the 4pt correlation function holds true for any value of $b$ and we will not get a exponential growth in the double trumpet. As another special case, we consider the case of $b=2\pi$. Using the exact form of the function $G$ in eq.\eqref{G1Weier}, we find that the OTOC in eq.\eqref{otoc1bdy} is of the form
	\begin{align}
		\frac{\langle  V^+(u_1)W^+(u_3)V^+(u_2) W^+(u_4) \rangle}{\langle W^+(u_3)W^+(u_4) \rangle \langle V^+(u_1)V^+(u_2)\rangle}=\alpha_0+\alpha_1 e^{it}+\alpha_2 e^{-it}\label{otoc1bdyval}
	\end{align}
	where $\alpha_0,\alpha_1,\alpha_2\sim \order{(1)}$.

	Now for the second correlator in eq.\eqref{otocpos} with operators on two boundaries, again using the vertex functions in eq.\eqref{dstiminpos}, we get
	\begin{align}
		&\frac{\langle V^+(u_1)W^+(u_3)V^-(u_2) W^-(u_4) \rangle}{\langle W^+(u_3)W^-(u_4) \rangle \langle V^+(u_1)V^-(u_2)\rangle}\nonumber\\
		&=\left\langle \pqty{\del_1\epsilon_1^+ + \frac{\epsilon_1^+\del_1 H_{12}}{H_{12}} }\pqty{\del_3\epsilon_3^+ + \frac{\epsilon_3^+\del_3 H_{34}}{H_{34}} }\right \rangle +\left\langle \pqty{\del_2\epsilon_2^+ + \frac{\epsilon_2^+\del_2 H_{12}}{H_{12}} }\pqty{\del_4\epsilon_4^+ + \frac{\epsilon_4^+\del_4 H_{34}}{H_{34}} }\right \rangle \nonumber\\
		&=-(\del_1^2\epsilon_{13}^+ 
		+\del_2^2\epsilon_{24}^+ +\tilde{H}_{12}\tilde{H}_{34}(\epsilon_{13}^+  + \epsilon_{24}^+)+(\tilde{H}_{12}+\tilde{H}_{34})(\del_{1}\epsilon_{13}^+ - \del_{2}\epsilon_{24}^-))\label{otoc2bdy}
	\end{align}
	Again letting the coordinates $u_i$ to take the values in eq.\eqref{minkucordss}, with $l_{\pm}=2\pi,b=2\pi$, we get
	\begin{align}
		&\frac{\langle V^+(u_1)W^+(u_3)V^-(u_2) W^-(u_4) \rangle}{\langle W^+(u_3)W^-(u_4) \rangle \langle V^+(u_1)V^-(u_2)\rangle}=\tilde{\alpha}_0+\tilde{\alpha}_1 t+\tilde{\alpha}_2 e^{-it}\label{otoc2bdyval}
	\end{align}
	where again $\tilde{\alpha}_0,\tilde{\alpha}_1,\tilde{\alpha}_2\sim \order{(1)}$. So, we again see that we do not get an exponential behaviour rather a phase factor, which is again tied to the presence of a hypergeometric function in eq.\eqref{postimprop}.
	
	Thus, we conclude that we do not have a chaotic growth of the OTOC's on a double trumpet geometry with two boundaries. Indeed the argument can be extended to an arbitrary higher genus surfaces with more than one boundary. As shown by working in the first order formalism in \cite{Saad:2019lba}  the partition function of the surfaces with higher genus or boundaries can be written as an integral over moduli with the integrand containing a factor of volume of the underlying bordered Reimann surface and products of Single trumpet partition function for each of the asymptotic boundaries, see eq.127 of \cite{Saad:2019lba}.
	Each of such single trumpet  partition function will have a time reparametrization mode whose propagator is of the form in eq.\eqref{postimprop} which when used in the computation of OTOC leads to a phase factor as in eq.\eqref{otoc2bdyval},\eqref{otoc1bdyval}. Finally, matter 4-pt functions in de Sitter are discussed in \cite{Maldacena:2019cbz,Cotler:2019nbi}.

	\subsection{Moduli stabilization and Saddle points}
	\label{modstab}
	
	In this subsection we shall show how to stabilize the double trumpet by inserting operators at the boundaries. Suppose we have $N$ species of scalar matter fields. Each of these $N$ species of matter has $2$-pt functions as given in eq.\eqref{G1Weier}. Now  consider $2n$ operators with $n$ of them inserted at each of the boundaries. We would like to analyze the $2n$-pt function of these operators. For the moment let us only examine the cross-boundary correlations. Even in this case, one has to do a bit of combinatorics to get the exact $2n$-pt correlator. Let us look at any one particular configuration. The $2n$-pt function in the double trumpet geometry will be given by an expression analogous to eq.\eqref{excro} for the 2-pt function and reads
	\begin{align}
		\langle O_1(u_1)\dots O_{2n}(u_{2n})\rangle =\int bdb\,\,e^{-\frac{b^2}{G_{\text{eff}}}}(\sqrt{\det(-\nabla^2)})^{-N} H(\Delta u)^n\label{crosgen}
	\end{align}
	where $G_{\text{eff}}=G_e l_{\text{eff}}$ with $G_e, l_{\text{eff}}$  as given in eq.\eqref{defgeff}, $\Delta u$ is the difference between the locations that are contracted, one on either boundary, which is taken to be the same for every pair of contraction.  The scalar determinant is given by
	\begin{align}
		\sqrt{\det(-\nabla^2)}=e^{b\over {24}}\eta\left(\frac{ib}{2\pi}\right)\label{detscal}
	\end{align}
	and so the correlator above can be written as
	\begin{align}
		\langle O_1(u_1)\dots O_{2n}(u_{2n})\rangle =\int_0^{\infty} db\,\exp(\ln b-\frac{b^2}{G_{\text{eff}}}-\frac{Nb}{24}-N\ln\eta +n \ln H)\label{corfulexp}
	\end{align}
	Let us denote the  full exponent above by $\text{Fexp}$ and is given by 
	\begin{align}
		\text{Fexp}=\ln b-\frac{b^2}{G_{\text{eff}}}-\frac{Nb}{24}-N\ln\eta +n \ln H\label{fexp}
	\end{align}
	The asymptotic forms of the Dedekind Eta function read
	\begin{align}
		\eta\left(\frac{ib}{2\pi}\right)=\begin{cases}
			\sqrt{\frac{2\pi}{b}}e^{-\frac{\pi^2}{6b}}\quad & b\rightarrow 0\\
			e^{-\frac{b}{24}} \quad & b\rightarrow \infty
		\end{cases}
		\label{etaasym}
	\end{align}
	For the case of $n=0$, we see that the integrand diverges near $b\rightarrow0$ due to  the $\eta$ function. The presence of cross-boundary contractions can control that divergence as we will see. The asymptotic behaviour of the $H(\Delta u)$ is given by 
	\begin{align}
		H(\Delta u)=\begin{cases}
			k_1 e^{-\frac{\pi^2}{b}}\quad &b\rightarrow 0\\
			-k_2 b^2 +\dots,\quad &b\rightarrow \infty, \Delta u=0,\\
			-k_2 b^2\exp(-k_3 b)+\dots,\quad &b\rightarrow \infty, \Delta u\neq 0
		\end{cases}
		\label{hfasym}
	\end{align}
	where $k_3,k_2>0,k_1$ are constants of $\order{(1)}$ that can be obtained from the full expression for $H$ in eq.\eqref{huincscderiv},\eqref{hucschdeff}.  We see from the $b\rightarrow 0$ behaviour above, that the full exponent has the behaviour in this limit given by 
	\begin{align}
		\text{Fexp}=-\pqty{n-\frac{N}{6}}\frac{\pi^2}{b}
		\label{fexpb0}
	\end{align}
	which renders the integrand finite for 
	\begin{align}
		n>\frac{N}{6}\equiv n_{th}\label{stabcond}
	\end{align}
	Having stabilized the $b$ integrand, one can look for possible saddle points. Just looking  at the asymptotic values of $\text{Fexp}$, we have
	\begin{align}
		\text{Fexp}=\begin{cases}
			-\pqty{n-\frac{N}{6}}\frac{\pi^2}{b}&\quad b\rightarrow 0\\
			-\frac{b^2}{G_{\text{eff}}}&\quad b\rightarrow\infty
		\end{cases}
		\label{fasym}
	\end{align}
So, the function $\text{Fexp}\rightarrow-\infty$ both near $b\rightarrow 0, \infty$ as long as eq.\eqref{stabcond}. 
	Thus there should be atleast one extremum for this function. Although it will be hard to exactly evaluate the saddle points, we shall be able to evaluate the saddle points analytically in the limits of $b\rightarrow 0, \infty$ where one expects various expressions to get simplified. 
	
	Let us first analyze the saddle around $b\sim 0$. Using the corresponding asymptotic forms for Dedekind Eta function in eq.\eqref{etaasym} and correlation function $H$ in eq.\eqref{hfasym}, the full exponent in eq.\eqref{fexp} in this limit becomes
	\begin{align}
		\text{Fexp}\simeq \left(1+\frac{N}{2}\right)\ln b-\frac{b^2}{G_{\text{eff}}}-\frac{Nb}{24}-\pqty{n-\frac{N}{6}}\frac{\pi^2}{b}\label{finzero}
	\end{align}
	Extremizing it we get
	\begin{align}
		\frac{\del \text{Fexp}}{\del b}=\left(1+\frac{N}{2}\right)\frac{1}{b}-\frac{2b}{G_{\text{eff}}}-\frac{N}{24}+\pqty{n-\frac{N}{6}}\frac{\pi^2}{b^2}\label{delfb}
	\end{align}
	Now considering the limit of $b^2\gg G_{\text{eff}}N$, we have that 
	\begin{align}
		\frac{\del \text{Fexp}}{\del b}=0\implies  &\pqty{n-\frac{N}{6}}\frac{\pi^2}{b^2}-\frac{2b}{G_{\text{eff}}}=0\nonumber\\
		b&=\pqty{G_{\text{eff}}\pqty{n-\frac{N}{6}}\frac{\pi^2}{2}}^{1\over 3}\label{b0sadd}
	\end{align}
	For the special case of $n=\frac{N}{6}$, we can still have a saddle point provided $G_{\text{eff}}N$ is sufficiently small so that the first two terms in eq.\eqref{delfb} become important for which we get the saddle point as
	\begin{align}
		b=\sqrt{\frac{G_{\text{eff}}}{2}\pqty{1+\frac{N}{2}}}\label{spb0sad}
	\end{align}
	Both these extrema are stable saddle points which is easy to either using by evaluating the second derivative of $\text{Fexp}$ or just by looking at the qualitative graph of the function. 
	
	Now we look for saddles in the region $b\gg 1$. In this region the function $\text{Fexp}$ has the following form after using eq.\eqref{etaasym},\eqref{hfasym}
	\begin{align}
		\text{Fexp}=(2n+1)\ln b-\frac{b^2}{G_{\text{eff}}}-n k_3 b\label{Fbinf}
	\end{align}
	Extremizing it gives
	\begin{align}
		\frac{\del \text{Fexp}}{\del b}=\frac{2n+1}{b}-\frac{2b}{G_{\text{eff}}}-n k_3=0,\label{dFbinf}
	\end{align}
	Since $k_3>0$ for generic $\Delta u\neq 0$ and $b\gg 1$, it is not possible to obtain an extrema in this region.
	 However, for the special case of $k_3=0$ as in the case of $\Delta u=0$, we do find an extrema, whose location is given by 
	\begin{align}
		b\simeq \sqrt{\pqty{n+\half}G_{\text{eff}}}\label{binfsad}
	\end{align}
	It is easy to see that this too corresponds to a stable saddle point. 
	
	One should analyze the full form of Dedekind Eta function and $H$ function to understand the saddle points for other generic values of $b$, which is not analytically tractable, as far as we can say. However, one interesting observation one can make is for the case when we have saddle points both in $b\rightarrow 0$ and $b\gg 1$ regions, which as mentioned earlier correspond to stable  saddles. Thus, there must be atleast one more saddle point in between these two saddle, which is unstable. 
	
	We shall now repeat the analysis for the case of same boundary insertions. The strategy remains the same except that now all the operators are inserted on the same boundary and so the analog  of eq.\eqref{crosgen} now reads
	\begin{align}
		\langle O_1(u_1)\dots O_{2n}(u_{2n})\rangle =\int bdb\, e^{-\frac{b^2}{G_{\text{eff}}}}(\sqrt{\det(-\nabla^2)})^{-N}G(\Delta u)^n\label{samegen}
	\end{align}
	where the function $G(u)$ is given in eq.\eqref{gsinhderiv}
	The full exponent in this case is  given by 
	\begin{align}
		\text{Fexp}=\ln b-\frac{b^2}{G_{\text{eff}}}-\frac{Nb}{24}-N\ln\eta +n \ln G\label{fexpg}
	\end{align}
	
	From eq.\eqref{gsinhderiv}, we find 
	\begin{align}
		G(\Delta u)=\begin{cases}
			-\frac{1}{4\pi}\csc(\Delta u)^2 +\order{( e^{-\frac{1}{b}})}\quad &b\rightarrow 0\\
			-\tilde{k}_2 b^2\exp(-\tilde{k}_3 b)+\dots,\quad &b\rightarrow \infty, \Delta u\neq 0
		\end{cases}
		\label{gfasym}
	\end{align}
	where $\tilde{k}_2,\tilde{k}_3$ are again constants of order unity. We see from the above  that the $b\rightarrow 0$ behaviour of the function G is a constant which cannot counteract the divergence due to the scalar determinant. The expression eq.\eqref{fexpg} then becomes
	\begin{align}
		\text{Fexp}\simeq \left(1+\frac{N}{2}\right)\ln b-\frac{b^2}{G_{\text{eff}}}-\frac{Nb}{24}+\frac{N}{6}\frac{\pi^2}{b}-n\ln (4\pi\sin^2(\Delta u))\label{fgb0}
	\end{align}
	Extremizing this gives
	\begin{align}
		-\frac{N\pi^2}{6b^2}+\pqty{1+\frac{N}{2}}\frac{1}{b}-\frac{N}{24}-\frac{2b}{G_{\text{eff}}}=0\label{fgb0ex}
	\end{align}
	for which no extrema are possible. 
	Let us now analyze the region $b\gg1$. Comparing eq.\eqref{gfasym} and eq.\eqref{hfasym}, we see that both $G(\Delta u)$ and $H(\Delta u)$ have similar behaviour in this region for $\Delta u\neq 0$ and so we can rely on our earlier analysis to conclude that there will be no extrema in this region for the same boundary correlators. 	Although it is hard to prove that there are no extrema in the intermediate region, the numerical plots that we had obtained have shown no  presence of an extrema. 
	

	\section{dS double trumpet }
	\label{fdsdt}
	\subsection{Schwarzian action calculation}
	\label{scactcal}
	In this section, we shall derive various results related to the de Sitter double trumpet. First let us extend the calculations related to the $b$ dependence in eq.\eqref{jhbh}. The boundary  term in -AdS$_2$ is given by 
	\begin{align}
		S_{-AdS,i}=-\frac{\phi_{B_i}}{8\pi G}\int_{\del_i} ds (K-1)\label{adsbtterm}
	\end{align}
	The metric for -AdS$_2$ double trumpet is taken to be
	\begin{align}
		ds^2=-\pqty{\frac{dr^2}{r^2+1}+(r^2+1)d\theta^2},\quad \theta\sim \theta+b\label{adsdbmet}
	\end{align}
	The outward pointing normal vector at the boundaries located at $r\gg 1$, normalized as $n^\mu n_\mu=-1$,  are given by 
	\begin{align}
		n^r_+=\sqrt{r^2+1},\quad n^r_-=-\sqrt{r^2+1}\label{nroutward}
	\end{align}
	The extrinsic curvature at the boundaries is then given by 
	\begin{align}
		K_\pm=\nabla_\mu n^\mu_\pm=\del_r n^r_\pm=\pm \frac{r}{\sqrt{r^2+1}}\label{exkadsdt}
	\end{align}
	The line element on the boundary is given by 
	\begin{align}
		ds=\sqrt{\abs{g_{\theta\theta}}}d\theta=\sqrt{1+r^2}\,d\theta\label{bdliele}
	\end{align}
	The on-shell action at the right boundary $r\rightarrow r_+>0$  is given by 
\begin{align}
	S_{-AdS,+}=&-\frac{\phi_{B_+}}{8\pi G}\int_{\del_{+}} d\theta \,\sqrt{\gamma}(K-1)\nonumber\\
			=&\frac{\phi_{B_+}b}{16\pi Gr_+}\nonumber\\
		=&\frac{b^2}{16\pi GJ l_+}\label{sosadsdt}
	\end{align}
where we used that $r_+=\frac{2\pi }{b\epsilon_+}$.
Analytic continuation gives
\begin{align}
	S_{dS,+}=\mp\frac{ib \,\phi_{B_+}}{16\pi Gr_+}=\mp\frac{ib^2}{16\pi GJ l_+} ,\quad r_+\rightarrow \pm i r_+  \label{sosdsdtn}
\end{align}
Similarly, at the other boundary $r\rightarrow r_-<0$, we get
\begin{align}
	S_{-AdS,-}=&-\frac{\phi_{B_-}}{8\pi G}\int_{\del_{-}} d\theta \,\sqrt{\gamma}(K-1)\nonumber\\
	=&-\frac{\phi_{B_-}b}{16\pi Gr_-}\nonumber\\
	=&\frac{b^2}{16\pi GJ l_-}\label{sopsadsdt}
\end{align}
where we have used the fact that at the boundaries $r_-=-\frac{2\pi}{b\epsilon_- }$. Analytic continuation gives
\begin{align}
		S_{dS,-}=\pm\frac{ib \,\phi_{B_-}}{16\pi Gr_-}=\mp\frac{ib^2}{16\pi GJ l_-} ,\quad r_-\rightarrow \pm i r_-  \label{sosdsdtn}
\end{align}

%
%

	\subsection{Matter in dS double trumpet}
	\label{mdsdt}
	In this subsection we shall show some steps detailing the calculation of on-shell action for matter fields in  dS spacetime. 
	We shall follow the same template of first working in -AdS spacetime and then analytically continuting to the dS spacetime.  To begin with, consider the double trumpet geometry in (0,2) signature AdS spacetime whose metric is given by eq.\eqref{adsdbmet}
	The action for a massless scalar field in the (0,2) AdS DT is given by 
	\begin{align}
		S_{M}=-\frac{1}{2}\int d^2x \sqrt{g} (\nabla\phi)^2\label{nadsmat}
	\end{align}
	Note the presence of an additional minus sign in the action above compared to the conventional action in the (2,0) signature metrics. This minus sign is required to render the path integral well defined in the (0,2) signature. The on-shell action after imposing the equations of motion, noting the outward normals given in eq.\eqref{nroutward}, to leading order in $\abs{r}\gg 1$ at the asymptotic boundaries  becomes
	\begin{align}
		S_{M}^{OS}=\frac{1}{2}\int_{ \del_+ } r^2 d\theta \varphi(r)\del_r \varphi(r)-\frac{1}{2}\int_{ \del_-  } r^2 d\theta \varphi(r)\del_r\varphi(r)\label{nadsmtact}
	\end{align}
	Now doing the analytic continuation in 
	\begin{align}
		r_-\rightarrow  i r_-  , \, r_+  \rightarrow -i r_+  \label{pfunianacon}
	\end{align}
	gives the matter on-shell action as
	\begin{align}
		S_{M}^{OS}=-\frac{i}{2}\int_{ \del_+ } r^2 d\theta \,\varphi(r)\del_r \varphi(r)-\frac{i}{2}\int_{ \del_-  } r^2 d\theta \,\varphi(r)\del_r\varphi(r)\label{2unidt}
	\end{align}
	For the other analytic continuation corresponding to
	\begin{align}
		r_-\rightarrow - i r_-  , \, r_+  \rightarrow -i r_+  \label{2unianacon}
	\end{align}
	we get the matter action to be
	\begin{align}
		S_{M}^{OS}=-\frac{i}{2}\int_{ \del_+ } r^2 d\theta \varphi(r)\del_r \varphi(r)+\frac{i}{2}\int_{ \del_-  } r^2 d\theta \varphi(r)\del_r\varphi(r)\label{p2fmtact}
	\end{align}
	
	We shall now use these results to compute the  dependence of the wavefunction on the boundary  configuration of the matter field. For now, the matter fields are taken to have periodic boundary conditions along the $\theta$ direction,i.e.
	\begin{align}
		\varphi(r,\theta+b)=\varphi(r,\theta)\label{pbcscal}
	\end{align}
	First consider the solution of the equation of motion for a matter field in the AdS Double trumpet geometry. The most general solution is given by 
	\begin{align}
		\varphi(r,\theta)=\sum_{k}e^{i\tilde{k}\theta}\varphi_k(r),\qquad \varphi_k(r)=c_1\pqty{\frac{r+i}{r-i}}^{\frac{i\tilde{k}}{2}}+c_2\pqty{\frac{r+i}{r-i}}^{-\frac{i\tilde{k}}{2}}\label{nadsmtsol}
	\end{align}
This expansion is only valid for $k\neq 0$. For $k =0$ mode, the most general solution is given by
\begin{align}
	\varphi(r,\theta)=\varphi_0(r)=a_1+a_2 \tan^{-1}(r)\label{phi0}
\end{align}
We shall first calculation the on-shell action for modes $k\neq 0$. 
	Expanding this solution near $r\rightarrow\pm\infty$ following the phase conventions
	\begin{align}
		\ln z=\ln \abs{z}+i \text{Arg}(z),\quad \text{Arg}(z)\in[-\pi,\pi]\label{phcon}
	\end{align}	
	we find
	\begin{align}
		\varphi_k&=c_1+c_2+\frac{\tilde{k}}{r}(c_2-c_1),\qquad &r\rightarrow\infty\nonumber\\
		&=c_1 e^{-\tilde{k}\pi}+c_2e^{\tilde{k}\pi}+\frac{\tilde{k}}{r}(c_2e^{\tilde{k}\pi}-c_1e^{-\tilde{k}\pi}),\qquad &r\rightarrow -\infty\label{nadsrinf}
	\end{align}
	In the de Sitter region, we would like to impose the boundary conditions
	\begin{align}
		\lim_{r\rightarrow r_+  }\varphi_k={\varphi}^+_k\nonumber\\
		\lim_{r\rightarrow r_-  }\varphi_k={\varphi}^-_k\label{dsdtmbd}
	\end{align}
	Let us first consider the case of past to future transition. Doing the analytic continuations as in eq.\eqref{pfunianacon}, we get
	\begin{align}
		\varphi_k&=c_1+c_2+\frac{i\tilde{k}}{r}(c_2-c_1),\qquad &r\rightarrow\infty\nonumber\\
		&=c_1 e^{-\tilde{k}\pi}+c_2e^{\tilde{k}\pi}-\frac{i\tilde{k}}{r}(c_2e^{\tilde{k}\pi}-c_1e^{-\tilde{k}\pi}),\qquad &r\rightarrow -\infty\label{matdsdtexp}
	\end{align}
	and hence the boundary conditions eq.\eqref{dsdtmbd} read
	\begin{align}
		c_1+c_2+\frac{i\tilde{k}}{r_+}(c_2-c_1)&={\varphi}^+_k\nonumber\\
		c_1 e^{-\tilde{k}\pi}+c_2e^{\tilde{k}\pi}-\frac{i\tilde{k}}{r_-}(c_2e^{\tilde{k}\pi}-c_1e^{-\tilde{k}\pi})&={\varphi}^-_k\label{0to2bdcon}
	\end{align}
	Solving these for $c_1,c_2$ and plugging them back to find the solution for the matter field, we find
	\begin{align}
		\varphi_k&=\hat{\varphi}^+_k-\frac{i\tilde{k}}{\sinh\tilde{k}\pi}(\varphi^+_k\cosh\tilde{k}\pi-{\varphi}^-_k)\pqty{\frac{1}{r}-\frac{1}{r_+}},\qquad &r\rightarrow\infty\nonumber\\
		&={\varphi}^-_k+\frac{i\tilde{k}}{\sinh\tilde{k}\pi}({\varphi}^+_k-{\varphi}^-_k \cosh\tilde{k}\pi)\pqty{\frac{1}{r}-\frac{1}{r_-}},\qquad &r\rightarrow -\infty\label{dsdt02msol}
	\end{align}
	Now using these results to compute the on-shell matter action in eq.\eqref{2unidt} as a function of $\varphi_k^+$ and $\varphi_k^-$, we get
	\begin{align}
		S_M=\frac{b}{2}\sum_{k\neq 0}\tilde{k}\left(({\varphi}^+_{-k}{\varphi}^+_{k}+{\varphi}^-_{-k}{\varphi}^-_{k})\coth\tilde{k}\pi-2{\varphi}^+_{-k}{\varphi}^-_{k}\csch\tilde{k}\pi\right)\label{dsdt02mact2}
	\end{align}
Now, for the $k=0$ modes, expanding the solution eq.\eqref{phi0} near large $\abs{r}$ gives
\begin{align}
	\varphi_0(r)&=a_1+ a_2\left(\frac{\pi}{2}-\frac{1}{r}\right),\quad r\rightarrow\infty\nonumber\\
	&=a_1- a_2\left(\frac{\pi}{2}+\frac{1}{r}\right),\quad r\rightarrow-\infty
	 \label{phi0p2f}
\end{align}
which after analytic continuations  eq.\eqref{pfunianacon} becomes
\begin{align}
		\varphi_0(r)&=a_1+ a_2\left(\frac{\pi}{2}-\frac{i}{r}\right),\quad r\rightarrow\infty\nonumber\\
	&=a_1- a_2\left(\frac{\pi}{2}-\frac{i}{r}\right),\quad r\rightarrow-\infty\label{phi0p2fs}
\end{align}
The boundary conditions then mean that
\begin{align}
	a_1+\frac{\pi a_2}{2}-\frac{i a_2}{r_+}=\varphi_0^+\nonumber\\
		a_1-\frac{\pi a_2}{2}+\frac{i a_2}{r_-}=\varphi_0^-\label{phi0bcp2f}
\end{align}
The on-shell action for these modes using eq.\eqref{2unidt} then becomes
\begin{align}
	S_0=\frac{b}{2\pi}(\varphi_0^+-\varphi_0^-)^2\label{s0p2fact}
\end{align}
So, the net action, combining eq.\eqref{dsdt02mact2} and \eqref{s0p2fact} then becomes
\begin{align}
		S_M=\frac{b}{2}\sum_{k}\tilde{k}\left(({\varphi}^+_{-k}{\varphi}^+_{k}+{\varphi}^-_{-k}{\varphi}^-_{k})\coth\tilde{k}\pi-2{\varphi}^+_{-k}{\varphi}^-_{k}\csch\tilde{k}\pi\right)\label{dsdt02mact3}
\end{align}
where the sum is now over all integers, with the $k=0$ term understood to be taken as a limit.

	Now, we shall redo the same analysis for the case of  transition from nothing to two expanding universes. Again, starting from the  nAdS solution in eq.\eqref{nadsrinf} and eq.\eqref{phi0p2f}, we now have to do the analytic continuation in eq.\eqref{2unianacon}. Imposing then the boundary conditions in eq.\eqref{dsdtmbd} we find, the analog of eq.\eqref{0to2bdcon}  as
	\begin{align}
		c_1+c_2+\frac{i\tilde{k}}{r_+}(c_2-c_1)&={\varphi}^+_k\nonumber\\
		c_1 e^{-\tilde{k}\pi}+c_2e^{\tilde{k}\pi}-\frac{i\tilde{k}}{r_-}(c_2e^{\tilde{k}\pi}-c_1e^{-\tilde{k}\pi})&={\varphi}^-_k\nonumber\\
			a_1+\frac{\pi a_2}{2}-\frac{i a_2}{r_+}&=\varphi_0^+\nonumber\\
		a_1-\frac{\pi a_2}{2}-\frac{i a_2}{r_-}&=\varphi_0^-
		\label{ptofbdcon}
	\end{align}
	which upon solving leads to
	\begin{align}
		\varphi_k&={\varphi}^+_k-\frac{i\tilde{k}}{\sinh\tilde{k}\pi}({\varphi}^+_k\cosh\tilde{k}\pi-{\varphi}^-_k)\pqty{\frac{1}{r}-\frac{1}{r_+}},\qquad &r\rightarrow\infty\nonumber\\
		&={\varphi}^-_k-\frac{i\tilde{k}}{\sinh\tilde{k}\pi}({\varphi}^+_k-{\varphi}^-_k \cosh\tilde{k}\pi)\pqty{\frac{1}{r}-\frac{1}{r_-}},\qquad &r\rightarrow -\infty\nonumber\\
		\varphi_0&=\varphi_0^+ -\frac{i}{\pi}(\varphi_0^+ - \varphi_0^-)\pqty{\frac{1}{r}-\frac{1}{r_+}} &r\rightarrow\infty\nonumber\\
		&=\varphi_0^- - \frac{i}{\pi}(\varphi_0^+ - \varphi_0^-)\pqty{\frac{1}{r}-\frac{1}{r_-}} \quad &r\rightarrow -\infty
		\label{dsdtpfmsol}
	\end{align}
	Inserting this solution in eq.\eqref{p2fmtact}, we find that the on-shell action is still given by eq.\eqref{dsdt02mact3}. 
	
	One can repeat the above steps to compute the on-shell action for the complex scalar field with the twisted boundary condition eq.\eqref{twistedbc}. The solution for the complex scalar field now is still given by eq.\eqref{nadsmtsol} but with $\tilde{k}\rightarrow \tilde{k}+\tilde{\alpha}$.  The action for the complex scalar field is taken to be eq.\eqref{comsc} which has an extra factor of $2$ compared to the real scalar field. These facts combine to give the final on-shell action to be that in eq.\eqref{dsdt02mact3} with $\tilde{k}$ replaced by $\tilde{k}+\tilde{\alpha}$ and an additional factor of $2$, and so the final expression reads as given in eq.\eqref{onacttwi}. 
	
	\subsection{Initial state}
	\label{initstate}
	In this subsection, we shall evaluate the inner product of an appropriate initial state with a field eigenstate. 
	The natural initial state in the past is vacuum state with respect  to the coordinates in eq.\eqref{milnezpm} for which we have carried out the semi-classical analysis in section \ref{dssemicla}. The  scalar field $\varphi$ will have the mode expansion
	\begin{align}
		\varphi=\varphi_0-\frac{ib}{4\pi}\pi_0 \ln (z_+ z_-)+\sum_{n=-\infty, n\neq 0}^{\infty}(a_+(n)z_+^{-n}+a_-(n)z_-^{-n})\label{scalmdexp}
	\end{align}
	where $z_{\pm}$ are related to $r_*,\theta$ as
	\begin{align}
		z_{\pm}=\exp(\frac{2\pi{i}}{b}(r_*\pm \theta))\label{zpmpastvac}
	\end{align}
	The initial state is the vacuum state annihilated by the modes $a_{\pm}(n)$, i.e
	\begin{align}
		a_{\pm}(n)|0\rangle =0 \qquad\forall n>0\label{initdsdts}
	\end{align}
	Defining an operator $A_n$ 
	\begin{align}
		A_n=\begin{cases}
			a_+(n),\qquad n>0,\\
			a_-(-n),\qquad n<0
		\end{cases}
		\label{Andef}
	\end{align}
	in terms of which the condition of vacuum state in eq.\eqref{initdsdts} becomes
	\begin{align}
		A_n|0\rangle =0 \qquad \forall n\neq 0\label{Anvac}
	\end{align}
	However, the Hamiltonian $H$, which involves derivatives acting on $\varphi$, will be independent of  $\varphi_0$. Thus a  general vacuum will be a one-parameter state $|p\rangle$ satisfying eq.\eqref{initdsdts} and also $\pi_0|p\rangle=p |p\rangle $, with $p$ being a continuous eigenvalue. For now, we consider the case of $p=0$ and the corresponding vacuum state is denoted $|0\rangle$. 
	The scalar field mode expansion becomes
	\begin{align}
		\varphi=\varphi_0+\pi_0 r_*+\sum_{n\neq 0}A_n\exp(-\frac{2\pi i}{b}(n\theta+E_n r_*))\label{dsdtsinA}
	\end{align}
	where $E_n=\abs{n}$. The vacuum state condition then becomes
	\begin{align}
		\pqty{\varphi_n+\frac{ib}{2\pi E_n}\del_{r_*}\varphi_n}|0\rangle=0\label{phivaceq}
	\end{align}
	where $\varphi_n$ is the Fourier transform of $\varphi(\theta)$ and is given by 
	\begin{align}
		\varphi_m=\frac{1}{b}\int d\theta e^{\frac{2i\pi m\theta}{b}}\varphi(\theta)\label{matfourer}
	\end{align}
	The equation eq.\eqref{phivaceq} should now be expressed in terms of the field $\varphi$ and its canonically conjugate momentum $\Pi_{\varphi}=\del_{r_*}
	\varphi$.
	Imposing the canonical equal-time commutation relations 
	\begin{align}
		[\varphi(\theta),\Pi_\varphi(\theta')]=i\delta(\theta-\theta')\Rightarrow [\varphi_m,\Pi_{\varphi,n}]=\frac{i}{b}\delta_{m+n,0}\label{mtcommu}
	\end{align}
	Thus the operator $\Pi_{\varphi,n}$ in terms of  $\varphi_m$ is given by $\Pi_{\varphi,n}=-\frac{i}{b}\del_{\varphi_{-m}}$.
	Taking inner product of eq.\eqref{phivaceq} with field eigenstate $\langle \varphi|$ gives
	\begin{align}
		\pqty{\frac{\del\quad}{\del\varphi_{-n}}+{2\pi E_n}\varphi_{n}}\langle\varphi_{n}|0\rangle=0\label{mtvaceq}
	\end{align}
	which has the solution
	\begin{align}
		\langle\varphi_{n}|0\rangle=\exp(-{2\pi}E_n\varphi_{-n}\varphi_{n})\label{fegvacmt}
	\end{align}
	and so it follows
	\begin{align}
		\langle {\varphi}^-|0\rangle =\exp(-2\pi\sum_{n\neq 0}{E_{n}}{\varphi}^-_{n}{\varphi}^-_{n})\label{fullfegvac}
	\end{align}
A more general vacuum state with a non-zero eigenvalue for $\pi_0$ will  have the sum in the exponent above to also include $n=0$ term with $E_0$ being a function of that eigenvalue.

The above calculations can be generalized to the case of twisted boundary conditions eq.\eqref{twistedbc}. For  the complex scalar field, the mode expansion is still given by 
		\begin{align}
		\varphi=\sum_{n=-\infty}^{\infty}(a_+(n)z_+^{-(n-\alpha)}+a_-(n)z_-^{-(n+\alpha)})\label{scalmdexp}
	\end{align}
The annihilation operators that define the vacuum are now given by 
	\begin{align}
	A_n=\begin{cases}
		a_+(n),\qquad n>0,\\
		a_-(n),\qquad n\geq 0
	\end{cases}
	\label{Andef}
\end{align}
The vacuum state condition still is given by eq.\eqref{phivaceq} but with $E_n$ now  given by $E_n=\abs{n+\alpha}$. Noting that $\Pi_{\varphi^\dagger}=\del_{r_*}\varphi$ and that upon imposing canonical commutation relations 
\begin{align}
	[\varphi(\theta),\Pi_\varphi(\theta')]&=i\delta(\theta-\theta')\Rightarrow[\varphi_m,\Pi_{\varphi,n}]=\frac{i}{b}\delta_{m,n}\nonumber\\
	[\varphi^\dagger(\theta),\Pi_\varphi^\dagger(\theta')]&=i\delta(\theta-\theta')\Rightarrow[\varphi^\dagger_m,\Pi_{\varphi^\dagger,n}]=\frac{i}{b}\delta_{m,n}\label{twcac}
\end{align}
where now $\varphi^\dagger_n, \Pi_{\varphi^\dagger,n}$ are now defined to be
{
\begin{align}
		\varphi_n&=\frac{1}{b}\int_0^b d\theta e^{-i(\tilde{n}+\tilde{\alpha})\theta}\varphi(\theta),\nonumber\\
	\Pi_{\varphi,n}&=\frac{1}{b}\int_0^b d\theta e^{i(\tilde{n}+\tilde{\alpha})\theta}\Pi_\varphi(\theta)\nonumber\\
	(\varphi^\dagger)_n&=\frac{1}{b}\int_0^b d\theta e^{-i(\tilde{n}+\tilde{\alpha})\theta}\varphi^\dagger(\theta),\nonumber\\
		\Pi_{\varphi^\dagger,n}&=\frac{1}{b}\int_0^b d\theta e^{i(\tilde{n}+\tilde{\alpha})\theta}\Pi_{\varphi^\dagger}(\theta),\label{twmoinf}
\end{align}}
Further noting that $\Pi_{\varphi,n}={-\frac{i}{b}\frac{\del}{\del{\varphi^\dagger_n}}}$, we get the analog of eq.\eqref{mtvaceq} to be
\begin{align}
		\pqty{\frac{\del\quad}{\del\varphi^\dagger_{n}}+{2\pi E_n}\varphi_{n}}\langle\varphi_{n}|0\rangle=0\label{mtvaceqco}
\end{align}
and so we get
\begin{align}
		\langle {\varphi}^-|0\rangle =\exp(-2\pi\sum_{n=-\infty }^{\infty}{E_{n}}{(\varphi^\dagger)}^-_{n}{\varphi}^-_{n})\label{fullfegvacco}
\end{align}
where now $E_n=\abs{n+\alpha}$.
	
	\section{Double trumpet SFF}
	\label{sffapp}
	
	In this appendix, we shall show expliclity in detail, the way the integral  eq.\eqref{jhbh} is evaluated.
	Consider the integral
	\begin{align}
		\hat{Z}(b) & = \int_C dx {e^{S(x)}\over\sqrt{1+x^2} }\nonumber\\
		S(x) & =  2ET x\pqty{1-\frac{b^2}{{b}_0^2}\frac{1}{1+x^2}}\label{dtsadact2}
	\end{align}
	Here $C$ is the Bromwich contour in terms of the variable $x={\beta \over T}$. Our aim in this appendix is  to show that the integral in eq.\eqref{dtsadact2} is finite in the region 
	\begin{align}
		b^2 EG_e^{-1}\geq 1\label{beggg1}
	\end{align}
	
	It is difficult to do the $x$-integral exactly in eq.\eqref{dtsadact2} and so we will carry out the analysis in the saddle point method wherever possible.  	We shall see that we can further split the region eq.\eqref{beggg1} into various sub-regions. In some such sub-regions saddle point analysis leads to a good estimate of the integral. In regions not amenable to saddle point analysis, we shall use other arguments to justify that the value of the double trumpet SFF is finite in that region.  The various sub-regions are as follows
	\begin{align}
		&b> b_0,\nonumber\\
		&{b-{b}_0\over {b}_0}\ll  1,\nonumber\\
		&b^2 EG_e^{-1}\geq 1 \,\,\& \,\, b<b_0\label{bregions}
	\end{align}

	As can be seen from eq.\eqref{dtsadact2}, the integrand has a square root branch cut. To compute the saddle point and the corresponding on-shell action, we first adopt the following conventions where we work with two Riemann sheets with the following definitions 
	\begin{align}
		1^{st} \text{sheet}\Rightarrow\sqrt{(1+x^2)}=\abs{\sqrt{(1+x^2)}}e^{\frac{1}{2}\arg{(1+x^2)}}\nonumber\\
		2^{nd} \text{sheet}\Rightarrow\sqrt{(1+x^2)}=-\abs{\sqrt{(1+x^2)}}e^{\frac{1}{2}\arg{(1+x^2)}}\label{sqrtconv}
	\end{align}
	where $\arg{x}\in [-\pi,\pi]$.
	
	The saddle point equation for the $x$ integral obtained from eq.\eqref{dtsadact2} is given by 
	\begin{align}
		1-\frac{b^2}{{b}_0^2}\frac{1-x^2}{(1+x^2)^2}+\frac{1}{2{b}_0\sqrt{E G_e^{-1}}}\frac{x}{1+x^2}=0\label{dtsaddeq}
	\end{align}
	For the saddle point analysis to be a good approximation, we require 
	\begin{align}
		ET\gg 1\label{etlarge}
	\end{align}
	, which in effect plays  the role of $\hbar^{-1}$. It is easy to see from eq.\eqref{valb0} and eq.\eqref{condbc} that $b_0\gg b_c$ when eq.\eqref{etlarge} is satisfied. 
	Since we are working in the regime where eq.\eqref{enc},\eqref{etlarge}  is satisfied,
	it is easy to see from the saddle point eq.\eqref{dtsaddeq} that we can neglect the last term and so the saddle point equation becomes
	
	\begin{align}
		1-\frac{b^2}{{b}_0^2}\frac{1-x^2}{(1+x^2)^2}=0\label{lgngsdeq}
	\end{align}
	The saddle points are given by 
	\begin{align}
		y=x^2&=-\pqty{1+\frac{b^2}{2{b}_0^2}}\pm \frac{b}{2{b}_0}\sqrt{\frac{b^2}{{b}_0^2}+8}
		\label{sadlgng}
	\end{align}
	For $b>{b}_0$, two of the saddle points in eq.\eqref{sadlgng} are along the imaginary axis and two along the real axis of the complex x-plane. For the case $b<{b}_0$, all the four saddle points are along the imaginary axis. At $b={b}_0$, two of the saddle points are at the origin while the other two are along the imaginary axis. Following are the values of the saddle points in the extreme limits
	\begin{align}
		x\simeq\begin{cases}
			\pm i\pqty{1+\frac{b}{\sqrt{2}{b}_0}},\pm i \pqty{1-\frac{b}{\sqrt{2}{b}_0}}\qquad &b\ll {b}_0\\
			\pm i\pqty{\frac{b}{{b}_0}+\frac{3}{2}\frac{{b}_0}{b}},\pm \pqty{1-2\frac{{b}_0^2}{b^2}}\qquad &b\gg {b}_0\\
			0,0,\pm i \sqrt{3}\qquad &b={b}_0
		\end{cases}
		\label{saddatext}
	\end{align}
	The saddle points have a small positive real part when the contribution of the last term in eq.\eqref{dtsaddeq} is treated perturbatively. 
	The saddle point value is a good approximation only when the following condition is met
	\begin{align}
		\abs{{S'''(x_*)\over S''(x_*)}(x-x_*)}\ll 1\Rightarrow\abs{\frac{S'''(x_*)}{S''(x_*)^{\frac{3}{2}}}}\ll 1\label{dtsadconscond}
	\end{align}
	where $x_*$ is one of the saddle points in eq.\eqref{sadlgng} through with the contour is deformed to pass. Computing the second and third derivatives of the action from eq.\eqref{dtsadact} by ignoring the log term for reasons mentioned around eq.\eqref{etlarge}, we get
	\begin{align}
		S''(x)&=-\frac{4 ET b^2 x \left(x^2-3\right)}{{{b}_0}^2 \left(x^2+1\right)^3}\nonumber\\
		S'''(x)&=-\frac{12ET b^2 \left(x^4-6 x^2+1\right)}{{{b}_0}^2 \left(x^2+1\right)^4}\label{s2ands3dt}
	\end{align}
	using which the consistency condition eq.\eqref{dtsadconscond} becomes
	\begin{align}
		\abs{\frac{S'''(x)}{ S''(x)^{\frac{3}{2}}}}=\frac{3 {{b}_0}  \left(x^4-6 x^2+1\right)\sqrt{\left(x^2+1\right)}}{2 b\sqrt{ET}  \left({ x \left(x^2-3\right)}\right)^{3/2}}\label{dtsadconsis}
	\end{align}
	It is easy to see from the above expression and using the saddle point values in the limiting cases given in eq.\eqref{saddatext}, that the saddle point condition fails in the region $b\simeq b_0$ or alternately $\abs{q}\ll 1$. The failure of the saddle point analysis in this region  is due to the smallness of the second derivative as a result of two of the saddle points coming close to each other  around the origin. Away from this region, the saddle point approximation is good and we shall show then that the value of the $b$ integral is finite in those regions. 
	
	\begin{figure}[h!]

		\tikzset{every picture/.style={line width=0.75pt}} 
		
		\begin{tikzpicture}[x=0.75pt,y=0.75pt,yscale=-1,xscale=0.55]
			
			\draw  (100,217.74) -- (486.5,217.74)(294.29,52) -- (294.29,390) (479.5,212.74) -- (486.5,217.74) -- (479.5,222.74) (289.29,59) -- (294.29,52) -- (299.29,59)  ;
			\draw   (289.17,74.04) -- (298.87,81.23) -- (288.92,80.86) -- (298.62,88.06) -- (288.67,87.69) -- (298.37,94.89) -- (288.41,94.52) ;
			\draw   (288.41,94.52) -- (298.12,101.72) -- (288.16,101.35) -- (297.87,108.55) -- (287.91,108.18) -- (297.62,115.38) -- (287.66,115.01) ;
			\draw   (289.92,53.55) -- (299.63,60.75) -- (289.67,60.38) -- (299.38,67.58) -- (289.42,67.21) -- (299.13,74.4) -- (289.17,74.04) ;
			\draw   (287.34,115.17) -- (297.55,121.63) -- (287.59,122) -- (297.81,128.46) -- (287.85,128.83) -- (298.06,135.29) -- (288.11,135.66) ;
			\draw   (289.84,303.07) -- (299.15,310.43) -- (289.96,310.6) -- (299.26,317.96) -- (290.07,318.12) -- (299.38,325.49) -- (290.19,325.65) ;
			\draw   (290.19,325.65) -- (299.5,333.02) -- (290.31,333.18) -- (299.62,340.54) -- (290.43,340.71) -- (299.74,348.07) -- (290.55,348.24) ;
			\draw   (290.55,348.24) -- (299.86,355.6) -- (290.67,355.76) -- (299.98,363.13) -- (290.79,363.29) -- (300.09,370.66) -- (290.9,370.82) ;
			\draw   (290.91,370.82) -- (300.21,378.18) -- (291.02,378.35) -- (300.33,385.71) -- (291.14,385.88) -- (300.45,393.24) -- (291.26,393.4) ;
			\draw [color={rgb, 255:red, 208; green, 2; blue, 27 }  ,draw opacity=1 ]   (113.3,192.65) .. controls (250.3,187.65) and (357.34,181.02) .. (294.11,138.66) ;
			\draw [shift={(232.57,186.09)}, rotate = 534.8399999999999] [fill={rgb, 255:red, 208; green, 2; blue, 27 }  ,fill opacity=1 ][line width=0.08]  [draw opacity=0] (8.93,-4.29) -- (0,0) -- (8.93,4.29) -- cycle    ;
			\draw  [fill={rgb, 255:red, 0; green, 0; blue, 0 }  ,fill opacity=1 ] (296.81,138.66) .. controls (296.81,137.17) and (295.6,135.96) .. (294.11,135.96) .. controls (292.62,135.96) and (291.41,137.17) .. (291.41,138.66) .. controls (291.41,140.15) and (292.62,141.36) .. (294.11,141.36) .. controls (295.6,141.36) and (296.81,140.15) .. (296.81,138.66) -- cycle ;
			\draw  [fill={rgb, 255:red, 0; green, 0; blue, 0 }  ,fill opacity=1 ] (295.24,303.07) .. controls (295.24,301.58) and (294.03,300.37) .. (292.54,300.37) .. controls (291.05,300.37) and (289.84,301.58) .. (289.84,303.07) .. controls (289.84,304.56) and (291.05,305.77) .. (292.54,305.77) .. controls (294.03,305.77) and (295.24,304.56) .. (295.24,303.07) -- cycle ;
			\draw  [fill={rgb, 255:red, 0; green, 0; blue, 0 }  ,fill opacity=1 ] (315.11,155.66) .. controls (315.11,154.17) and (313.9,152.96) .. (312.41,152.96) .. controls (310.92,152.96) and (309.71,154.17) .. (309.71,155.66) .. controls (309.71,157.15) and (310.92,158.36) .. (312.41,158.36) .. controls (313.9,158.36) and (315.11,157.15) .. (315.11,155.66) -- cycle ;
			\draw [color={rgb, 255:red, 208; green, 2; blue, 27 }  ,draw opacity=1 ]   (117.3,246.65) .. controls (205.3,248.65) and (367.3,246.65) .. (295.24,303.07) ;
			\draw [shift={(235.75,250.87)}, rotate = 4.57] [fill={rgb, 255:red, 208; green, 2; blue, 27 }  ,fill opacity=1 ][line width=0.08]  [draw opacity=0] (8.93,-4.29) -- (0,0) -- (8.93,4.29) -- cycle    ;
			\draw  [fill={rgb, 255:red, 0; green, 0; blue, 0 }  ,fill opacity=1 ] (315.2,278) .. controls (315.2,276.51) and (313.99,275.3) .. (312.5,275.3) .. controls (311.01,275.3) and (309.8,276.51) .. (309.8,278) .. controls (309.8,279.49) and (311.01,280.7) .. (312.5,280.7) .. controls (313.99,280.7) and (315.2,279.49) .. (315.2,278) -- cycle ;
			\draw [color={rgb, 255:red, 223; green, 13; blue, 13 }  ,draw opacity=1 ][line width=3]    (981,30) -- (1020.5,30) ;
			\draw [color={rgb, 255:red, 126; green, 211; blue, 33 }  ,draw opacity=1 ][line width=3]    (980.5,67) -- (1020.5,67) ;
			\draw  (549,218.74) -- (935.5,218.74)(743.29,53) -- (743.29,391) (928.5,213.74) -- (935.5,218.74) -- (928.5,223.74) (738.29,60) -- (743.29,53) -- (748.29,60)  ;
			\draw   (738.17,75.04) -- (747.87,82.23) -- (737.92,81.86) -- (747.62,89.06) -- (737.67,88.69) -- (747.37,95.89) -- (737.41,95.52) ;
			\draw   (737.41,95.52) -- (747.12,102.72) -- (737.16,102.35) -- (746.87,109.55) -- (736.91,109.18) -- (746.62,116.38) -- (736.66,116.01) ;
			\draw   (738.92,54.55) -- (748.63,61.75) -- (738.67,61.38) -- (748.38,68.58) -- (738.42,68.21) -- (748.13,75.4) -- (738.17,75.04) ;
			\draw   (736.34,116.17) -- (746.55,122.63) -- (736.59,123) -- (746.81,129.46) -- (736.85,129.83) -- (747.06,136.29) -- (737.11,136.66) ;
			\draw   (738.84,304.07) -- (748.15,311.43) -- (738.96,311.6) -- (748.26,318.96) -- (739.07,319.12) -- (748.38,326.49) -- (739.19,326.65) ;
			\draw   (739.19,326.65) -- (748.5,334.02) -- (739.31,334.18) -- (748.62,341.54) -- (739.43,341.71) -- (748.74,349.07) -- (739.55,349.24) ;
			\draw   (739.55,349.24) -- (748.86,356.6) -- (739.67,356.76) -- (748.98,364.13) -- (739.79,364.29) -- (749.09,371.66) -- (739.9,371.82) ;
			\draw   (739.91,371.82) -- (749.21,379.18) -- (740.02,379.35) -- (749.33,386.71) -- (740.14,386.88) -- (749.45,394.24) -- (740.26,394.4) ;
			\draw [color={rgb, 255:red, 208; green, 2; blue, 27 }  ,draw opacity=1 ]   (744.24,304.07) .. controls (837.3,267.65) and (810.3,149.65) .. (745.81,139.66) ;
			\draw [shift={(804.48,220.48)}, rotate = 452.05] [fill={rgb, 255:red, 208; green, 2; blue, 27 }  ,fill opacity=1 ][line width=0.08]  [draw opacity=0] (8.93,-4.29) -- (0,0) -- (8.93,4.29) -- cycle    ;
			\draw  [fill={rgb, 255:red, 0; green, 0; blue, 0 }  ,fill opacity=1 ] (745.81,139.66) .. controls (745.81,138.17) and (744.6,136.96) .. (743.11,136.96) .. controls (741.62,136.96) and (740.41,138.17) .. (740.41,139.66) .. controls (740.41,141.15) and (741.62,142.36) .. (743.11,142.36) .. controls (744.6,142.36) and (745.81,141.15) .. (745.81,139.66) -- cycle ;
			\draw  [fill={rgb, 255:red, 0; green, 0; blue, 0 }  ,fill opacity=1 ] (744.24,304.07) .. controls (744.24,302.58) and (743.03,301.37) .. (741.54,301.37) .. controls (740.05,301.37) and (738.84,302.58) .. (738.84,304.07) .. controls (738.84,305.56) and (740.05,306.77) .. (741.54,306.77) .. controls (743.03,306.77) and (744.24,305.56) .. (744.24,304.07) -- cycle ;
			\draw  [fill={rgb, 255:red, 0; green, 0; blue, 0 }  ,fill opacity=1 ] (807.2,220) .. controls (807.2,218.51) and (805.99,217.3) .. (804.5,217.3) .. controls (803.01,217.3) and (801.8,218.51) .. (801.8,220) .. controls (801.8,221.49) and (803.01,222.7) .. (804.5,222.7) .. controls (805.99,222.7) and (807.2,221.49) .. (807.2,220) -- cycle ;
			\draw [color={rgb, 255:red, 126; green, 211; blue, 33 }  ,draw opacity=1 ]   (600.41,179.31) .. controls (606.71,194.96) and (705.41,133.31) .. (743.11,139.66) ;
			\draw [shift={(670.98,158.39)}, rotate = 337.51] [fill={rgb, 255:red, 126; green, 211; blue, 33 }  ,fill opacity=1 ][line width=0.08]  [draw opacity=0] (8.93,-4.29) -- (0,0) -- (8.93,4.29) -- cycle    ;
			\draw [color={rgb, 255:red, 126; green, 211; blue, 33 }  ,draw opacity=1 ]   (601.3,252.65) .. controls (631.3,246.65) and (668.3,305.65) .. (738.84,304.07) ;
			\draw [shift={(667.87,284.52)}, rotate = 210.07] [fill={rgb, 255:red, 126; green, 211; blue, 33 }  ,fill opacity=1 ][line width=0.08]  [draw opacity=0] (8.93,-4.29) -- (0,0) -- (8.93,4.29) -- cycle    ;
			\draw [color={rgb, 255:red, 126; green, 211; blue, 33 }  ,draw opacity=1 ]   (100.3,180.65) .. controls (150.3,179.65) and (215.6,136.95) .. (291.41,138.66) ;
			\draw [shift={(195.89,155.48)}, rotate = 342.12] [fill={rgb, 255:red, 126; green, 211; blue, 33 }  ,fill opacity=1 ][line width=0.08]  [draw opacity=0] (8.93,-4.29) -- (0,0) -- (8.93,4.29) -- cycle    ;
			\draw [color={rgb, 255:red, 126; green, 211; blue, 33 }  ,draw opacity=1 ]   (113.3,263.65) .. controls (163.3,262.65) and (214.3,300.65) .. (289.84,303.07) ;
			\draw [shift={(201.5,285.27)}, rotate = 198.5] [fill={rgb, 255:red, 126; green, 211; blue, 33 }  ,fill opacity=1 ][line width=0.08]  [draw opacity=0] (8.93,-4.29) -- (0,0) -- (8.93,4.29) -- cycle    ;
			\draw  [fill={rgb, 255:red, 0; green, 0; blue, 0 }  ,fill opacity=1 ] (317.11,110.66) .. controls (317.11,109.17) and (315.9,107.96) .. (314.41,107.96) .. controls (312.92,107.96) and (311.71,109.17) .. (311.71,110.66) .. controls (311.71,112.15) and (312.92,113.36) .. (314.41,113.36) .. controls (315.9,113.36) and (317.11,112.15) .. (317.11,110.66) -- cycle ;
			\draw  [fill={rgb, 255:red, 0; green, 0; blue, 0 }  ,fill opacity=1 ] (319.11,319.66) .. controls (319.11,318.17) and (317.9,316.96) .. (316.41,316.96) .. controls (314.92,316.96) and (313.71,318.17) .. (313.71,319.66) .. controls (313.71,321.15) and (314.92,322.36) .. (316.41,322.36) .. controls (317.9,322.36) and (319.11,321.15) .. (319.11,319.66) -- cycle ;
			\draw  [fill={rgb, 255:red, 0; green, 0; blue, 0 }  ,fill opacity=1 ] (679.2,219) .. controls (679.2,217.51) and (677.99,216.3) .. (676.5,216.3) .. controls (675.01,216.3) and (673.8,217.51) .. (673.8,219) .. controls (673.8,220.49) and (675.01,221.7) .. (676.5,221.7) .. controls (677.99,221.7) and (679.2,220.49) .. (679.2,219) -- cycle ;
			\draw  [fill={rgb, 255:red, 0; green, 0; blue, 0 }  ,fill opacity=1 ] (758.2,65) .. controls (758.2,63.51) and (756.99,62.3) .. (755.5,62.3) .. controls (754.01,62.3) and (752.8,63.51) .. (752.8,65) .. controls (752.8,66.49) and (754.01,67.7) .. (755.5,67.7) .. controls (756.99,67.7) and (758.2,66.49) .. (758.2,65) -- cycle ;
			\draw  [fill={rgb, 255:red, 0; green, 0; blue, 0 }  ,fill opacity=1 ] (758.2,400) .. controls (758.2,398.51) and (756.99,397.3) .. (755.5,397.3) .. controls (754.01,397.3) and (752.8,398.51) .. (752.8,400) .. controls (752.8,401.49) and (754.01,402.7) .. (755.5,402.7) .. controls (756.99,402.7) and (758.2,401.49) .. (758.2,400) -- cycle ;
			\draw [color={rgb, 255:red, 245; green, 166; blue, 35 }  ,draw opacity=1 ]   (405,59) -- (402,392) ;
			\draw [shift={(403.5,225.5)}, rotate = 90.52] [fill={rgb, 255:red, 245; green, 166; blue, 35 }  ,fill opacity=1 ][line width=0.08]  [draw opacity=0] (8.93,-4.29) -- (0,0) -- (8.93,4.29) -- cycle    ;
			\draw [color={rgb, 255:red, 245; green, 166; blue, 35 }  ,draw opacity=1 ]   (857,63) -- (854,396) ;
			\draw [shift={(855.5,229.5)}, rotate = 90.52] [fill={rgb, 255:red, 245; green, 166; blue, 35 }  ,fill opacity=1 ][line width=0.08]  [draw opacity=0] (8.93,-4.29) -- (0,0) -- (8.93,4.29) -- cycle    ;
			\draw [color={rgb, 255:red, 245; green, 166; blue, 35 }  ,draw opacity=1 ][line width=3]    (980.5,111) -- (1020.5,111) ;
			
			\draw (1032,21) node [anchor=north west][inner sep=0.75pt]   [align=left] {1st Sheet};
			\draw (1030,58) node [anchor=north west][inner sep=0.75pt]   [align=left] {2nd Sheet};
			\draw (322,146.4) node [anchor=north west][inner sep=0.75pt]  [font=\large]  {$x_{2}$};
			\draw (323,309.4) node [anchor=north west][inner sep=0.75pt]  [font=\large]  {$x_{4}$};
			\draw (809,194.4) node [anchor=north west][inner sep=0.75pt]  [font=\large]  {$x_{3}$};
			\draw (673,17.4) node [anchor=north west][inner sep=0.75pt]    {$b >b_{0}$};
			\draw (216,15.4) node [anchor=north west][inner sep=0.75pt]    {$b< b_{0}$};
			\draw (322,266.4) node [anchor=north west][inner sep=0.75pt]  [font=\large]  {$x_{3}$};
			\draw (321,98.4) node [anchor=north west][inner sep=0.75pt]  [font=\large]  {$x_{1}$};
			\draw (766,55.4) node [anchor=north west][inner sep=0.75pt]  [font=\large]  {$x_{1}$};
			\draw (681.2,222.4) node [anchor=north west][inner sep=0.75pt]  [font=\large]  {$x_{2}$};
			\draw (760,375.4) node [anchor=north west][inner sep=0.75pt]  [font=\large]  {$x_{4}$};
			\draw (1032,104) node [anchor=north west][inner sep=0.75pt]   [align=left] {Original contour };
			\draw (1032,120) node [anchor=north west][inner sep=0.75pt]   [align=left] {(in First sheet)};

		\end{tikzpicture}
		\caption{Double trumpet contours}
		\centering
		\label{dtsfffig}
	\end{figure}
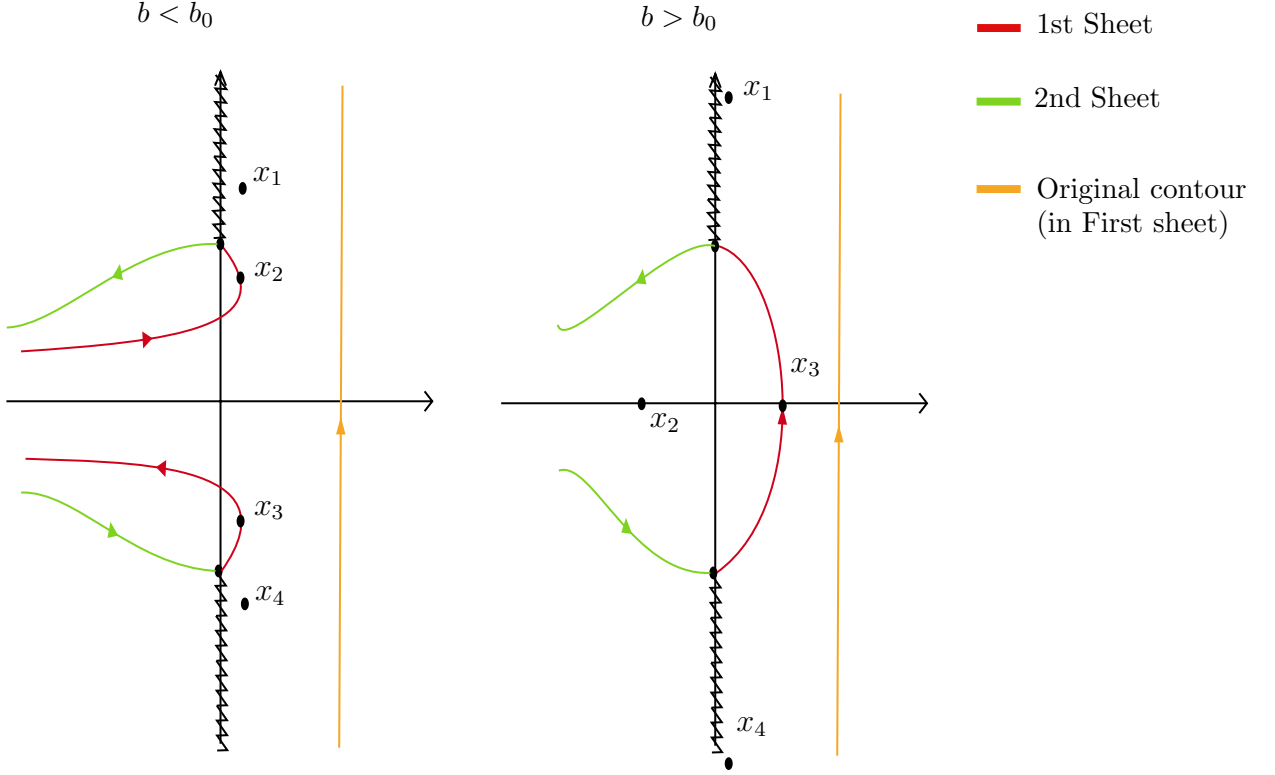

	\subsection{$b>{b}_0$}
	\label{qge1}
	The relevant saddle point in this region is given by
	\begin{align}
		x_*= \sqrt{ \frac{b}{2{b}_0}\sqrt{\frac{b^2}{{b}_0^2}+8}-\pqty{1+\frac{b^2}{2{b}_0^2}}}\label{qg1sad}
	\end{align}
	The range of the saddle point for $b>{b}_0$ is $x_*\in (0,1)$.  The appropriate contour that passes through this saddle point is shown in fig.\ref{dtsfffig}. The $x$-integral in eq.\eqref{dtsadact2} gives
	\begin{align}
		\hat{Z}(b)={e^{S(x_*)+i\frac{\pi}{2}}\over \sqrt{(1+x_*^2)\abs{S''(x_*)}}}\label{xintzhat}
	\end{align} 
	where the phase factor of $\pi\over 2$ is due to the steepest descent direction. Now, the $b$ integral becomes
	\begin{align}
		Z_{DT}(E,T)\simeq \int_{(1+q)b_0}^{\infty} db\, \frac{bZ_M[b]}{{G_e }}{e^{S(x_*)+i\frac{\pi}{2}}\over \sqrt{(1+x_*^2)\abs{S''(x_*)}}}\label{qg1zdt}
	\end{align} 
	where $q=\frac{b-b_0}{b_0}>0$ but is not very small compared to unity. As it is hard to do the $b$-integral exactly for $b$ in the range under consideration, we shall argue for the finiteness of the integral indirectly. First notice that the value of the integrand is finite near the lower limit as long as $q$ is not sufficiently close to zero.  Moreover the integrand is a smooth function of $b$. Let us now look at the upper end of the integral.   In this  region,  $b\gg {b}_0$ and the saddle point gets closer to unity,
	\begin{align}
		x_*\simeq 1-\frac{2 b_0^2}{b^2}\label{bggb0sadle}
	\end{align}
	This region in comparitively easier to analyze. We shall see that the integrand is an exponentially decaying function of $b$ in this region. Combined with the finiteness at the lower end and smoothness of the integrand in the range under consideration, this will allow us to conclude that the full integral is finite.  
	The on-shell action and its second derivative are given by 
	\begin{align}
		S(x)\simeq-\frac{b^2}{{b}_0}\sqrt{EG_e^{-1}},\quad S''(x)\simeq \frac{b^2}{{b}_0}\sqrt{EG_e^{-1}}\label{bggb0os}
	\end{align}
	Doing the $x$ integral and using the asymptotic form of the matter contribution from eq.\eqref{defzm},  then gives, upto numerical factors,
	\begin{align}
		Z_{DT}(E,T)\simeq&i\sqrt{\frac{T}{2 G_e}}\int db \, e^{-\frac{b^2}{\tilde{b}_0}\sqrt{EG_e^{-1}}}
		\label{bggb0res}
	\end{align}
	As mentioned earlier, the integrand for the $b$-integral is exponentially decaying in this region and hence there will be no divergence arising from this region. This result combined with arguments mentioned earlier suffice to argue that the net value of the $b$-integral is finite in the region $q\geq 1$.  Let us now analyze the region around $b-b_0\ll b_0$.

	\subsection{$\frac{b-b_0}{b_0}\ll 1 $}
	\label{qeq1}
	As discussed earlier after eq.\eqref{dtsadconsis}, the saddle point method fails in this region for the saddles corresponding to the contour deformations in fig.\ref{dtsfffig} and so we shall argue for the finitiness of the second term in  eq.\eqref{bu} in a different way. The x-integral in eq.\eqref{dtsadact2}, exactly at $b=b_0$ becomes
	\begin{align}
		\hat{Z}_b=\int dx \frac{e^{2ET \frac{x^3}{x^2+1}}}{\sqrt{x^2+1}}\label{xintbeqb0}
	\end{align}
	Taking $x=\gamma+i u$, with $\gamma\ll 1$ and dropping all the $\gamma$ dependent terms, we get
	\begin{align}
		\hat{Z}_b=\int_{-\infty}^\infty du \frac{e^{-2iET \frac{u^3}{1-u^2}}}{\sqrt{1-u^2}}\label{uintbeqb0}
	\end{align}
	Away from $u=0$, the phase of the above integral is wildly oscillating due to the large factor of $ET$ in the exponent which leads to a suppression of the total integral.  In particular for $u\gg 1$, the phase is wildly oscillating and the magnitude is also suppressed. Near $u=0$, the integrand is well behaved. The only region which appears to be troublesome is the region near $u\simeq \pm 1 $.  Let us consider the region around $u=1$. Let $u=1+\epsilon$ and hence the above integral becomes
	\begin{align}
		\hat{Z}_b\simeq \int d\epsilon  \frac{e^{\frac{i ET}{\epsilon}}}{\sqrt{-2\epsilon}}\label{zhbineps}
	\end{align}
	This integral is convergent around $\epsilon=0$ as can be easily checked by  doing a variable transformation $\epsilon=\frac{1}{s}$ and looking at the region $s\gg 1$. 
	So, in all, the above integral eq.\eqref{xintbeqb0} is well-behaved and  leads to a finite answer.
	Let us now analyze the last of the regions in eq.\eqref{bregions}. 


	\subsection{$b^2 EG_e^{-1}\gg 1 \,\,\&\,\, b<b_0$}
	\label{bsmbnl}
	This  region is amenable to saddle point analysis as we shall detail below. The region under consideration is
	\begin{align}
		\frac{1}{\sqrt{EG_e^{-1}}}\ll b< {b}_0\label{brangebggege}
	\end{align}
	The contour can be deformed to go through two saddle points in the first sheet and none in the second, which is shown in the first of the plots in fig.\ref{dtsfffig}. The corresponding saddle points through which the contour passes are complex conjugates of each other, labelled $x_2, x_3$ in fig.\ref{dtsfffig}, and are given by 
	\begin{align}
		x_*=\pm \sqrt{ \frac{b}{2{b}_0}\sqrt{\frac{b^2}{{b}_0^2}+8}-\pqty{1+\frac{b^2}{2{b}_0^2}}}\label{blb0sads}
	\end{align}
	The $x$-integral then gives the analog of eq.\eqref{xintzhat} with the contribution from both the saddle points included. The $b$-integral is hard to do exactly. 
	But, since the range of the $b$ integral is finite, and the matter contribution in eq.\eqref{bu} is also finite as $b$ is away from zero,  the value of the $b$ integral in this range will be finite. 
	However, we can  show that the $b$-integral is dominated by the  contribution from the lower end of the region eq.\eqref{brangebggege}. Hence, we shall just focus on the region $\frac{1}{\sqrt{EG_e^{-1}}}\ll b\ll 1$. In this region, the matter contribution increases towards smaller $b$  which results in the $b$-integral dominated by its lower limit as we shall show in detail below. 
	 In this region the saddle point, 
	the corresponding on-shell action and its higher derivatives are 
	\begin{align}
		x&\simeq \pm i \pqty{1-\frac{b}{\sqrt{2}{b}_0}}\nonumber\\
		S(x)&\simeq\pm i 2{b}_0\sqrt{EG_e^{-1}}\pqty{1-\frac{\sqrt{2}b}{{b}_0}}\nonumber\\
		S''(x)&\simeq\pm 4i \sqrt{2}\frac{{b}_0^2\sqrt{EG_e^{-1}}}{{b}}\nonumber\\
		S'''(x_*)&\simeq \frac{6b^2\sqrt{EG_e^{-1}}}{{b}_0}\pqty{\frac{\sqrt{2}{b}_0}{b}}^4
		\label{bllb0bg}
	\end{align}
	The steepest descent angles at the saddle points are $\frac{\pi}{4},\frac{3\pi}{4}$. The $b$-integral,upto numerical factors,  then reads
	\begin{align}
		Z_{DT}(E,T)\simeq\int db \frac{b^{{2}} e^{\frac{\pi^2}{6b}}}{2\pi G_e {b}_0(EG_e^{-1})^{\frac{1}{4}}}\pqty{e^{2i\sqrt{EG_e^{-1}}(\tilde{b}_0-\sqrt{2}b)+\frac{i\pi}{4}}+e^{-2i\sqrt{EG_e^{-1}}({b}_0-\sqrt{2}b)+\frac{3i\pi}{4}}}\label{dtbllb0bgein}
	\end{align}
	The condition for the saddle point estimate to be good is given by 
	\begin{align}
		\frac{1}{\sqrt{b}(EG_e^{-1})^{\frac{1}{4}}}\ll 1\label{bgegg1conscond}
	\end{align}
	which is satisfied in the range eq.\eqref{brangebggege}.
	So, we need to do the integrals of the form 
	\begin{align}
		I_{\alpha}=\int_{y_0}^{\alpha} dy\,\, e^{\pm i y+\frac{\alpha}{y}},\qquad \alpha\gg 1\label{ialpha}
	\end{align}	
	where $y_0$ is such that it satisfies $1\ll y_0\ll \alpha$, with $\alpha\sim \order{(\sqrt{EG_e^{-1}})}$ and the variable $y$ is related to $b$ as $y=b\sqrt{EG_e^{-1}}$. It is easy to see that the above integral gets maximum contribution from the lower limit of the integral, i.e. near $y=y_0$
	\begin{align}
		I_\alpha\sim e^{\frac{\alpha}{y_0}}\label{llestialph}
	\end{align}
	So, we are forced to estimate the value of the integral when $b$ is even smaller than the regime mentioned in eq.\eqref{brangebggege}. This region has already been  analyzed with the result in eq.\eqref{bpertzhat}. Thus we see that the contribution to the spectral form factor from the double trumpet geometry diverges.

	\newpage
	\bibliographystyle{JHEP}
	\bibliography{refs}

\providecommand{\href}[2]{#2}\begingroup\raggedright\begin{thebibliography}{10}

\bibitem{Sachdev:1992fk}
S.~Sachdev and J.~Ye, \emph{{Gapless spin fluid ground state in a random,
  quantum Heisenberg magnet}},
  \href{http://dx.doi.org/10.1103/PhysRevLett.70.3339}{\emph{Phys. Rev. Lett.}
  {\bf 70} (1993) 3339}, [\href{http://arxiv.org/abs/cond-mat/9212030}{{\tt
  cond-mat/9212030}}].

\bibitem{Kitaev-talks:2015}
A.~Kitaev, \emph{{A simple model of quantum holography}}, {\emph{Talks at KITP}
  (2015) }.

\bibitem{JACKIW1985343}
R.~Jackiw, \emph{Lower dimensional gravity},
  \href{http://dx.doi.org/https://doi.org/10.1016/0550-3213(85)90448-1}{\emph{Nuclear
  Physics B} {\bf 252} (1985) 343 -- 356}.

\bibitem{Teitelboim:1983ux}
C.~Teitelboim, \emph{{Gravitation and Hamiltonian Structure in Two Space-Time
  Dimensions}},
  \href{http://dx.doi.org/10.1016/0370-2693(83)90012-6}{\emph{Phys. Lett.} {\bf
  126B} (1983) 41--45}.

\bibitem{nayak}
P.~Nayak, A.~Shukla, R.~M. Soni, S.~P. Trivedi and V.~Vishal, \emph{{On the
  Dynamics of Near-Extremal Black Holes}},
  \href{http://arxiv.org/abs/1802.09547}{{\tt 1802.09547}}.

\bibitem{Moitra:2019bub}
U.~Moitra, S.~K. Sake, S.~P. Trivedi and V.~Vishal, \emph{{Jackiw-Teitelboim
  Gravity and Rotating Black Holes}},
  \href{http://dx.doi.org/10.1007/JHEP11(2019)047}{\emph{JHEP} {\bf 11} (2019)
  047}, [\href{http://arxiv.org/abs/1905.10378}{{\tt 1905.10378}}].

\bibitem{Moitra:2018jqs}
U.~Moitra, S.~P. Trivedi and V.~Vishal, \emph{{Extremal and near-extremal black
  holes and near-CFT$_{1}$}},
  \href{http://dx.doi.org/10.1007/JHEP07(2019)055}{\emph{JHEP} {\bf 07} (2019)
  055}, [\href{http://arxiv.org/abs/1808.08239}{{\tt 1808.08239}}].

\bibitem{Ghosh:2019rcj}
A.~Ghosh, H.~Maxfield and G.~J. Turiaci, \emph{{A universal Schwarzian sector
  in two-dimensional conformal field theories}},
  \href{http://dx.doi.org/10.1007/JHEP05(2020)104}{\emph{JHEP} {\bf 05} (2020)
  104}, [\href{http://arxiv.org/abs/1912.07654}{{\tt 1912.07654}}].

\bibitem{Maldacena:2019cbz}
J.~Maldacena, G.~J. Turiaci and Z.~Yang, \emph{{Two dimensional Nearly de
  Sitter gravity}},  \href{http://arxiv.org/abs/1904.01911}{{\tt 1904.01911}}.

\bibitem{Cotler:2019nbi}
J.~Cotler, K.~Jensen and A.~Maloney, \emph{{Low-dimensional de Sitter quantum
  gravity}},  \href{http://arxiv.org/abs/1905.03780}{{\tt 1905.03780}}.

\bibitem{Saad:2019lba}
P.~Saad, S.~H. Shenker and D.~Stanford, \emph{{JT gravity as a matrix
  integral}},  \href{http://arxiv.org/abs/1903.11115}{{\tt 1903.11115}}.

\bibitem{Moitra:2021uiv}
U.~Moitra, S.~K. Sake and S.~P. Trivedi, \emph{{Jackiw-Teitelboim gravity in
  the second order formalism}},
  \href{http://dx.doi.org/10.1007/JHEP10(2021)204}{\emph{JHEP} {\bf 10} (2021)
  204}, [\href{http://arxiv.org/abs/2101.00596}{{\tt 2101.00596}}].

\bibitem{Gao:2021uro}
P.~Gao, D.~L. Jafferis and D.~K. Kolchmeyer, \emph{{An effective matrix model
  for dynamical end of the world branes in Jackiw-Teitelboim gravity}},
  \href{http://dx.doi.org/10.1007/JHEP01(2022)038}{\emph{JHEP} {\bf 01} (2022)
  038}, [\href{http://arxiv.org/abs/2104.01184}{{\tt 2104.01184}}].

\bibitem{Blommaert:2021fob}
A.~Blommaert, L.~V. Iliesiu and J.~Kruthoff, \emph{{Gravity factorized}},
  \href{http://arxiv.org/abs/2111.07863}{{\tt 2111.07863}}.

\bibitem{Stanford:2020wkf}
D.~Stanford, \emph{{More quantum noise from wormholes}},
  \href{http://arxiv.org/abs/2008.08570}{{\tt 2008.08570}}.

\bibitem{Stanford:2021bhl}
D.~Stanford, Z.~Yang and S.~Yao, \emph{{Subleading Weingartens}},
  \href{http://arxiv.org/abs/2107.10252}{{\tt 2107.10252}}.

\bibitem{Vilenkin:2021awm}
G.~Fanaras and A.~Vilenkin, \emph{{Jackiw-Teitelboim and Kantowski-Sachs
  quantum cosmology}},  \href{http://arxiv.org/abs/2112.00919}{{\tt
  2112.00919}}.

\bibitem{Vilenkin:1986cy}
A.~Vilenkin, \emph{{Boundary Conditions in Quantum Cosmology}},
  \href{http://dx.doi.org/10.1103/PhysRevD.33.3560}{\emph{Phys. Rev. D} {\bf
  33} (1986) 3560}.

\bibitem{Witten:2021nzp}
E.~Witten, \emph{{A Note On Complex Spacetime Metrics}},
  \href{http://arxiv.org/abs/2111.06514}{{\tt 2111.06514}}.

\bibitem{Suzuki:2021zbe}
K.~Suzuki and T.~Takayanagi, \emph{{JT gravity limit of Liouville CFT and
  matrix model}}, \href{http://dx.doi.org/10.1007/JHEP11(2021)137}{\emph{JHEP}
  {\bf 11} (2021) 137}, [\href{http://arxiv.org/abs/2108.12096}{{\tt
  2108.12096}}].

\bibitem{Castro:2021fhc}
A.~Castro, J.~F. Pedraza, C.~Toldo and E.~Verheijden, \emph{{Rotating 5D Black
  Holes: Interactions and deformations near extremality}},
  \href{http://dx.doi.org/10.21468/SciPostPhys.11.6.102}{\emph{SciPost Phys.}
  {\bf 11} (2021) 102}, [\href{http://arxiv.org/abs/2106.00649}{{\tt
  2106.00649}}].

\bibitem{Narayan:2020pyj}
K.~Narayan, \emph{{Aspects of two-dimensional dilaton gravity, dimensional
  reduction, and holography}},
  \href{http://dx.doi.org/10.1103/PhysRevD.104.026007}{\emph{Phys. Rev. D} {\bf
  104} (2021) 026007}, [\href{http://arxiv.org/abs/2010.12955}{{\tt
  2010.12955}}].

\bibitem{Hartman:2020khs}
T.~Hartman, Y.~Jiang and E.~Shaghoulian, \emph{{Islands in cosmology}},
  \href{http://dx.doi.org/10.1007/JHEP11(2020)111}{\emph{JHEP} {\bf 11} (2020)
  111}, [\href{http://arxiv.org/abs/2008.01022}{{\tt 2008.01022}}].

\bibitem{Balasubramanian:2020xqf}
V.~Balasubramanian, A.~Kar and T.~Ugajin, \emph{{Islands in de Sitter space}},
  \href{http://dx.doi.org/10.1007/JHEP02(2021)072}{\emph{JHEP} {\bf 02} (2021)
  072}, [\href{http://arxiv.org/abs/2008.05275}{{\tt 2008.05275}}].

\bibitem{Chen:2020tes}
Y.~Chen, V.~Gorbenko and J.~Maldacena, \emph{{Bra-ket wormholes in
  gravitationally prepared states}},
  \href{http://dx.doi.org/10.1007/JHEP02(2021)009}{\emph{JHEP} {\bf 02} (2021)
  009}, [\href{http://arxiv.org/abs/2007.16091}{{\tt 2007.16091}}].

\bibitem{Witten:2020ert}
E.~Witten, \emph{{Deformations of JT Gravity and Phase Transitions}},
  \href{http://arxiv.org/abs/2006.03494}{{\tt 2006.03494}}.

\bibitem{Stanford:2020qhm}
D.~Stanford and Z.~Yang, \emph{{Finite-cutoff JT gravity and self-avoiding
  loops}},  \href{http://arxiv.org/abs/2004.08005}{{\tt 2004.08005}}.

\bibitem{Betzios:2020nry}
P.~Betzios and O.~Papadoulaki, \emph{{Liouville theory and Matrix models: A
  Wheeler DeWitt perspective}},
  \href{http://dx.doi.org/10.1007/JHEP09(2020)125}{\emph{JHEP} {\bf 09} (2020)
  125}, [\href{http://arxiv.org/abs/2004.00002}{{\tt 2004.00002}}].

\bibitem{Mirbabayi:2020grb}
M.~Mirbabayi, \emph{{Uptunneling to de Sitter}},
  \href{http://dx.doi.org/10.1007/JHEP09(2020)070}{\emph{JHEP} {\bf 09} (2020)
  070}, [\href{http://arxiv.org/abs/2003.05460}{{\tt 2003.05460}}].

\bibitem{Cotler:2019dcj}
J.~Cotler and K.~Jensen, \emph{{Emergent unitarity in de Sitter from matrix
  integrals}}, \href{http://dx.doi.org/10.1007/JHEP12(2021)089}{\emph{JHEP}
  {\bf 12} (2021) 089}, [\href{http://arxiv.org/abs/1911.12358}{{\tt
  1911.12358}}].

\bibitem{Fernandes:2019ige}
K.~Fernandes, K.~S. Kolekar, K.~Narayan and S.~Roy, \emph{{Schwarzschild de
  Sitter and extremal surfaces}},
  \href{http://dx.doi.org/10.1140/epjc/s10052-020-08437-2}{\emph{Eur. Phys. J.
  C} {\bf 80} (2020) 866}, [\href{http://arxiv.org/abs/1910.11788}{{\tt
  1910.11788}}].

\bibitem{Hikida:2021ese}
Y.~Hikida, T.~Nishioka, T.~Takayanagi and Y.~Taki, \emph{{Holography in de
  Sitter Space via Chern-Simons Gauge Theory}},
  \href{http://arxiv.org/abs/2110.03197}{{\tt 2110.03197}}.

\bibitem{Aguilar-Gutierrez:2021bns}
S.~E. Aguilar-Gutierrez, A.~Chatwin-Davies, T.~Hertog, N.~Pinzani-Fokeeva and
  B.~Robinson, \emph{{Islands in multiverse models}},
  \href{http://dx.doi.org/10.1007/JHEP11(2021)212}{\emph{JHEP} {\bf 11} (2021)
  212}, [\href{http://arxiv.org/abs/2108.01278}{{\tt 2108.01278}}].

\bibitem{Forste:2021roo}
S.~Forste, H.~Jockers, J.~Kames-King and A.~Kanargias, \emph{{Deformations of
  JT gravity via topological gravity and applications}},
  \href{http://dx.doi.org/10.1007/JHEP11(2021)154}{\emph{JHEP} {\bf 11} (2021)
  154}, [\href{http://arxiv.org/abs/2107.02773}{{\tt 2107.02773}}].

\bibitem{Blommaert:2020tht}
A.~Blommaert, \emph{{Searching for butterflies in dS JT gravity}},
  \href{http://arxiv.org/abs/2010.14539}{{\tt 2010.14539}}.

\bibitem{PhysRevD.28.2960}
J.~B. Hartle and S.~W. Hawking, \emph{Wave function of the universe},
  \href{http://dx.doi.org/10.1103/PhysRevD.28.2960}{\emph{Phys. Rev. D} {\bf
  28} (Dec, 1983) 2960--2975}.

\bibitem{Stanford:2017thb}
D.~Stanford and E.~Witten, \emph{{Fermionic Localization of the Schwarzian
  Theory}}, \href{http://dx.doi.org/10.1007/JHEP10(2017)008}{\emph{JHEP} {\bf
  10} (2017) 008}, [\href{http://arxiv.org/abs/1703.04612}{{\tt 1703.04612}}].

\bibitem{Iliesiu:2020zld}
L.~V. Iliesiu, J.~Kruthoff, G.~J. Turiaci and H.~Verlinde, \emph{{JT gravity at
  finite cutoff}},
  \href{http://dx.doi.org/10.21468/SciPostPhys.9.2.023}{\emph{SciPost Phys.}
  {\bf 9} (2020) 023}, [\href{http://arxiv.org/abs/2004.07242}{{\tt
  2004.07242}}].

\bibitem{Moitra:2019xoj}
U.~Moitra, S.~K. Sake, S.~P. Trivedi and V.~Vishal, \emph{{Jackiw-Teitelboim
  Model Coupled to Conformal Matter in the Semi-Classical Limit}},
  \href{http://dx.doi.org/10.1007/JHEP04(2020)199}{\emph{JHEP} {\bf 04} (2020)
  199}, [\href{http://arxiv.org/abs/1908.08523}{{\tt 1908.08523}}].

\bibitem{Hennauxjt}
M.~Henneaux, \emph{Quantum gravity in two dimensions: Exact solution of the
  jackiw model},
  \href{http://dx.doi.org/10.1103/PhysRevLett.54.959}{\emph{Phys. Rev. Lett.}
  {\bf 54} (Mar, 1985) 959--962}.

\bibitem{Louis-Martinez:1993bge}
D.~Louis-Martinez, J.~Gegenberg and G.~Kunstatter, \emph{{Exact Dirac
  quantization of all 2-D dilaton gravity theories}},
  \href{http://dx.doi.org/10.1016/0370-2693(94)90463-4}{\emph{Phys. Lett. B}
  {\bf 321} (1994) 193--198}, [\href{http://arxiv.org/abs/gr-qc/9309018}{{\tt
  gr-qc/9309018}}].

\bibitem{Strobl:1993yn}
T.~Strobl, \emph{{Quantization and the issue of time for various
  two-dimensional models of gravity}},
  \href{http://dx.doi.org/10.1142/S0218271894000460}{\emph{Int. J. Mod. Phys.
  D} {\bf 3} (1994) 281--284}, [\href{http://arxiv.org/abs/hep-th/9308155}{{\tt
  hep-th/9308155}}].

\bibitem{Saad:2018bqo}
P.~Saad, S.~H. Shenker and D.~Stanford, \emph{{A semiclassical ramp in SYK and
  in gravity}},  \href{http://arxiv.org/abs/1806.06840}{{\tt 1806.06840}}.

\bibitem{Eynard:2007fi}
B.~Eynard and N.~Orantin, \emph{{Weil-Petersson volume of moduli spaces,
  Mirzakhani's recursion and matrix models}},
  \href{http://arxiv.org/abs/0705.3600}{{\tt 0705.3600}}.

\bibitem{Mirzakhani:2006fta}
M.~Mirzakhani, \emph{{Simple geodesics and Weil-Petersson volumes of moduli
  spaces of bordered Riemann surfaces}},
  \href{http://dx.doi.org/10.1007/s00222-006-0013-2}{\emph{Invent. Math.} {\bf
  167} (2006) 179--222}.

\bibitem{GOLDMAN1984200}
W.~M. Goldman, \emph{The symplectic nature of fundamental groups of surfaces},
  \href{http://dx.doi.org/https://doi.org/10.1016/0001-8708(84)90040-9}{\emph{Advances
  in Mathematics} {\bf 54} (1984) 200--225}.

\bibitem{wolperteleform}
S.~Wolpert, \emph{The fenchel-nielsen deformation}, {\emph{Annals of
  Mathematics} {\bf 115} (1982) 501--528}.

\bibitem{wolpertonweil}
S.~Wolpert, \emph{On the weil-petersson geometry of the moduli space of
  curves}, {\emph{American Journal of Mathematics} {\bf 107} (1985) 969--997}.

\bibitem{Anninos:2021eit}
D.~Anninos and B.~M\"uhlmann, \emph{{The semiclassical gravitational path
  integral and random matrices (toward a microscopic picture of a dS$_{2}$
  universe)}}, \href{http://dx.doi.org/10.1007/JHEP12(2021)206}{\emph{JHEP}
  {\bf 12} (2021) 206}, [\href{http://arxiv.org/abs/2111.05344}{{\tt
  2111.05344}}].

\bibitem{Almheiri:2014cka}
A.~Almheiri and J.~Polchinski, \emph{{Models of AdS$_{2}$ backreaction and
  holography}}, \href{http://dx.doi.org/10.1007/JHEP11(2015)014}{\emph{JHEP}
  {\bf 11} (2015) 014}, [\href{http://arxiv.org/abs/1402.6334}{{\tt
  1402.6334}}].

\bibitem{Jensen:2016pah}
K.~Jensen, \emph{{Chaos in AdS$_2$ Holography}},
  \href{http://dx.doi.org/10.1103/PhysRevLett.117.111601}{\emph{Phys. Rev.
  Lett.} {\bf 117} (2016) 111601}, [\href{http://arxiv.org/abs/1605.06098}{{\tt
  1605.06098}}].

\bibitem{Maldacena:2016upp}
J.~Maldacena, D.~Stanford and Z.~Yang, \emph{{Conformal symmetry and its
  breaking in two dimensional Nearly Anti-de-Sitter space}},
  \href{http://dx.doi.org/10.1093/ptep/ptw124}{\emph{PTEP} {\bf 2016} (2016)
  12C104}, [\href{http://arxiv.org/abs/1606.01857}{{\tt 1606.01857}}].

\bibitem{Engelsoy:2016xyb}
J.~Engelsöy, T.~G. Mertens and H.~Verlinde, \emph{{An investigation of
  AdS$_{2}$ backreaction and holography}},
  \href{http://dx.doi.org/10.1007/JHEP07(2016)139}{\emph{JHEP} {\bf 07} (2016)
  139}, [\href{http://arxiv.org/abs/1606.03438}{{\tt 1606.03438}}].

\bibitem{Harlow:2018tqv}
D.~Harlow and D.~Jafferis, \emph{{The Factorization Problem in
  Jackiw-Teitelboim Gravity}},  \href{http://arxiv.org/abs/1804.01081}{{\tt
  1804.01081}}.

\bibitem{Blommaert:2018iqz}
A.~Blommaert, T.~G. Mertens and H.~Verschelde, \emph{{Fine Structure of
  Jackiw-Teitelboim Quantum Gravity}},
  \href{http://arxiv.org/abs/1812.00918}{{\tt 1812.00918}}.

\bibitem{Yang:2018gdb}
Z.~Yang, \emph{{The Quantum Gravity Dynamics of Near Extremal Black Holes}},
  \href{http://dx.doi.org/10.1007/JHEP05(2019)205}{\emph{JHEP} {\bf 05} (2019)
  205}, [\href{http://arxiv.org/abs/1809.08647}{{\tt 1809.08647}}].

\bibitem{Blommaert:2019hjr}
A.~Blommaert, T.~G. Mertens and H.~Verschelde, \emph{{Clocks and Rods in
  Jackiw-Teitelboim Quantum Gravity}},
  \href{http://arxiv.org/abs/1902.11194}{{\tt 1902.11194}}.

\bibitem{Mertens:2017mtv}
T.~G. Mertens, G.~J. Turiaci and H.~L. Verlinde, \emph{{Solving the Schwarzian
  via the Conformal Bootstrap}},
  \href{http://dx.doi.org/10.1007/JHEP08(2017)136}{\emph{JHEP} {\bf 08} (2017)
  136}, [\href{http://arxiv.org/abs/1705.08408}{{\tt 1705.08408}}].

\bibitem{Kitaev:2017awl}
A.~Kitaev and S.~J. Suh, \emph{{The soft mode in the Sachdev-Ye-Kitaev model
  and its gravity dual}},
  \href{http://dx.doi.org/10.1007/JHEP05(2018)183}{\emph{JHEP} {\bf 05} (2018)
  183}, [\href{http://arxiv.org/abs/1711.08467}{{\tt 1711.08467}}].

\bibitem{Lin:2018xkj}
J.~Lin, \emph{{Entanglement entropy in Jackiw-Teitelboim Gravity}},
  \href{http://arxiv.org/abs/1807.06575}{{\tt 1807.06575}}.

\bibitem{Stanford:2019vob}
D.~Stanford and E.~Witten, \emph{{JT Gravity and the Ensembles of Random Matrix
  Theory}},  \href{http://arxiv.org/abs/1907.03363}{{\tt 1907.03363}}.

\bibitem{Mertens:2019tcm}
T.~G. Mertens and G.~J. Turiaci, \emph{{Defects in Jackiw-Teitelboim Quantum
  Gravity}},  \href{http://arxiv.org/abs/1904.05228}{{\tt 1904.05228}}.

\bibitem{Iliesiu:2019xuh}
L.~V. Iliesiu, S.~S. Pufu, H.~Verlinde and Y.~Wang, \emph{{An exact
  quantization of Jackiw-Teitelboim gravity}},
  \href{http://arxiv.org/abs/1905.02726}{{\tt 1905.02726}}.

\bibitem{Mertens:2019bvy}
T.~G. Mertens, \emph{{Towards Black Hole Evaporation in Jackiw-Teitelboim
  Gravity}},  \href{http://arxiv.org/abs/1903.10485}{{\tt 1903.10485}}.

\bibitem{Blommaert:2020seb}
A.~Blommaert, \emph{{Dissecting the ensemble in JT gravity}},
  \href{http://arxiv.org/abs/2006.13971}{{\tt 2006.13971}}.

\bibitem{Lin:2019qwu}
H.~W. Lin, J.~Maldacena and Y.~Zhao, \emph{{Symmetries Near the Horizon}},
  \href{http://arxiv.org/abs/1904.12820}{{\tt 1904.12820}}.

\bibitem{Maldacena:2018lmt}
J.~Maldacena and X.-L. Qi, \emph{{Eternal traversable wormhole}},
  \href{http://arxiv.org/abs/1804.00491}{{\tt 1804.00491}}.

\bibitem{Mefford:2020vde}
E.~Mefford and K.~Suzuki, \emph{{Jackiw-Teitelboim quantum gravity with defects
  and the Aharonov-Bohm effect}},
  \href{http://dx.doi.org/10.1007/JHEP05(2021)026}{\emph{JHEP} {\bf 05} (2021)
  026}, [\href{http://arxiv.org/abs/2011.04695}{{\tt 2011.04695}}].

\bibitem{Suh:2020lco}
S.~J. Suh, \emph{{Dynamics of black holes in Jackiw-Teitelboim gravity}},
  \href{http://dx.doi.org/10.1007/JHEP03(2020)093}{\emph{JHEP} {\bf 03} (2020)
  093}, [\href{http://arxiv.org/abs/1912.00861}{{\tt 1912.00861}}].

\bibitem{Saad:2019pqd}
P.~Saad, \emph{{Late Time Correlation Functions, Baby Universes, and ETH in JT
  Gravity}},  \href{http://arxiv.org/abs/1910.10311}{{\tt 1910.10311}}.

\bibitem{Xian:2019qmt}
Z.-Y. Xian and L.~Zhao, \emph{{Wormholes and the Thermodynamic Arrow of Time}},
  \href{http://dx.doi.org/10.1103/PhysRevResearch.2.043095}{\emph{Phys. Rev.
  Res.} {\bf 2} (2020) 043095}, [\href{http://arxiv.org/abs/1911.03021}{{\tt
  1911.03021}}].

\bibitem{Grumiller:2020fbb}
D.~Grumiller and R.~McNees, \emph{{Universal flow equations and chaos bound
  saturation in 2d dilaton gravity}},
  \href{http://dx.doi.org/10.1007/JHEP01(2021)112}{\emph{JHEP} {\bf 01} (2021)
  112}, [\href{http://arxiv.org/abs/2007.03673}{{\tt 2007.03673}}].

\bibitem{Haehl:2017pak}
F.~M. Haehl and M.~Rozali, \emph{{Fine Grained Chaos in $AdS_2$ Gravity}},
  \href{http://dx.doi.org/10.1103/PhysRevLett.120.121601}{\emph{Phys. Rev.
  Lett.} {\bf 120} (2018) 121601}, [\href{http://arxiv.org/abs/1712.04963}{{\tt
  1712.04963}}].

\bibitem{PhysRevResearch.2.043310}
A.~M. Garc\'{\i}a-Garc\'{\i}a and S.~Zacar\'{\i}as, \emph{Quantum
  jackiw-teitelboim gravity, selberg trace formula, and random matrix theory},
  \href{http://dx.doi.org/10.1103/PhysRevResearch.2.043310}{\emph{Phys. Rev.
  Research} {\bf 2} (Dec, 2020) 043310}.

\bibitem{Jafferis:2019wkd}
D.~L. Jafferis and D.~K. Kolchmeyer, \emph{{Entanglement Entropy in
  Jackiw-Teitelboim Gravity}},  \href{http://arxiv.org/abs/1911.10663}{{\tt
  1911.10663}}.

\bibitem{Constantinidis:2008ty}
C.~P. Constantinidis, O.~Piguet and A.~Perez, \emph{{Quantization of the
  Jackiw-Teitelboim model}},
  \href{http://dx.doi.org/10.1103/PhysRevD.79.084007}{\emph{Phys. Rev. D} {\bf
  79} (2009) 084007}, [\href{http://arxiv.org/abs/0812.0577}{{\tt 0812.0577}}].

\bibitem{Gaikwad:2018dfc}
A.~Gaikwad, L.~K. Joshi, G.~Mandal and S.~R. Wadia, \emph{{Holographic dual to
  charged SYK from 3D Gravity and Chern-Simons}},
  \href{http://arxiv.org/abs/1802.07746}{{\tt 1802.07746}}.

\bibitem{Anninos:2018svg}
D.~Anninos, D.~A. Galante and D.~M. Hofman, \emph{{De Sitter horizons \&
  holographic liquids}},
  \href{http://dx.doi.org/10.1007/JHEP07(2019)038}{\emph{JHEP} {\bf 07} (2019)
  038}, [\href{http://arxiv.org/abs/1811.08153}{{\tt 1811.08153}}].

\bibitem{Anninos:2020cwo}
D.~Anninos and D.~A. Galante, \emph{{Constructing AdS$_{2}$ flow geometries}},
  \href{http://dx.doi.org/10.1007/JHEP02(2021)045}{\emph{JHEP} {\bf 02} (2021)
  045}, [\href{http://arxiv.org/abs/2011.01944}{{\tt 2011.01944}}].

\bibitem{Griguolo:2021wgy}
L.~Griguolo, R.~Panerai, J.~Papalini and D.~Seminara, \emph{{Nonperturbative
  effects and resurgence in Jackiw-Teitelboim gravity at finite cutoff}},
  \href{http://dx.doi.org/10.1103/PhysRevD.105.046015}{\emph{Phys. Rev. D} {\bf
  105} (2022) 046015}, [\href{http://arxiv.org/abs/2106.01375}{{\tt
  2106.01375}}].

\bibitem{Horowitz:2002mw}
G.~T. Horowitz and J.~Polchinski, \emph{{Instability of space - like and null
  orbifold singularities}},
  \href{http://dx.doi.org/10.1103/PhysRevD.66.103512}{\emph{Phys. Rev. D} {\bf
  66} (2002) 103512}, [\href{http://arxiv.org/abs/hep-th/0206228}{{\tt
  hep-th/0206228}}].

\bibitem{Liu:2002ft}
H.~Liu, G.~W. Moore and N.~Seiberg, \emph{{Strings in a time dependent
  orbifold}},
  \href{http://dx.doi.org/10.1088/1126-6708/2002/06/045}{\emph{JHEP} {\bf 06}
  (2002) 045}, [\href{http://arxiv.org/abs/hep-th/0204168}{{\tt
  hep-th/0204168}}].

\bibitem{Liu:2002kb}
H.~Liu, G.~W. Moore and N.~Seiberg, \emph{{Strings in time dependent
  orbifolds}},
  \href{http://dx.doi.org/10.1088/1126-6708/2002/10/031}{\emph{JHEP} {\bf 10}
  (2002) 031}, [\href{http://arxiv.org/abs/hep-th/0206182}{{\tt
  hep-th/0206182}}].

\bibitem{Wolpert1981AnEF}
S.~A. Wolpert, \emph{An elementary formula for the fenchel-nielsen twist},
  {\emph{Commentarii Mathematici Helvetici} {\bf 56} (1981) 132--135}.

\bibitem{Susskind:2021esx}
L.~Susskind, \emph{{Entanglement and Chaos in De Sitter Holography: An SYK
  Example}},  \href{http://arxiv.org/abs/2109.14104}{{\tt 2109.14104}}.

\bibitem{whittaker_watson_1996}
E.~T. Whittaker and G.~N. Watson, \emph{A Course of Modern Analysis}.
\newblock Cambridge Mathematical Library. Cambridge University Press, 4~ed.,
  1996,
  \href{http://dx.doi.org/10.1017/CBO9780511608759}{10.1017/CBO9780511608759}.

\bibitem{Maldacena:2019ufo}
J.~Maldacena and A.~Milekhin, \emph{{SYK wormhole formation in real time}},
  \href{http://dx.doi.org/10.1007/JHEP04(2021)258}{\emph{JHEP} {\bf 04} (2021)
  258}, [\href{http://arxiv.org/abs/1912.03276}{{\tt 1912.03276}}].

\bibitem{COLEMAN1988643}
S.~Coleman, \emph{Why there is nothing rather than something: A theory of the
  cosmological constant},
  \href{http://dx.doi.org/https://doi.org/10.1016/0550-3213(88)90097-1}{\emph{Nuclear
  Physics B} {\bf 310} (1988) 643--668}.

\bibitem{Christensen:1977jc}
S.~M. Christensen and S.~A. Fulling, \emph{{Trace Anomalies and the Hawking
  Effect}}, \href{http://dx.doi.org/10.1103/PhysRevD.15.2088}{\emph{Phys. Rev.}
  {\bf D15} (1977) 2088--2104}.

\end{thebibliography}\endgroup

\end{document}